\documentclass[aps,rmp,notitlepage]{revtex4-1}
\makeatletter
\def\@bibstyle{apsrev\substyle@post}%
\def\@bibdataout@aps{%
\immediate\write\@bibdataout{%
 @CONTROL{%
  apsrev41Control,author="08",editor="1",pages="0",title="2",year="1",eprint="1"%
 }%
}%
\if@filesw
 \immediate\write\@auxout{\string\citation{apsrev41Control}}%
\fi
}%
\makeatother 

\usepackage{times}
\usepackage{epsfig}
\usepackage{url}
\usepackage{soul}
\usepackage{graphicx,graphics}
 \usepackage[usenames,dvipsnames]{xcolor}
\usepackage{amssymb,amsmath,amsthm,bbm}
\usepackage{latexsym}
\usepackage{subfigure}
\usepackage{wrapfig}
\usepackage[makeroom]{cancel}
\usepackage[justification=justified, small,bf]{caption}
\usepackage{framed}
\usepackage[colorlinks={true},breaklinks]{hyperref}  
\usepackage{ulem}
\hypersetup{
  colorlinks,
  citecolor=PineGreen,
  linkcolor=MidnightBlue,
  urlcolor=MidnightBlue}
\usepackage{breakurl}

\newcommand{\smallfrac}[2]{\mbox{ $\frac{#1}{#2} $ }}

\newcommand{\tr}{{\rm tr\thinspace}}
\newcommand{\bra}[1]{\ensuremath{\left\langle{#1}\right\vert}}
\newcommand{\ket}[1]{\ensuremath{\left\vert{#1}\right\rangle}}

\newcommand{\expt}[1]{\langle {#1} \rangle}
\newcommand{\op}[2]{\left |{#1}\right\rangle\! \!\left \langle {#2}\right |}

\newcommand{\expect}[1]{\ensuremath{\left\langle{#1}\right\rangle}}

\newcommand{\dg}{^{\dagger}}

\renewcommand{\Re}{\operatorname{Re}}

\newcommand{\half}{\smallfrac{1}{2}}
\newcommand{\SLH}{\textit{(S,L,H)}}
\newcommand{\vSLH}{({\bf S},{\bf L},\textit{H})}

\newcounter{aside}[section]

\renewcommand{\theaside}{\thesection.\arabic{aside}}

\usepackage[framemethod=TikZ]{mdframed}
\newenvironment{aside}[1][]{%
    \refstepcounter{aside}
    \begin{mdframed}[%
        frametitle={\centerline{Example \theaside\ : ~#1} },
   leftmargin=1em,rightmargin=1em,
        backgroundcolor=gray!20  
    ]%
}{\vspace{0.1cm}
    \end{mdframed} 
}

\newcommand{\nn}{\nonumber}
\newcommand{\eg}{\textit{e.g.},~}
\newcommand{\ie}{\textit{i.e.},~}
\newcommand{\etal}{\textit{et al.}~}
\newcommand{\emn}[1]{ \mathbbm{E}_{#1} }

\newcommand{\bin}{b_{{\rm in}}}
\newcommand{\bout}{b_{{\rm out}}}
\newcommand{\Bin}{B_{{\rm in}}}

\newcommand{\Isf}{I_{\rm SF}}

\def\R{\mathbb{R}} 
\def\C{\mathbb{C}}

\newcommand{\SNL}{Digital \& Quantum Information Systems,
Sandia National Laboratories, Livermore, CA 94550, USA}
\newcommand{\IQC}{Institute for Quantum Computing and Department of Applied Mathematics, University of Waterloo, Waterloo, Ontario N2L 3G1, Canada}
\newcommand{\PI}{Perimeter Institute for Theoretical Physics, 31 Caroline St. N, Waterloo, Ontario N2L 2Y5, Canada}
\newcommand{\HRL}{HRL Laboratories, LLC, 3011 Malibu Canyon Road, Malibu, CA 90265, USA}
\newcommand{\UQ}{Centre for Engineered Quantum Systems, School of Mathematics and Physics, University of Queensland, Brisbane, QLD, Australia}

\theoremstyle{definition} \newtheorem{SLHrule}{Rule}
\theoremstyle{definition} \newtheorem{Remark}{Remark}
\theoremstyle{definition} \newtheorem*{Remark*}{Remark}

\usepackage{cleveref} 
\crefformat{equation}{Eq.~(#2#1#3)} 
\crefmultiformat{equation}{Eqs.~(#2#1#3)}{ and~(#2#1#3)}{, (#2#1#3)}{ and~(#2#1#3)}
\Crefformat{equation}{Equation~(#2#1#3)}
\crefformat{section}{Sec.~#2#1#3}
\Crefformat{section}{Section~#2#1#3}
\crefformat{figure}{Fig.~#2#1#3}
\Crefformat{figure}{Figure~#2#1#3}
\crefformat{aside}{Example~#2#1#3}
\Crefformat{aside}{Example~#2#1#3}
\crefmultiformat{aside}{Examples~(#2#1#3)}{ and~(#2#1#3)}{, (#2#1#3)}{ and~(#2#1#3)}
\Crefmultiformat{aside}{Examples~(#2#1#3)}{ and~(#2#1#3)}{, (#2#1#3)}{ and~(#2#1#3)}
\crefformat{Remark}{Rem.~#2#1#3}
\Crefformat{Remark}{Remark~#2#1#3}
\crefformat{SLHrule}{SLH rule~#2#1#3}

\setcitestyle{numbers,square}
\begin{document}

\title{The SLH framework for modeling quantum input-output networks}

\author{Joshua Combes }
\email{joshua.combes@gmail.com}
\address{\IQC}
\address{\PI}
\address{\UQ}
\author{Joseph Kerckhoff}
\email{jakerckhoff@hrl.com}
\address{\HRL}
\author{Mohan Sarovar}
\email{mnsarov@sandia.gov}
\address{\SNL}

\date{\today}

\begin{abstract}
Many emerging quantum technologies demand precise engineering and control over networks consisting of quantum mechanical degrees of freedom connected by propagating electromagnetic fields, or \emph{quantum input-output networks}. 
Here we review recent progress in theory and experiment related to such quantum input-output networks, with a focus on the \emph{SLH framework}, a powerful modeling framework for networked quantum systems that is naturally endowed with properties such as modularity and hierarchy.
We begin by explaining the physical approximations required to represent any individual node of a network, \eg atoms in cavity or a mechanical oscillator, and its coupling to quantum fields by an operator triple \SLH. Then we explain how these nodes can be composed into a network with arbitrary connectivity, including coherent feedback channels, using algebraic rules, and how to derive the dynamics of network components and output fields. The second part of the review discusses several extensions to the basic SLH framework that expand its modeling capabilities, and the prospects for modeling integrated implementations of quantum input-output networks.
In addition to summarizing major results and recent literature, we discuss the potential applications and limitations of the SLH framework and quantum input-output networks, with the intention of providing context to a reader unfamiliar with the field.

\end{abstract}


\maketitle

\tableofcontents
 
\section{Introduction}\label{sec:intro}

Large scale communication and computing technologies are integral to modern life. The ubiquity of these technologies is largely due to the emergence of large scale integrated electronic circuits, which in turn are enabled by mature and powerful tools for electronic circuit design automation and analysis (\eg SPICE, gEDA). Quantum technologies for communication and computation are being developed on several physical platforms and have the potential to one day upend aspects of these foundational information processing tasks \cite{mikeandike}. Impressive progress in superconducting circuits, integrated quantum optics, integrated semiconductor devices, and integrated atom trapping devices \cite{Lad.Jel.etal-2010} have led to many demonstrations of high fidelity control, measurement and state preparation in assemblies of quantum coherent systems on all of these platforms. The next step, realizing large scale quantum technologies, will require the development of sophisticated modeling and analysis tools for quantum hardware that have the same enabling capabilities as existing electronic circuit design automation and analysis tools -- \eg these tools should incorporate useful abstractions, such as modularity, networks and hierarchy, and enable coordination between high-level software and algorithmic needs and low-level hardware design. 

In this review we summarize progress in developing a modeling framework, known as the \emph{SLH framework}, that is capable of incorporating many of the useful abstractions listed above, and thus has the potential to form the foundation for developing tools that enable design and analysis of large scale assemblies of quantum coherent systems. The SLH framework was initially developed to model specific quantum optical networks composed of localized components that interact via itinerant, quantum bosonic fields, which we will term \emph{quantum input-output networks} (QIONs). It is naturally a modular framework since each localized component is treated as a black box that the propagating fields scatter off with some pre-specified input-output behavior. In addition, it incorporates the quantum nature of the itinerant fields and any quantum dynamics in the localized components. An important aspect of using the SLH framework to model QIONs is that it enables control-theoretic analysis of such networks of quantum coherent systems, and thus facilitates the use of natural generalizations of techniques and tools from classical control theory. In particular, feedback and feedforward are naturally incorporated into the framework, which are sometimes difficult to capture within other modeling approaches. In fact, this framework is sometimes referred to in the literature as coherent quantum feedback control (CQFC) theory because the notion of modeling coherent feedback was central to its development. However, as the methodology and associated tools have developed over the past decade, it has grown into a more general framework for modeling complex networks of quantum or semi-classical systems interconnected via coherent fields (possibly involving feedback). For this reason we prefer the name SLH to refer to the framework, and the term quantum input-output network to refer to the physical apparatus being modeled.

\begin{figure}[]
\includegraphics[width=\columnwidth]{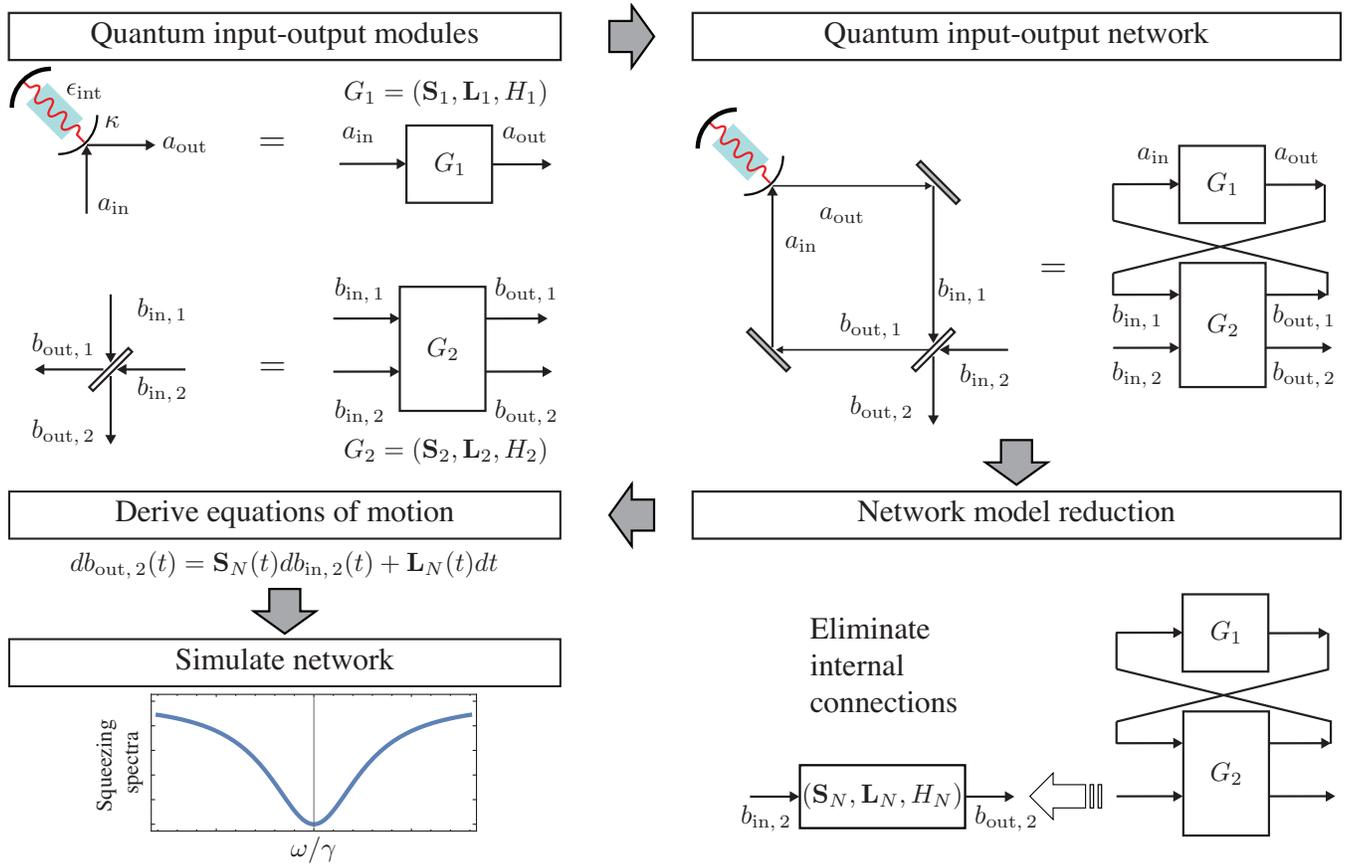}
\caption{An example of the type of workflow for modeling complex quantum networks that is enabled by the SLH framework. Each step of the workflow is covered in this review.}\label{fig:workflow}
\end{figure}

\cref{fig:workflow} depicts an example of the typical workflow enabled by the SLH framework. A major goal of this review is to enable a reader new to the field to step through this workflow themselves.  At the first step, individual quantum modules or components (\eg a nonlinear optical cavity and optical beamsplitter in the upper left) are specified by triples $(\mathbf{S},\mathbf{L},H)$ that completely capture how the component and its internal degrees of freedom interact with input (incident) and output (scattered) fields. As stand alone modules, this description is sufficient to systematically derive equations of motion for both the field modes and internal degrees of freedom. More complex behavior and functionality is generated by connecting these components into networks, where a connection is defined as routing the output field of a module into the input field of another module, \eg \cref{fig:workflow}  upper right. The SLH framework provides machinery to eliminate internal connections between the modules, resulting in a simpler, reduced model for the entire interconnected network, \eg \cref{fig:workflow} lower right.  This reduced network model has the same representation as the original components, in that it is described by a single $(\mathbf{S},\mathbf{L},H)$ triple that captures how internal degrees of freedom interact with the remaining incident and scattered fields (not the itinerant fields making internal connections, which have been eliminated).  As the entire network is now describable in the same format as its constituent components, equations of motion for the entire network may also be derived systematically from this model, such as the relationship between incident and scattered fields depicted in \cref{fig:workflow} lower left. A key benefit of the universal SLH triple description of network components is that it enables many aspects of this workflow to be automated and implemented in software, thus reducing the computational burden on the user and facilitating automated design and analysis. Some fairly complicated examples of this workflow have been examined in the literature, see \eg Refs. \cite{KercNurdNurd10,KercPavlPavl11}.

In addition to modeling networks for quantum technologies, the SLH framework is useful for modeling classical information processing networks where quantum noise in signals or components cannot be ignored. 
For example, at optical frequencies, an attojoule pulse contains just a few photons and thus photon shot noise cannot be ignored when modeling the interaction of such pulses with optical components. 
Low power classical information processing networks are becoming increasingly important as uncontrollable heat dissipation and power usage emerge as hard obstacles to scaling up the complexity of conventional integrated electronic circuits. Two important low power classical information processing applications that could benefit from SLH modeling are all-optical computers and optical interconnects \cite{Miller:2009bm, Miller:2000kb}.

The aim of this review is to present SLH techniques for modeling QIONs from a physics perspective. Some of the original derivations of QION results are heavily mathematical and in this review we attempt to motivate these results from physical considerations. Consequently, we will emphasize physical content and intuition, sometimes at the expense of rigorous mathematical proofs. The hope is that by emphasizing the physics, this will introduce SLH to a wider community and thus encourage wider adoption of this  useful methodology and associated tools. For alternative reviews of QION theory and SLH, we refer the reader to Refs. \cite{Kerckhoff:2011tn,Goug12,ZhanJame12}.

The structure of the remainder of the paper is as follows. We begin in  \cref{sec:applications} with an overview of the historical development of the SLH framework, as well as a survey of the literature on QION applications with an emphasis on some key advantages to utilizing coherent interconnects over classical signals, especially in the context of feedback systems.  \cref{sec:cascade} reviews the basic ingredients for the development of a theory of networked quantum systems: input-output theory and the notion of cascading outputs from one system into another.  \cref{sec:qsdes} summarizes the quantum stochastic calculus constructs that naturally describe propagating fields and their interaction with localized components in a QION. The idea is not to provide a formal treatment of quantum stochastic calculus but to lay out the essential concepts with physical insight.  \cref{sec:slhformalism} presents the main modeling constructs of the SLH framework, including the representation of components and rules for developing models of arbitrary networks of components. \cref{sec:linear} summarizes the treatment of a subclass of quantum networks, known as linear networks, for which a number of simplifications enable the formulation of powerful analysis tools.
 \cref{sec:generalizations} reviews a number of important extensions to the basic SLH framework that have been developed to expand the applicability of this modeling approach. The discussion in this section also reveals some of the key limitations of the SLH framework.  \cref{sec:challenges_integ_implem} examines the application of the SLH framework to the modeling of integrated QION implementations, and some of the associated issues. In particular, we discuss in detail the application of SLH to two leading integrated platforms for quantum technologies: silicon photonics and superconducting microwave circuits. Finally,  \cref{sec:outlook} concludes with a discussion of the outlook for QIONs and QION modeling using the SLH framework.

\section{History, applications and advantages}
\label{sec:applications}
In this section we summarize some of the historical developments and motivations in applied QION research, focusing on the literature related to the SLH framework.  The literature on applications of this theory is already quite large and while we cannot survey all of it, we will attempt to point out the results that we believe will be of most interest to applied physicists.  While this context is not strictly necessary for the technical discussions in the remainder of the paper, we hope that readers new to the field may find it illuminating.

\subsection{History}
A critical ingredient in many practical theories of complex networks is modularity. That is, one must be able to model each component of the circuit independently, and then be able to develop a model for a network of connected components without resorting to an approach where the entire network is modeled from first-principles. In electrical networks this is enabled by the lumped element treatment of electrical components, where each component is characterized by simple properties, \eg resistance, capacitance, and the properties derived from the connectivity of the network are modeled by circuit theory without having to resort to Maxwell's equations. However, such a treatment is not possible for optical circuits because a lumped element description of conventional optical or even integrated photonics components are invalid (optical wavelengths are much smaller than the size of optical circuit components). However, one can recover modularity by turning to scattering theory, and modeling the effect of each optical circuit component as a scattering transformation from input modes to output modes. 

While such a scattering approach is routine in quantum field theory (\eg Lehman-Symanzik-Zimmerman reduction), it was not until the formulation of \emph{input-output theory} (IOT) by Gardiner and Collett \cite{CollGard84,GardColl85} in the mid-1980s that such an approach became popular in quantum optics. As discussed more in section~\ref{sec:IOrelations}, the Gardiner-Collett input-output relations relate the far-field (asymptotically free) output fields in terms of transformations of the (asymptotically free) input fields, which models an interaction with a localized system. This in turn allows one to connect the output field of one localized system to the input field of another, in a \emph{cascade} configuration. This was originally realized by Gardiner \cite{Gard93} and Carmichael \cite{Carm93} in 1993, and is an important starting point for the SLH framework.  The physical models that result from cascading components with interconnecting coherent fields are described in more detail in Section \ref{sec:CascadeSys}. Much of the motivation for developing these models was to understand the dynamics of quantum optical systems driven by non-classical light, \eg \cite{Gard93, GardPark94, Cla.Pen.etal-2003}. There were a number of parallel developments in the 1980s closely related to the above formulations of input-output theory and cascaded quantum systems. In the physics community, Yurke and Denker~\cite{YurkDenk84} developed their own version of quantum network theory for quantized electrical circuits in the lumped-element approximation and a formalism for composing components based on electrical circuit theory. This theory has some similarities to input-output theory, and shares many of the same motivations. In addition, Kolobov and Sokolov~\cite{KoloSoko87} developed, from input-output theory, a special case of cascading. 
At the same time, a group of applied mathematicians independently discovered the mathematical objects that underlie the theory, namely quantum stochastic differential equations {as rigorously developed by Hudson and Parthasarathy}~\cite{Hudson:1984wt}, and analyzed their properties. See Refs.~\cite{Khol91,Part12,AccaLuVolo13,BoutHandJame07} for reviews of this mathematical physics approach. 
{The work of Hudson and Parthasarathy was motivated by the desire to write down a description of system-environment evolution (sometimes called a \emph{dilation}) sufficient for generating Markovian semigroup evolution of the system. This description resulted in a coupling of the system to broadband bosonic fields that can be interpreted as the input and output fields described by Gardiner, Collett and Carmichael. This mathematical description of localized systems interacting with itinerant bosonic fields developed by Hudson and Parthasarathy was vital for the development of the SLH framework.}

The first analysis of a QION that went beyond simple cascade interconnections dates to Wiseman and Milburn in 1994 \cite{WiseMilb94}.  They compared systems in which optical signals from a nonlinear cavity are either measured, producing a photocurrent that then controls the classical electro-modulation of the nonlinear crystal in the source cavity (measurement-based feedback control, which they termed electro-optic control), or directly routed back to the source crystal, modulating it optically and coherently (coherent feedback control, which they termed all-optical control).  They discovered that the measurement-based feedback scheme was fundamentally the same as the coherent scheme when the crystal couples to only a single optical quadrature (\eg the electric or magnetic field).  This equivalence breaks down, however, when the crystal couples to both quadratures, in which case the coherent feedback scheme yields optical squeezing dynamics unseen in the measurement-based scheme.  This early theoretical result suggested that quantum optical coherent feedback networks are more general than measurement-based feedback networks and that coherent feedback potentially offers new capabilities. Additionally Wiseman and Milburn's work contained the first instance of how to algebraically model coherent feedback.

Lloyd first coined the term ``coherent quantum feedback'' \cite{Lloy00} in his comparison of measurement-based feedback of a few-ion spin system with a fully-quantum feedback system in which one ion controls another set of ions.  Lloyd observed that through a feedback protocol, an ion-controller was capable of generating entanglement between ions that never interact directly, while the analogous classical controller was not.  This particular application was well-known, but Ref. \cite{Lloy00} was the first to articulate that this also demonstrates that coherent feedback schemes are fundamentally more capable than measurement-based feedback control of quantum systems. While both Refs. \cite{WiseMilb94} and \cite{Lloy00} are early examples of coherent feedback control, in this work we reserve the term quantum input-output network for systems such as Ref.~\cite{WiseMilb94} in which itinerant bosonic fields stitch together subsystems separated by many optical (or microwave) wavelengths. These are the systems that possess the type of modularity assumed by the SLH framework.  

A series of papers by Yanagisawa and Kimura in 2003 \cite{Yan.Kim-2003,Yan.Kim-2003a} marks the first attempt to formalize the modeling of quantum networks using control theoretic tools. The authors focused on linear quantum networks (see \cref{sec:linear} for a formal definition) and described representations for network components that enable the formulation of algebraic rules for composing multiple components in series, parallel, and feedback configurations. 

Then in 2009, building on the mathematical physics approach of Hudson and Parthasarathy (and somewhat motivated by IOT and Gardiner's cascaded systems theory), Gough and James developed the fundamentals of the SLH framework as a means to compose, model and analyze networks of arbitrary (not necessary linear) components \cite{GougJame09,GougJame09a}. Following this initial formulation, there has been rapid progress in extending the SLH framework in various directions, including: relaxing approximations to make it more widely applicable, integrating control-theoretic and systems-theoretic analysis tools into the framework, and developing practical tools and software to apply the framework. In addition, there have been numerous applications of the SLH framework to model and analyze quantum and semi-classical networks. In the remainder of this review, we will describe many of these extensions and applications.

\subsection{Applications and advantages}
An important motivation for developing the SLH framework was the desire to efficiently model and analyze coherent feedback networks, like the one considered in Ref.~\cite{WiseMilb94}, from a control theoretic-perspective (\eg \cite{dHel06,Jame08,NurdJamePete09,Jame10}).  Because of this heritage, much of the literature that employs SLH models tends to focus on networks with coherent feedback, and what makes it different from measurement-based control systems, even though the SLH framework is not restricted to modeling feedback systems.  For example, some of the earliest insights from applying the framework were that coherent feedback networks can fundamentally outperform measurement-based feedback networks with the same control goals.  This was first observed in Ref.~\cite{NurdJamePete09}, which considered a linear-quadratic-Gaussian (LQG) control problem of a linear quantum optical system with Gaussian noise inputs (\ie quantum noise on input optical fields).  Nurdin \etal formulated a coherent feedback controller design that achieved a lower cost (a quadratic function on the magnitude of both the plant'Âs and controller's fields) than the provably optimal, measurement-based feedback design \cite{NurdJamePete09}. Following on from this work, Ref.~\cite{Hame12} identified the physical mechanism that enables this superior performance, namely, that coherent feedback controllers are capable of simultaneously processing both non-commuting quadratures of plant output field.  By contrast, measurement-based controllers can only measure one quadrature of this output field, and thus necessarily inject additional noise when the control goal requires knowledge of both quadratures. The identification of this fundamental advantage echoes some of the insights in Ref.~\cite{WiseMilb94}.  Similarly, Yamamoto \cite{Yama14} derived a handful of no-go theorems proving that linear QIONs with coherent feedback are more capable than measurement-based systems. The tasks that coherent quantum feedback enables include the generation of backaction evading measurements, generation of quantum non-demolished variables, and generation of decoherence-free subsystems (when such systems or variables are absent in the plant without the controller) \cite{Yama14}.

Since the development of the SLH framework for QIONs, there have been a handful of experimental realizations demonstrating its validity.  To date, these experiments have been implemented in free space optical and superconducting microwave systems.  Ref.~\cite{Mabu08} initiated such experimental SLH studies, demonstrating the successful implementation of a fully coherent feedback loop between two free space optical cavities.  The controller cavity's dynamic response was systematically designed to reject broadband laser disturbances injected into the plant cavity.  This closed loop, all-optical system was completely linear and classical, but still encapsulated much of the new, analytic machinery of the SLH model of coherent feedback networks.  Extending this study, Refs.~\cite{Iida12,Cris13} demonstrated that experimental coherent feedback networks can also be implemented in the quantum regime, with linear free space optical networks that modified and enhanced the quantum squeezing of optical signals through coherent feedback, successfully modeled and designed using the SLH framework.  Similarly, Ref. \cite{Kerc12} (inspired by the proposal Ref.~\cite{Mabu11a}) demonstrated the validity of coherent control and SLH in a superconducting microwave context with classical, digital components.  Ref.~\cite{Zhif12} demonstrated digital logic gates using coherent feedback in a free space atomic and optical system.  Finally, Ref.~\cite{Kerc13} demonstrated that new capabilities (tunable quality factors of cavities) could be added to the superconducting electromechanical toolbox by interconnecting two standard, ``off-the-shelf'' modules in a coherent quantum network.

While the fundamental benefits of utilizing QION, especially in applications that require feedback, motivate further attention and understanding of these systems, actual, future adoption of QION feedback as a common practical technique will likely depend on technical advantages, costs, and conveniences.  Quantum systems are delicate and in their technological infancy, while classical controller technology tends to be mature, accessible, and commercially available.  Thus, coherent feedback control --- in which both plant and controller are immature technologies --- and complex QIONs in general, often do not provide a clear advantage today.  Most experimental applications of coherent feedback control today are more analogous to systems considered by Lloyd in Ref. \cite{Lloy00}, in which one ion controls another set of adjacent ions through near-field interactions, rather than the scattering networks exemplified by Ref. \cite{WiseMilb94}, that interact via asymptotically free fields.  High-profile examples of these include the first repeated quantum error correction in an ion trap \cite{Schi11}, the first demonstration of quantum error correction in superconducting circuits \cite{Reed12}, and the near-ubiquitous use of sideband cooling in quantum optomechanics \cite{Aspe14}.  These are instances in which adjacent quantum ions, circuits, or mechanical oscillators in cavities, are coupled directly or via near-field interactions (as opposed to via asymptotically free fields).    Coherent control techniques were used because of the technical expediency, rather than fundamental advantages.  While classical microprocessors are capable of far more sophisticated control laws, interfacing quantum plants with a large, remote measurement-based feedback controller proved more burdensome than coupling them to technologically-similar, quantum controllers that were readily integrated with their plants. More recently, Ref. \cite{Liu16} experimentally studied the various technical advantages, such as feedback latency and hardware overhead, of coherent- over measurement-based control in a superconducting microwave qubit system. The QIONs described in this article rely on itinerant bosonic fields to mediate interactions between quantum subsystems, with physical separations of many optical (or microwave) wavelengths.  As a consequence, the interactions between plant and controller are more separable and more modular, but are also more susceptible to decoherence in the itinerant fields (\eg photon loss) than the direct coherent interactions (mediated by virtual photons) \cite{Schi11,Reed12,Aspe14,Liu16}, and are thus more difficult to implement today.  However, as quantum engineering matures, QIONs have the potential to offer the modularity and flexibility of measurement-based controllers and the integrability, speed, and fundamental advantages of direct coherent interaction schemes. 

Today, proposals for QION systems typically emphasize applications in either quantum information systems, or classical information systems operating at such low energies that quantum effects become important. For example, Refs. \cite{KercNurdNurd10,KercPavlPavl11,Nguy15} build off of direct coherent control error correction experiments such as Refs. \cite{Schi11,Reed12} and continuous-time, measurement-based quantum error correction proposals \cite{Ahn02, Sar.Ahn.etal-2004} to construct autonomous, error corrected quantum memories.  Emphasizing the potential to combine the natural integrability and speed of coherent feedback control with the flexibility of a networked, modular design these proposals formulate quantum networks that implement quantum error corrected memories using the 3-qubit bit/phase flip code \cite{KercNurdNurd10}, the 9-qubit Bacon-Shor~\cite{KercPavlPavl11}, and a large-scale surface code ~\cite{Nguy15}.  Other quantum information applications include proposals to generate remotely entangled pairs of photons~\cite{Shi2015} or qubits~\cite{MotzHalpWang15}, with greater robustness to parameter uncertainty and interconnection loss by virtue of coherent feedback control.

Some researchers argue that despite the excitement over potential quantum information applications, QIONs could find more near term success in ultra-low energy classical information systems.  Integrated photonic circuits have shown increasing promise in the past two decades for classical information processing applications.  However, to be competitive with many electronic information systems, these photonic circuits will have to work at such low energies that fundamental quantum fluctuations (\eg photon shot noise) will contribute significantly to signal noise and uncertainties.  Many fundamental classical logic operations, such as latches and comparators, are based on integrated feedback at the hardware level, and extending such models to the ultra-low power regime demands a well-developed theory of coherent feedback networks.  Many interesting questions have been recently considered in this vein including how to design digital logic gates~\cite{Mabu11a}, suppress the effects of fundamental, quantum noise sources~\cite{Mabu11}, and accurately approximate the relevant dynamics of a large scale photonic network performing classical information tasks, but operating at quantum energy scales~\cite{Sant14}.  

{Finally, the concept of QIONs and the SLH modeling framework has proven to be useful for modeling fundamental properties of materials and light-matter interactions. For example, Kockum \etal have considered how to model the light-matter interaction, using the SLH framework, when atoms cannot be treated as pointlike objects \cite{FrisDelsJoha14}. This treatment has also been experimentally studied in Ref. \cite{GustArefKock14}. The SLH framework has also been used to construct a QION for performing QND detection of a propagating microwave photon \cite{FanJohaComb14}. Another example is recent work by Brod {\em et al.}~\cite{BrodComb16,BrodCombGeaB16}, which found a counter example to the claim that single photon cross Kerr nonlinearites cannot aid quantum computation \cite{Shap06,GeaB10}. Brod {\em et al.} used the SLH framework to construct a QION that models a distributed Kerr medium using a finite number of cross Kerr interaction sites.} 

\section{Input-output theory and cascaded quantum systems}
\label{sec:cascade}
We begin the technical portion of this review by summarizing early work on modeling quantum optical networks, which can be seen as the first steps towards SLH as a more general theory of QION. Throughout this review we work in units such that $\hbar=1$.

\subsection{Input-output relations}\label{sec:IOrelations}
\begin{figure}[]
\includegraphics[width=\columnwidth]{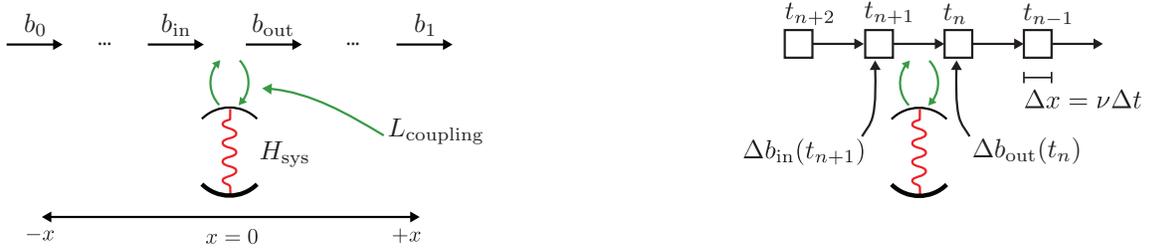}
\caption{ (Left) Schematic showing the relationship between the bosonic fields defined in the text and an arbitrary localized system, in this case depicted as a cavity located at $x=0$. (Right) A discrete time input-output model ~\cite{Brun02,Yana10,BoutHandJame09}.  Here the boxes representing discrete input field modes that are labeled by the time they interact with the localized system \eg $\Delta \bin(t_n)$, while the arrows indicate the directionality of the field propagation. The modes propagate at speed $\nu$ and have length $\Delta x = \nu \Delta t$. As the interaction is effectively instantaneous the field mode $\Delta \bin(t_n)$ gets mapped to the output field as depicted at time $t_n$. After the interaction the system has inprinted information on the field mode via the usual input-output relation $\Delta \bout(t_n) = \Delta \bin(t_n)+ L(t_n) \Delta t$ . 
}\label{fig:simple_inout}
\end{figure}

The starting point to modeling QIONs is quantum IOT, which captures the relation between asymptotic (or far-field) input and output fields that interact with a localized system, see \cref{fig:simple_inout} (a). We begin with a fully Hamiltonian description of a quantized bosonic field interacting with an arbitrary localized system:
\begin{align}
H_{\rm tot} &= H_{\rm sys} + H_{\rm B} + H_{\rm int}\label{eq:HSysBath}
\end{align} 
where $H_B$ is the Hamiltonian for the bosonic field in isolation, $H_{\rm sys}$ is the Hamiltonian for the localized system's internal dynamics, and $H_{\rm int}$ represents the interaction between the localized system and the field.  While $H_{\rm sys}$ may remain unspecified for the moment, the field has a dense spectrum Hamiltonian in the lab frame of
\begin{align}
H_{\rm B} = \int_{0}^\infty d\omega \omega b\dg(\omega) b(\omega)
\end{align}
where $b(\omega)$ are bosonic annihilation operators for the quantized field modes with units of $\sqrt{\text{time}}$, and satisfying canonical commutation relations $[b(\omega), b\dg(\omega')]=\delta(\omega-\omega')$.  Usually, this bosonic field models a guided-wave, optical-frequency electromagnetic field mode, but IOT has also found success in modeling other systems such as microwave electrical signals and vibrational phononic modes with sufficiently low loss and at sufficiently low temperatures \footnote{We note that there are formulations of IOT for fermionic itinerant quanta \cite{Milb01,Gard04,ApplHuds84}, and even a unified formal treatment of bosonic and fermionic theory \cite{HudsPart86}, however we will only focus on the bosonic case in this review.}. The interaction term is assumed to take linear form
\begin{align}\label{eq:Hint_linear}
H_{\rm int} &= i\int_{0}^\infty d\omega \kappa(\omega) [b(\omega) + b\dg(\omega)][c - c\dg],
\end{align}
where $c$ is a system operator and $\kappa(\omega)$ is a coupling strength between the system and field.  This form of interaction is very common, \eg in the dipole approximation of light-matter interaction \cite{GardZoll-2004}. Note that we are suppressing tensor product operators for conciseness, \ie $b\dg c \equiv b\dg\otimes c$, etc.

The first assumption of IOT is that the system and bath are weakly coupled. This assumption implies that we can approximate $H_{\rm int}$ with a simpler form. To explain this approximation, we first transform the Hamiltonian into an interaction frame with respect to the bare Hamiltonian $H_0 = H_{\rm sys} + H_{\rm B}$. In this frame the interaction Hamiltonian becomes:
\begin{align}
\tilde{H}_{\rm int}(t) &= i\int_{0}^\infty d\omega \kappa(\omega) [b(\omega)e^{-i\omega t} + b\dg(\omega)e^{i\omega t}][\tilde{c}(t) - \tilde{c}(t)\dg],
\end{align}
where the tilde denotes operators in the interaction frame. We require that this interaction frame system operator take the form $\tilde{c}=e^{iH_0t}c e^{-iH_0t} = c e^{\pm i\Omega t}$, for some frequency $\Omega >0$. The most obvious set of operators that satisfy this relation are operator off-diagonal in the eigenbasis of $H_{\rm sys}$; \ie $c \propto \ket{\epsilon_i}\bra{\epsilon_j}$, where $\ket{\epsilon_{i/j}}$ are eigenstates of $H_{\rm sys}$. Given this form,
\begin{align}
\label{eq:int_ham}
\tilde{H}_{\rm int}(t) &= i\int_{0}^\infty d\omega \kappa(\omega) [cb\dg(\omega) e^{i(\omega-\Omega)t}-c\dg b(\omega)e^{-i(\omega-\Omega)t}] + i\int_{0}^\infty d\omega \kappa(\omega) [cb(\omega) e^{-i(\omega+\Omega)t}-c\dg b\dg(\omega)e^{i(\omega+\Omega)t}],
\end{align}
{We now make the \emph{rotating wave approximation} (RWA) and drop the counter-rotating terms (the second integral in the above expression) since the oscillating integrands imply that their contribution to the evolution of the system in time will be negligible. In addition, we observe that the first integral is dominated by terms around $\omega \approx \Omega$, and hence the interaction Hamiltonian can be well-approximated by:
\begin{align}
\tilde{H}_{\rm int}(t) \approx i\int_{\Omega-\zeta}^{\Omega+\zeta}d\omega \kappa(\omega) [cb\dg(\omega) e^{i(\omega-\Omega)t} - c\dg b(\omega)e^{-i(\omega-\Omega)t} ],
\end{align}
for some frequency range around $\Omega$ determined by $\zeta$. 

An additional assumption of IOT is that the coupling amplitude, $\kappa(\omega)$, has a sufficiently constant magnitude over the range $[\Omega-\zeta, \Omega+\zeta]$, so that we can approximate $\kappa(\omega)$ in the above expression as $\sqrt{\gamma/2\pi}$. This is known as the \emph{Markov approximation} since it ensures that the system couples uniformly to a broad band of field frequency modes, causing the field to act as a ``memoryless'' bath. This approximation is typically very good in systems with relatively weak system-bath interactions $\kappa(\omega)\ll\Omega$ such that the system-bath interaction is narrowband. We additionally assume that the dynamics induced by the interaction Hamiltonian is on timescales that are long compared to $1/\zeta$, due to the weakness of the interaction and large range of bath frequencies over which the Markov approximation is valid. Under this assumption we can take $\zeta\rightarrow \infty$ to yield:
\begin{align}
\tilde{H}_{\rm int}(t) \approx i\sqrt{\frac{\gamma}{2\pi}}\int_{-\infty}^{\infty}d\omega [cb\dg(\omega) e^{i(\omega-\Omega)t} - c\dg b(\omega)e^{-i(\omega-\Omega)t} ],
\end{align}
Finally, since most of the dynamics in this interaction frame is centered around the frequency $\Omega$, it's natural to transform the field degrees of freedom into a frame rotating at this frequency. Mathematically, this involves transforming into a frame defined by $U(t) = e^{i\left(\int d\omega (\Omega-\omega)b\dg(\omega)b(\omega)\right) t}$, and hence $b(\omega) \rightarrow b(\omega)e^{-i(\Omega-\omega)t}$. Performing this change of frame, we arrive at the final form of the interaction Hamiltonian:
\begin{align}
\tilde{H}_{\rm int}(t) \approx i\sqrt{\frac{\gamma}{2\pi}}\int_{-\infty}^{\infty}d\omega [cb\dg(\omega) - c\dg b(\omega) ].
\label{eq:final_int}
\end{align}
Also, due to the transformation of the field degrees of freedom into the rotating frame defined by $\Omega$, the field Hamiltonian becomes:
\begin{align}
	\tilde{H}_{\rm B} = \int_0^\infty d\omega (\omega-\Omega)b\dg(\omega)b(\omega) \approx  \int_{-\infty}^\infty d\omega (\omega-\Omega)b\dg(\omega)b(\omega),
	\label{eq:final_bath}
\end{align}
where in the second approximation we have formally extended the lower limit to $-\infty$ for later convenience. Note that this does not impact the dynamics of the system significantly since $\Omega \gg 0$ and as argued above, field modes far from $\Omega$ have negligible interaction with the system. Often we write the bath Hamiltonian as simply $\int d\omega \omega b\dg(\omega)b(\omega)$ with the understanding that these frequencies are all detunings from $\Omega$.
}

\begin{Remark}[\bf Linear coupling]
	Although restrictive, the linear form of the interaction in \cref{eq:Hint_linear} is consistent with the RWA and the assumption that $H_{\rm int}$ is very weak compared to $H_{\rm sys}$ and the relevant spectral components of $H_B$.  This is a particularly good approximation in optical systems in which $H_{\rm sys}$ and $H_B$ operate at 100s of THz and $H_{\rm int}$ typically has GHz or lower energy scales.  As a consequence, system-bath coupling Hamiltonians that are nonlinear in the bath operators (\eg $i(b\dg(\omega)^2c-b(\omega)^2c\dg$)) are ignored.  In practice, while such coupling interactions may be present, they are typically dominated by linear interactions such as \cref{eq:Hint_linear} in this weak coupling limit.  
\end{Remark}

\begin{Remark}[\bf Off-diagonal coupling]
	The other restriction in the above derivation is the demand that the elements in the coupling operator (\ie $c - c\dg$ above) are off-diagonal operators with respect to the system Hamiltonian. Although this form of the operator is restrictive, it is fairly common for light matter interactions, \eg the dipole approximation to the minimal coupling Hamiltonian from QED. To understand this restriction further, note that we can expand any system operator as $X = \sum_{ij} x_{ij}\ket{\epsilon_i}\bra{\epsilon_j}$, with $x_{ij} = \bra{\epsilon_i}X\ket{\epsilon_j}$, and $\ket{\epsilon_i}$ being eigenstates of $H_{\rm sys}$. In the interaction frame defined above, all components in this sum pick up rotating factors as required for the above derivation, except for the diagonal components $\ket{\epsilon_i}\bra{\epsilon_i}$ (or off-diagonal components if $\ket{\epsilon_i}$ and $\ket{\epsilon_j}$ are degenerate). The presence of such terms makes the approximations above difficult; in particular, in the presence of such terms the integrals in \cref{eq:int_ham} are dominated by terms around $\omega\approx 0$, which makes the extension of the lower limit of these integrals to $-\infty$ invalid because the unphysical terms with $\omega<0$ significantly influence dynamics. More general derivations within the Markov approximation that include diagonal terms in the system field coupling are possible, see {Refs.~\cite{Accardi:1624938,Gough:2006gu}.}
\end{Remark}

{
\begin{Remark}[\bf The interaction frame]
We derived the approximate interaction Hamiltonian in an interaction frame defined with respect to $H_{\rm sys} + \Omega \int b\dg(\omega)b(\omega)$. From here onwards we will dispense with the tilde notation since all operators will either be in this frame or the Heisenberg picture, with any exceptions specifically noted.
A consequence of working in this frame is that there is no Hamiltonian for the localized system in the interaction frame. However, it is common to see the interaction frame defined with respect to only some components of the system Hamiltonian, and in such cases, one would have a system Hamiltonian remaining in this frame. An example is if the system is a two-level atom, and $H_{\rm sys} = -\frac{\nu}{2}\sigma_z$ Then, if one chooses to define the interaction frame with respect to $-\frac{\nu'}{2}\sigma_z + \Omega \int b\dg(\omega)b(\omega)$, then the full system Hamiltonian in this frame would be composed of \cref{eq:final_int}, \cref{eq:final_bath}, and the term $-\frac{(\nu-\nu')}{2}\sigma_z$ (a detuning). In the following, any $H_{\rm sys}$ is assumed to be similarly defined, as the portion of the system Hamiltonian remaining after the interaction picture transformation.
\end{Remark}
}

Using these approximate Hamiltonians, Gardiner and Collett derive a quantum Langevin equation for the evolution of the arbitrary system operator $a$ in the Heisenberg picture \cite{GardColl85}:
\begin{align}
\dot{a}(t) &\approx - i[a(t), H_{\rm sys}]
- \left( [a(t),c\dg(t)] \left( \frac{\gamma}{2}c(t) + \sqrt{\gamma}\bin(t)\right)
  -\left(\frac{\gamma}{2}c\dg(t) + \sqrt{\gamma}\bin\dg(t)\right)[a(t),c(t)] \right). 
  \label{eq:Langevin1}
\end{align}
Note that here $a(t)$ is shorthand for $a(t)\otimes I_{\rm field}$. 
This equation quantifies the influence of an input field $\bin(t)$ on the dynamics of the system. The definition of the input field in terms of the field mode operators is 
\begin{align}
\bin(t) = \frac{1}{\sqrt{2\pi}} \int_{-\infty}^\infty d\omega e^{-i\omega(t-t_0)}b_0(\omega),
\label{eq:bin}
\end{align}
where $b_0(\omega)$ is the value of $b(\omega)$ at the initial time $t_0$. Colloquially, $\bin(t)$ is the portion of the field \emph{incident} on the localized system at time $t$. Canonical commutation relations for $b_0(\omega)$ imply that the commutation relations for $\bin(t)$ are also singular: 
\begin{align}
[\bin(t), \bin\dg(t')] = \delta(t-t').
\end{align} 
Note that the units of $\bin(t)$ are $(\text{time})^{-1/2}$.  The operators $\bin(t)$ and $\bin^\dagger(t)$ are called {\em quantum white noise operators} by analogy with classical stochastic processes where $\delta$-correlation in time implies a flat noise spectral density, \ie white noise. The singular commutation relation of the operators $\bin(t)$ and $\bin^\dagger(t)$ is  mathematically problematic, and to remedy this smoother quantum noise increments, \eg $dB_t= \int_t^{t+dt}ds\,\bin(s)$, will be introduced in \cref{sec:qsdes}, however we will not need these in this section.

It is often convenient to work with $\bin(\omega)$, the frequency domain representation of $\bin(t)$, which are related by
\begin{align} \label{eq:binFourier}
\bin(t) = \frac{1}{\sqrt{2\pi}}\int d\omega \, \bin (\omega)e^{-i\omega t}.
\end{align}
One can also define an output field as
\begin{align}
\bout(t) = \frac{1}{\sqrt{2\pi}} \int_{-\infty}^\infty d\omega e^{-i\omega(t-t_1)}b_1(\omega),\label{eq:bout}
\end{align}
where $b_1(\omega)$ is the value of $b(\omega)$ (in the Heisenberg picture) evolved to $t_1$ with $t_1>t$.  {$\bout(t)$ is the field at time $t$ immediately after its interaction with the localized system (more precisely, the field at some later time $t_1$, after its interaction with the localized system at time $t$, propagated back freely to time $t$)}.  Or, more colloquially, the portion of the field \emph{scattered} by the localized system at time $t$. Gardiner and Collett calculate the following critical relation between the input and output fields and the system \cite{GardColl85}
\begin{align}\label{eq:Binout}
\bout(t) = \bin(t) + \sqrt{\gamma}c(t).
\end{align}
This is typically called an input-output relation and models the effective scattering of the input modes to output modes through interaction with the localized system. Crucially, this relation allows one to calculate properties of the scattered output field once the input field and dynamics of the system operator, $c(t)$, are known.

The spatial properties of the itinerant field have not been emphasized in the above calculation. One can also derive the input-output relation in a space-time representation of the itinerant field \cite[Chap. 3.2]{GardZoll-2004}, in which case an excitation of the field, initially at position $x = -|x_0|$ at time $t_0$, propagates to the localized system, located at $x=0$, in time $t$. Loosely, excitations to the left of $x=0$ are inputs and excitations to the right are outputs, as illustrated in discrete time in \cref{fig:simple_inout}. The inputs and outputs are related by the boundary condition given in \cref{eq:Binout}.

\begin{aside}[IOT model for a single mode of an empty resonator]\label{eg:cav}
Consider a resonator (\eg a single-sided optical cavity, photonic microdisk, or microwave LC resonator) that supports a resonance with the center frequency $\omega_c$ that is detuned from the reference frequency $\Omega$ by $\Delta = \omega_c-\Omega$.  The resonator also decays into an itinerant guided wave or transmission line mode with energy decay rate $\gamma$.  Here, 
\begin{align}
H_{\rm sys} &= \Delta a\dg a
\end{align}
where $a$ is the annihilation operator of the resonator mode.  $H_{\rm sys}$ expresses that each photon adds energy $\Delta$ to the system in this frame. 
The interaction Hamiltonian (in the RWA) in this case is
\begin{align}
H_{\rm int} &= i\sqrt{\gamma/2\pi}\int_{-\infty}^\infty d\omega[ab\dg(\omega)-a\dg b(\omega)]
\end{align}
where $b(\omega)$ are annihilation operators for transmission line modes. $H_{\rm int}$ expresses that the annihilation of a resonator photon creates an itinerate photon, and vice versa.  From these definitions, applying \cref{eq:Langevin1,eq:Binout} gives the following equation of motion for the cavity mode and input-output relation: 
\begin{align}
\dot{a}(t) &= -(i\Delta+\frac\gamma2)a(t)-\sqrt{\gamma}b_{\rm in}(t) \label{eq:CavHeis} \\
b_{\rm out}(t) &= b_{\rm in}(t)+\sqrt{\gamma}a(t).
\label{eq:CavIOT}
\end{align}
\end{aside}

In much of the following, an important mathematical object will be the unitary propagator for the system, which generates evolution of any system operator (in the Heisenberg picture), $a(t) = U\dg(t) a U(t)$. For the dynamics described above, the propagator takes the form:
\begin{align}
U(t) = \mathcal{T} \exp \left\{ \int_{t_0}^t ds\left( -iH_{\rm sys}+ (L \bin\dg(s) - L\dg \bin(s)) \right) \right\}, \quad \textrm{with} \quad U(t_0) = \Isf.
\label{eq:U_single}
\end{align}
Here $\mathcal{T}$ denotes time ordering, $\Isf$ is shorthand for the identity operator on the system and field degrees of freedom (\ie $I_{\rm system}\otimes I_{\rm field}$), and we introduce the \emph{coupling operator} $L = \sqrt{\gamma}c$ (note that while $L$ is commonly referred to as an operator, it has units of $\rm{time}^{-1/2}$). One calculates the generator of this unitary, $K(t)$, as

\begin{align}
\label{eq:U_gen}
\dot{U}(t)  &= K(t) U(t) = \left[ -iH_{\rm sys} + (L\bin\dg(t) - L\dg \bin(t))\right] U(t),
\end{align} 
We prove this form for the propagator by calculating $\dot{a}(t) = \dot{U}\dg(t) a U(t) + U\dg(t) a \dot{U}(t)$, and showing agreement with \cref{eq:Langevin1}:
\begin{align}
\dot{a}(t) &= U\dg(t) \Big(-i[a,H_{\rm sys}]\Big)U(t) + U\dg(t) \bin\dg(t)  \Big( [a,L]\Big) U(t) -U\dg(t) \Big( [a,L\dg]\Big) \bin(t) U(t)
\label{eq:adot1}
\end{align} 
To proceed, we use the following identity \cite{GardColl85, GougJame16}:
\begin{align}
\label{eq:useful_id}
\bin(t) U(t) = U(t) \bin(t) + \frac{1}{2}L U(t). 
\end{align}
This identity is proven in Ref. \cite{GougJame16}, but we summarize the proof here for completeness. We begin by considering the commutator of the input process with the unitary propagator at the same time:

\begin{align}
[\bin(t), U(t)] &= \left[\bin(t), \int_0^t K(s) U(s) ds \right]
= \int_0^t [\bin(t), K(s)]U(s) ds + \int_0^t K(s) [\bin(t), U(s)] ds.
\end{align}
Using the form of the generator given in~\cref{eq:U_gen}, we get:
\begin{align}
\label{eq:id_1}
\int_0^t [\bin(t), K(s)]U(s) ds &= \int_0^t L U(s) \delta(t-s) ds = \frac{1}{2} L U(t).
\end{align}
Furthermore, 
\begin{align}
\label{eq:id_2}
\int_0^t K(s) [\bin(t), U(s)] ds = 0,
\end{align}
since the $U(s)$ in this integrand depends only on input fields at times $s < t$. Putting \cref{eq:id_1,eq:id_2} together yields the identity in \cref{eq:useful_id}. Finally, substituting this identity into \cref{eq:adot1}, and recalling that $L=\sqrt{\gamma}c$, exactly yields the Langevin equation in \cref{eq:Langevin1}, thus validating the form of the unitary propagator given above.

For those readers who want a more detailed account of input-output theory we recommend the following references: the original papers \cite{GardColl85,Gard93}, chapters 3, 5, and 11 of Ref. \cite{GardZoll-2004}, and the Appendix of Ref. \cite{ClerDevoGirv10}.

\subsection{Cascaded systems}\label{sec:CascadeSys}
Given input-output relations for localized components we can think about what happens when  the output field from one quantum optical system is routed into another. This problem was examined by Gardiner \cite{Gard93} and Carmichael \cite{Carm93} using different techniques.  Both authors considered a system in which the output of a driven optical cavity feeds into another, with both cavities containing separate, nonlinear quantum subsystems (\eg strongly coupled atoms).  In the following, we will summarize the results derived by Gardiner since they relate most directly to generalizations that will follow. 

\begin{figure}[]
\includegraphics[width=\columnwidth]{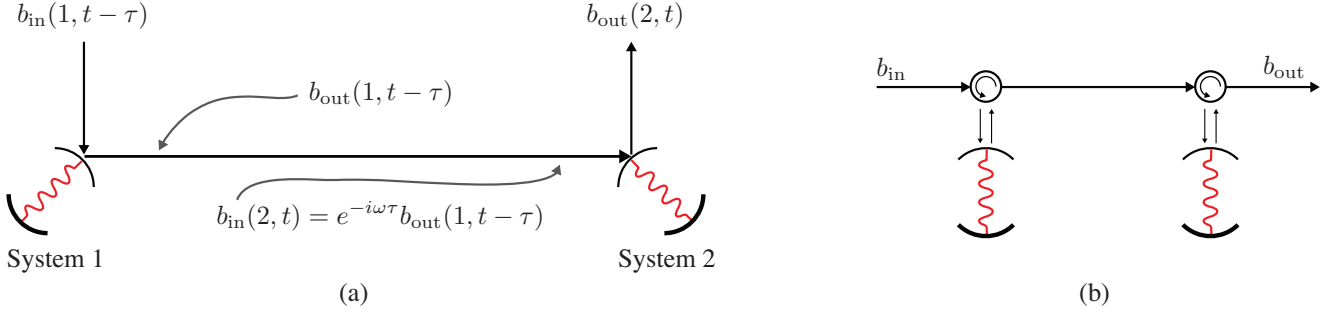}
\caption{The cascading of the output of one cavity into another. The top mirrors of each cavity are partially transmitting, while the bottom mirrors are perfectly reflecting and therefore each cavity has only one ``input" port. System 1 (2) is the cavity on the left (right). (a) is an idealized schematic, while (b) is a more experimentally accurate one that explicitly shows the circulators required to enforce unidirectional propagation of fields. The cavities could contain individual atoms or atomic ensembles, in which case the dynamics can become nonlinear. $b_{\rm in}(i,t)$ denotes the incident field interacting with system $i$ at time $t$, and $b_{\rm out}(i,t)$ denotes the scattered field that interacted with system $i$ at time $t$.}\label{fig:coupled_cavities}
\end{figure}

Consider the setup in  \cref{fig:coupled_cavities} where the reflected output of a single-port cavity is fed into the input port of another such cavity. This is referred to as ``cascading" the output from one system into another and is distinct from simple coupling because the probe field ($\bin(t)$) is assumed to be unidirectional with no back scattering from the second cavity (this can be ensured by inserting a circulator between the two cavities, for example). Gardiner begins by writing the Hamiltonian for the intra-cavity degrees of freedom and the propagating field (in an interaction frame, and assuming the weak coupling and Markov approximations discussed in \cref{sec:IOrelations}):
\begin{align}
H = H_{\rm sys,1} +H_{\rm sys,2} + \int_{-\infty}^\infty d\omega \omega b\dg(\omega) b(\omega) &+ i\sqrt{\gamma_1/2\pi} \int_{-\infty}^\infty d\omega \left[ c_1 b\dg(\omega) - c_1\dg b(\omega)\right] \nn \\
&+ i\sqrt{\gamma_2/2\pi} \int_{-\infty}^\infty d\omega \left[ c_2 b\dg(\omega)e^{-i\omega \tau} - c_2\dg b(\omega)e^{i\omega \tau}\right],
\label{eq:coupled_cavities_fullham}
\end{align}
where $H_{{\rm sys}, i}$ are free Hamiltonians for the intra-cavity degrees of freedom, $c_i$ is the arbitrary degree of freedom within cavity $i$ that couples to the propagating field, and $\tau$ is the propagation time between the two cavities.

Using this Hamiltonian description, Gardiner proceeds to derive a Langevin equation for an arbitrary intra-cavity degree of freedom, $a$:
\begin{widetext}
\begin{align}
\dot{a}(t) =& -i[a, H_{\rm sys,1}+H_{\rm sys,2}] - [a(t),c_1\dg(t)] \left\{\frac{\gamma_1}{2}c_1(t) + \sqrt{\gamma_1}b_{\rm in}(t)\right\} + \left\{\frac{\gamma_1}{2}c_1\dg(t) + \sqrt{\gamma_1}b\dg_{\rm in}(t)\right\}[a(t),c_1(t)] \nn \\
& - [a(t),c_2\dg(t)] \left\{\frac{\gamma_2}{2}c_2(t) + \sqrt{\gamma_1\gamma_2}c_1(t-\tau) + \sqrt{\gamma_2}b_{\rm in}(t-\tau)\right\} \nn \\
& + \left\{\frac{\gamma_2}{2}c_2\dg(t) + \sqrt{\gamma_1\gamma_2}c_1\dg(t-\tau) + \sqrt{\gamma_2}b\dg_{\rm in}(t-\tau)\right\}[a(t),c_2(t)], \label{eq:coupled_cavities_langevin_finite_tau}
\end{align}
where $\bin(t)$ is defined exactly as before as the asymptotic input field freely propagated to the interaction region, and $a$ is an operator representing a degree of freedom in cavity $1$ or $2$. We will now specialize to the limit of negligible propagation time between localized components, i.e. where $\tau \rightarrow 0$. This limit is most relevant for the more general treatments that follow in subsequent sections. In this zero delay limit, the above Langevin equation becomes:
\begin{align}
\dot{a}(t) =& -i[a, H_{\rm sys,1}+H_{\rm sys,2}] - [a(t),c_1\dg(t)] \left\{\frac{\gamma_1}{2}c_1(t) + \sqrt{\gamma_1}b_{\rm in}(t)\right\} + \left\{\frac{\gamma_1}{2}c_1\dg(t) + \sqrt{\gamma_1}b\dg_{\rm in}(t)\right\}[a(t),c_1(t)] \nn \\
& - [a(t),c_2\dg(t)] \left\{\frac{\gamma_2}{2}c_2(t) + \sqrt{\gamma_1\gamma_2}c_1(t) + \sqrt{\gamma_2}b_{\rm in}(t)\right\} + \left\{\frac{\gamma_2}{2}c_2\dg(t) + \sqrt{\gamma_1\gamma_2}c_1\dg(t) + \sqrt{\gamma_2}b\dg_{\rm in}(t)\right\}[a(t),c_2(t)],
\label{eq:coupled_cavities_langevin}
\end{align}
\end{widetext}
We note that the approximation that the propagation time $\tau$ may be taken to zero is consistent with the weak coupling approximations made earlier.  The dynamics of interest typically act on $1/\gamma_i$ time scales, and as long as $\tau$ is much shorter than these time scales of interest, the propagation delay may be ignored.  However, it is also true that relaxation of this assumption is occasionally required (\eg cm-scale propagation distances are significant when dynamics occur at GHz rates), which will be addressed in \cref{subsec:prop_delay}. 

Closer inspection of \cref{eq:coupled_cavities_langevin} highlights the key difference between cascading and simple Hamiltonian coupling, namely unidirectional flow of information. For example, if $a$ is an operator in the first cavity, all terms proportional to $[a,H_{\rm sys,2}]$, $[a,c_{2}]$, and $[a,c_{2}^{\dag}]$ drop out, leaving only terms proportional to $c_1$, $c_1^{\dag}$, and $\bin$.  Whereas, if $a$ is an operator in the second cavity terms proportional to $[a,c_{2}]$ and $[a,c_{2}^{\dag}]$ are potentially nonzero, so that the dynamics are potentially driven by $c_2$, $c_2^{\dag}$, $\bin$, $c_1$, and $c_1^{\dag}$. Therefore, system $2$ is affected by system $1$ but not vice versa, as we would expect if the probe field is unidirectional. Thus the cascaded coupling breaks time-reversal symmetry and establishes a clear direction of information flow in a network of components. 

This general, nonlinear cascaded system model allows one to model the two cascaded components as one effective component. As before, we can observe that the evolution of any operator within this component (in the Heisenberg picture) is generated by a unitary propagator, $a(t) = U\dg(t) a U(t)$. For this example, the unitary propagator takes the form $U(t) = \mathcal{T} \exp \left\{ \int_{t_0}^t ds K(s) \right\}$, with:
\begin{align}
K(s) &= -iH_{\rm sys,1} - iH_{\rm sys,2} - \frac{\sqrt{\gamma_1\gamma_2}}{2} (c_2\dg c_1 - c_1\dg c_2) + (\sqrt{\gamma_1}c_1 + \sqrt{\gamma_2}c_2)\bin\dg(s) - (\sqrt{\gamma_1} c_1\dg + \sqrt{\gamma_2} c_2\dg)\bin(s). 
\label{eq:coupled_cavities_prop}
\end{align}
As before, one can confirm this form of the propagator by deriving the equation of motion $\dot{a} = \dot{U}\dg a U + U\dg a \dot{U}$, and showing agreement with \cref{eq:coupled_cavities_langevin}. As part of this calculation, one requires the identity (derived using the same arguments as for the identity in \cref{eq:useful_id}):
\begin{align}
\bin(t)U(t) = U(t)\bin(t) + \frac{1}{2}(\sqrt{\gamma_1}c_1 + \sqrt{\gamma_2}c_2)U(t).
\end{align}

Comparing the form of the propagator in \cref{eq:coupled_cavities_prop} to \cref{eq:U_single} we see that an effective model for the component consisting of the two cavities can be specified by the following effective Hamiltonian and effective coupling operator:
\begin{align}
H_{\rm eff} &= H_{\rm sys,1} + H_{\rm sys,2} -i\frac{\sqrt{\gamma_1\gamma_2}}{2} (c_2\dg c_1 - c_1\dg c_2) \nn \\
L_{\rm eff} &= \sqrt{\gamma_1}c_1 + \sqrt{\gamma_2}c_2
\label{eq:Cascaded}
\end{align}

That is, by taking the inter-cavity propagation time to zero we are able to eliminate the field degrees of freedom between the cavities and treat the composite system as a single system with internal Hamiltonian $H_{\rm eff}$ and interacting the probe field with coupling operator $L_{\rm eff}$. Such effective descriptions of composite systems is the main aim of the framework we shall describe in the following sections. While in this case it was relatively straightforward to write the full Hamiltonian for the system, \cref{eq:coupled_cavities_fullham}, and derive the resulting dynamics, the framework we describe is capable of treating much more complex interconnected networks of components.  In particular, we will generalize this first-principles treatment of cascaded connections, while also formulating rules for other types of connections, including feedback, which significantly extends the richness of networked quantum dynamics.

\begin{aside}[Cascading one empty cavity after another]\label{ex:cascade}
Consider two empty optical resonators with IOT models \cref{eq:CavIOT}, cascaded as in \cref{fig:coupled_cavities}.  Let $a_{i}$, $\Delta_i$, and $\gamma_i$ be the annihilation operator, cavity frequency detunings, and energy decay rates for cavity modes $i$, respectively.  Using \cref{eq:Cascaded} to replace $H_{\rm sys}$ with $H_{\rm eff}$ and $L$ with $L_{\rm eff}$ in \cref{eq:adot1,eq:Binout} produces the IOT model for this cascade network
\begin{align}
\dot{a}_1(t) &= -i(\Delta_1+\frac{\gamma_1}{2})a_1(t)-\sqrt{\gamma_1}b_{\rm in}(t)\nn\\
\dot{a}_2(t) &= -i(\Delta_2+\frac{\gamma_2}{2})a_2(t)-\sqrt{\gamma_1\gamma_2}a_1(t)-\sqrt{\gamma_2}b_{\rm in}(t)\nn\\
b_{\rm out}(t) &= b_{\rm in}(t)+\sqrt{\gamma_1}a_1(t)+\sqrt{\gamma_2}a_2(t).
\label{eq:cascade_ex}
\end{align}
By inspection, one can see the effect of the unidirectional information flow: in the second equation, the first mode $a_1(t)$ drives the evolution of the second, $\dot{a}_2(t)$, but not vice versa. Also, the output field is composed of information from both cavities.
\end{aside}

\section{Quantum stochastic differential equations}
\label{sec:qsdes}
In the previous sections, we derived equations of motion for single and cascaded components interacting with probe fields, which produce dynamics when integrated.  It turns out, however, that proper integration is far from trivial, not just because the dynamics are complex, but because they are inherently stochastic.  In this section we will summarize the use of It\={o} calculus to calculate these stochastic quantum dynamics.  

 So far, we have been fairly cavalier (nevertheless, accurate) about dealing with the broadband input fields $\bin(t)$. The mathematical description of these fields is highly singular due to the canonical commutation relations $[\bin(t), \bin\dg(t')] = \delta(t-t')$. To sidestep such singularities, let us define the time-integrated quantities
 \begin{align}
\Bin(t) = \int_0^t ds\,\bin(s)\quad \text{and}\quad \Bin\dg(t) = \int_0^t ds\,\bin\dg (s),
  \end{align}
and consider increments in these fields
\begin{align}\label{eq:dB_def}
d\Bin(t) = \int_{t}^{t+dt} ds\, \bin(s), ~~ d\Bin\dg(t) = \int_{t}^{t+dt} ds \,\bin\dg(s). 
\end{align}
Note that the units of these increments are $\sqrt{\text{time}}$, and their commutation relations are $[d\Bin(t), d\Bin\dg(t')] = dt$ for $t=t'$ and zero otherwise.  These are quantum, non-commuting analogues of the classical Wiener process and are referred to as {\em quantum noise increments} or \emph{quantum stochastic increments}.

Further, by using the above singular commutation relations we can compute the following vacuum expectation values
\begin{align}\label{eq:first_ito_tab}
\expect{d\Bin(t) d\Bin(t')} = 0, &\quad \expect{d\Bin\dg(t) d\Bin\dg(t')} = 0\nn \\
 \expect{d\Bin\dg(t) d\Bin(t')} = 0, &\quad \expect{d\Bin(t) d\Bin\dg(t')} = dt, ~~~~~~ \textrm{for} ~~ t=t', ~~\textrm{and zero otherwise}
\end{align}
where $\expect{A} \equiv \tr(\rho_{\rm in} A)$, and $\rho_{\rm in}$ is the initial state of the asymptotic input field, which is assumed to be the vacuum state of all frequency modes. The vacuum expectation values above are somewhat surprising because they state that the average value of second order products of increments of the input fields can be proportional to a first-order time increment ($dt$). This bears resemblance to stochastic Wiener increments in classical stochastic theory \cite{Gar-2004}, and motivates us to think more deeply about how to integrate over such increments. Similar to classical stochastic increments, we define two types of integrals over the quantum stochastic increments $d\Bin(t)$:
\begin{align}
{\rm \mathbf I} \int_0^t g(s)d\Bin(s) &\equiv \lim_{n\rightarrow \infty} \sum_{i=0}^{n-1} g(t_i)[\Bin(t_{i+1}) - \Bin(t_i)], \nn \\
{\rm \mathbf S}\int_0^t g(s)d\Bin(s) &\equiv \lim_{n\rightarrow \infty} \sum_{i=0}^{n-1} g\left(\frac{1}{2}(t_{i+1}+t_i)\right) [\Bin(t_{i+1}) - \Bin(t_i)], \nn 
\end{align}
where the time interval $[0,t)$ has been discretized into $n$ segments, and $g$ is any operator in the system subspace. These two definitions of integration, the first of which is called an It\={o} integral and the second is called a Stratonovich integral, are equivalent in standard calculus where the increments are regular. However, since the quantum stochastic increments can vary wildly even in the $n\rightarrow \infty$ limit, these two integral definitions produce different results. As such, one must specify the type of integral a quantum stochastic differential equation (QSDE), such as \cref{eq:coupled_cavities_langevin} corresponds to. We refer the reader to Refs. \cite{GardColl85,GardParkZoll92, Zoller:2008vf} for a physics-based discussion of the differences between these two integral definitions. For a more mathematical treatment see Refs.~\cite{Hudson:1984wt,Khol91,Part12,AccaLuVolo13}.

In general, a QSDE derived from physical principles (\eg Heisenberg equations of motion) corresponds to the Stratonovich integral definition. 
To understand why this is, note that real physical noise is never exactly a white noise process. Instead, one uses (classical or quantum) white noise as an approximation of a real physical process in some limit (\eg white noise approximates the Ornstein-Uhlenbeck process in the vanishing correlation time limit). 
The Wong-Zakai theorem \cite{Wong:1965gk, Han.Sto.etal-2005a}, and its quantum generalization \cite{Gough:2006gu}, state that the behavior of a noise-driven physical system under this singular approximation of the real noise process is captured by a QSDE that is interpreted with respect to Stratonovich integration.
This is consistent with the fact that Stratonovich differentials are consistent with standard calculus rules, while It\={o} differentials obey a modified chain rule:
\begin{align}
\label{Itorule}
d(X(t) Y(t)) = dX(t) Y(t) + X(t) dY(t) + dX(t) dY(t),
\end{align}
where $X(t)$ and $Y(t)$ are arbitrary functions of operator valued stochastic variables and $dX(t)$ and $dY(t)$ are specified in terms of It\={o} QSDEs. The first two terms arise from the usual non-commutative chain rule and the third term is known as the ``It\={o} correction".

Therefore, the QSDEs we derived in the previous section for system operators (\eg \cref{eq:Langevin1,eq:coupled_cavities_langevin}) or unitary propagators (\eg \cref{eq:U_gen}) should be interpreted with respect to the Stratonovich integral (or more succinctly, we will refer to QSDEs being in Stratonovich or It\={o} ``form"). However, QSDEs in It\={o} form are often much easier to work with analytically and numerically \footnote{Fundamentally this is because the integrand $g(\cdot)$ is independent of the increment in the It\={o} definition (since the increment is in the future of where the integrand is evaluated). This independence makes certain manipulations significantly easier.}. Fortunately, there is a straightforward procedure to convert between QSDEs in Stratonovich and It\={o} forms, see \eg \cite{GardParkZoll92, GardZoll-2004}. 

Because it will be used heavily in later sections, we write the It\={o} form of \cref{eq:U_gen} here \cite{GardZoll-2004}:
\begin{align}\label{eq:dU_Ito}
dU(t) &= \left[- \big(iH_{\rm sys} + \half L\dg L\big)dt + Ld\Bin\dg(t) - L\dg d\Bin(t)\right] U(t), \quad {\rm with } \quad U(0)= \Isf,
\end{align}
where the term $- \frac{1}{2}L\dg Ldt$ arises from the conversion between Stratonovich and It\={o} forms (\ie the It\={o} correction).  We will often write the It\={o} propagator $U(t)$ as $U_t$ for convenience. 

\begin{Remark}[\bf QSDE notation]\label{remark:QSDEnotation}
By convention, QSDEs in It\={o} form are nearly always written in terms of increments (\eg an equation for $dU(t)$ and not $dU(t)/dt$). Stratonovich QSDEs are also sometimes written in terms of increments and in that case, it is customary to make explicit the Stratonovich interpretation by writing the product of a (possibly operator-valued) quantity $g(t)$ and an increment $dB(t)$ as: $g(t) \circ dB(t)$.
\end{Remark}

From here-onwards, we will work exclusively with QSDEs in It\={o} form since the original SLH framework was developed in this form.
Finally, for more detailed accounts of QSDEs from a physics perspective, we recommend the following references: Sec. 3.4 of Ref. \cite{WisePHD94}, Appendix A of Ref. \cite{JacoPHD98}, Section 3.2 and Chapter. 11 of Ref. \cite{GardZoll-2004}, and Section 3 of Ref. \cite{Han.Sto.etal-2005a}. Additionally, Cook's PhD thesis~ \cite{Cook13} provides a good bridge between the physics and the mathematics literature.\\

\begin{aside}[Deriving equations of motion in It\={o} form using quantum stochastic calculus]\label{ex:itoqsde}
One can derive equations of motion for operators of the localized systems using the propagator in \cref{eq:dU_Ito}.  Since our QSDEs are now It\={o} form, this requires taking differentials to second order, as done in \cref{Itorule}. To aid computations, we write down an It\={o} table, which prescribes the product of various quantum noise increments. Under the vacuum expectation, \ie \cref{eq:first_ito_tab} the rules for these products are given by the vacuum It\={o} table
\begin{align} \label{eq:itotab_singlemode}
\begin{tabular}{l |lll}
		$\times $        & $dB_t$ & $dB^{\dag }_t$ & $dt$ \\ \hline
		$dB_t$              & 0       & $dt$                  & $0$ \\
		$dB^{\dag }_t$ & 0        & 0                       & 0\\
		$dt$                & 0          & 0                       & 0 
\end{tabular} ,
\end{align}
where the product is understood as take the row and multiply by the column (row $\times$ column) to obtain the resulting product under vacuum.
To compute expression for the differential of a Heisenberg picture operator $O(t)$ we use a version of \cref{Itorule}:
\begin{align} \label{eq:diff}
		d(U_t \dg O U_t) = d U_t\dg O U_t + U_t\dg O dU_t + d U_t\dg O d U_t.
\end{align}
Consider again the single-sided resonator we treated in \cref{eg:cav}. The unitary propagator for this component takes the form in \cref{eq:dU_Ito} with $H_{\rm sys}= \Delta a\dg a$ and $L=\sqrt{\gamma}a$. Taking $O=a$ we can write down the It\={o} QSDE describing the dynamics of cavity mode using \cref{eq:diff}:
\begin{align}
da(t) &= U\dg(t)\left[- \big(-iH_{\rm sys} + \half L\dg L\big)dt + L\dg d\Bin(t) - L d\Bin\dg(t)\right] a \nn \\
& ~~~~ + a \left[- \big(iH_{\rm sys} + \half L\dg L\big)dt + Ld\Bin\dg(t) - L\dg d\Bin(t)\right]U(t) \nn \\
& ~~~~ + U\dg(t)\left[- \big(-iH_{\rm sys} + \half L\dg L\big)dt + L\dg d\Bin(t) - L d\Bin\dg(t)\right] a ~ \times \nn \\
&~~~~~~~~~~~~~~~~~~  \left[- \big(iH_{\rm sys} + \half L\dg L\big)dt + Ld\Bin\dg(t) - L\dg d\Bin(t)\right]U(t) \nonumber
\end{align}
Next we expand the terms in this equation. Several terms drop out according to the prescription for products of It\={o} increments given by \cref{eq:itotab_singlemode}. The most complicated term is the last one (corresponding to $d U_t\dg a d U_t$), so we write this out explicitly:
\begin{align} 
 d U_t\dg a d U_t 
 &=  U\dg(t)\left[- \big(-iH_{\rm sys} + \half L\dg L\big)dt + L\dg d\Bin(t) - L d\Bin\dg(t)\right] a ~ \times \nn \\
 &~~~~~~~~~~~~~~~~~~  \left[- \big(iH_{\rm sys} + \half L\dg L\big)dt + Ld\Bin\dg(t) - L\dg d\Bin(t)\right]U(t) \nonumber \\
&=  U\dg(t)\left[  L\dg d\Bin(t) - L d\Bin\dg(t)\right] a  \left[ Ld\Bin\dg(t) - L\dg d\Bin(t)\right]U(t) \label{eq:line1rand}\\
&=  U\dg(t) \left[ L\dg a L\cancelto{dt}{ d\Bin(t)d\Bin\dg(t)} - L a L \cancelto{0}{d\Bin\dg(t) d\Bin\dg(t)}-L\dg a L\dg \cancelto{0}{d\Bin(t)d\Bin(t)} \right. \nn \\
& ~~~~~~~~~~~~~~~~~~ \left. + L a L\dg \cancelto{0}{d\Bin\dg(t)d\Bin(t)} \right] U(t)\nonumber\\
&=  L\dg(t) a(t) L(t) dt = \gamma a(t)\dg a(t)^2 dt. \label{eq:line3rand}
\end{align}
In \cref{eq:line1rand} we dropped all terms of order $dt$ as their It\={o} products with any other increment is zero from \cref{eq:itotab_singlemode}. On the next line we expanded out the product, normally ordered then applied the It\={o} table rules. After computing the remaining terms (and normally ordering system and field operators) we arrive at 
\begin{align} 
\nn da(t) =& - \left (i \Delta + \frac{\gamma}{2}\right )a(t) dt -\sqrt{\gamma}dB_{\rm in}(t),
\end{align}
which is the It\={o} form of the equation of motion in \cref{eq:CavHeis}.
\end{aside}

\section{General quantum input-output networks and the SLH framework}
\label{sec:slhformalism}
Despite the success of input-output theory and the cascaded approach to networked open quantum systems, the approach sketched out in \cref{sec:cascade} can only be used to construct networks with a small number of components due to the difficult symbolic manipulations required. Thankfully a powerful elaboration of the cascaded approach, developed by Gough and James \cite{GougJame09,GougJame09a}, allows for description of large networked quantum systems using easy algebraic manipulations. In this section we will review the Gough-James formalism, which is commonly referred to as the \emph{SLH framework} or \emph{SLH formalism}. 

The general philosophy of the SLH framework is that the dynamics of an arbitrary local quantum system interacting with an input-output channel is described by a QSDE for the propagator for the system and field degrees of freedom. This QSDE is parameterized by a triple of operators \SLH. The mathematics behind this is described in \cref{sec:time_evo_op}. The power of the SLH formalism lies in its ability to compose the propagator for local components according to how they are connected in a network. The mathematical rules that govern the combining of SLH systems are given in \cref{sec:slh}. From the combined propagator one can derive Heisenberg equations of motion and input-output relations for the entire network, as described in \cref{sec:Network_heis_IO}. In addition, a master equation describing the evolution of the internal state of all local components in the network can be derived, see  \cref{sec:Network_master_eq}.

\subsection{The SLH time evolution operator and SLH triple}\label{sec:time_evo_op}
In \cref{sec:cascade} we saw that under the weak coupling and Markov approximations, the dynamics of an arbitrary local component, or more generally the dynamics of a cascaded system, interacting with input and output fields can be represented by a unitary propagator with some $H_{\rm eff}$ and $L_{\rm eff}$. This turns out to be widely applicable to all components where the weak coupling and Markov approximations can be made. The SLH framework is built around a slightly more general unitary propagator than \cref{eq:U_gen} or \cref{eq:dU_Ito}, which is traditionally written in its It\={o} form as:
\begin{align}   
\label{eq:QsdeU}
dU(t) = \Big\{ - (iH +\half & L\dg L)dt   + L d\Bin\dg(t) - L\dg  S d\Bin(t)+ (S - I )d\Lambda_{\rm in}(t)  \Big\} U(t), \quad {\rm with\ } U(0)= \Isf,
\end{align}
for some (interaction picture) operators $S$, $L$, and $H$ on the localized system Hilbert space. In the remainder of this work $I$ with no subscript will denote the identity operator on the system Hilbert space. This equation is often referred to as the Hudson-Parthasarathy equation \cite{Hudson:1984wt}. Consistent with the previous sections, we interpret $L$ as the system operator that couples to the external field (who's increments are $d\Bin(t)$), $H$ is the system Hamiltonian, and $S$ is a new object known as a scattering operator. The operator, $d\Lambda(t)$, is an increment in the field's number operator and its integral $\Lambda(t)$ is sometimes called the \emph{gauge process} in the literature. The gauge process and its increment are formally defined as:
\begin{align}\label{eq:dLambda_in}
\Lambda_{\rm in}(t)= \int_0^{t } ds \, \bin^\dag(s) \bin(s), \quad\quad d\Lambda_{\rm in}(t)= \int_t^{t + d t} ds \, \bin^\dag(s) \bin(s),
\end{align}
and are both unitless. Note that when $S=I$, \cref{eq:QsdeU} is exactly \cref{eq:dU_Ito}, the It\={o} form of the generator we specified in \cref{eq:U_gen}. Although we did not encounter $S\neq I$ in our discussion in \cref{sec:cascade}, it is necessary to describe localized components that impart phase shifts on the input field, \eg a mirror that adds a $\pi$ phase shift to an itinerant field.  A more general interpretation of $S$ and the gauge process will be given shortly when we consider multi-mode generalizations where each localized component may interact with multiple input and output fields. From \cref{eq:QsdeU} it is clear that a localized component is completely specified by an operator triple $(S,L,H)$, often termed the SLH triple. The formal solution to \cref{eq:QsdeU} is ~\cite{Khol91,Hole96,Hole03,Gough:2006gu}
\begin{align}
\label{eq:qsde_formal_soln}
U(t) = \mathcal{T} \exp \left\{ \int_{0}^t \left[ - (iH +\half L\dg L)ds  + L d\Bin\dg(s) - L\dg  S d\Bin(s)+ (S - I )d\Lambda_{\rm in}(s) \right] \right\}.
\end{align}
Here we use the symbol $U(t)$ to denote the system-field propagator in the time interval $[0,t)$. On occasion we will need to specify a different initial time, in which case the propagator is denoted $U(t,t_0)$ -- \ie $U(t) \equiv U(t,0)$.

\begin{Remark}\label{remark:approx}
	The physical approximations that are needed to specify a component using the unitary propagator in \cref{eq:QsdeU} are similar to the ones required to arrive at the unitary propagator we derived from physical arguments in \cref{sec:cascade} -- \ie weak, linear coupling of system degrees of freedom to itinerant fields, and the Markov approximation. 
\end{Remark}

When constructing large networks we will need to consider components interacting with multiple input-output modes. Multiple input-output modes may model the orthogonal free field polarization, spatial, or frequency modes that interact with localized components.  Multiple input-output modes may also model bosonic free fields of different physical origins, with a single localized component coupling to itinerant optical, microwave electrical, or even vibrational phononic modes, as appropriate.  When considering multimode QIONs we will suppress the subscript ``$\rm{in}$" on the input fields in favor of the mode labels $i,j,...$ for notational convenience.  Furthermore we will suppress time dependence of field operators unless it is essential for clarity.  The QSDE description of a localized system interacting with multiple ($n$) input fields is given by (using Einstein summation convention, with the sum ranging over $1,...,n$):
\begin{align}  
\label{eq:QSDE_multi}
		dU(t) =& \Big\{  - (i H +\half L_i^{\dagger} L_i  )dt  + L_i dB_i^\dagger  - L_{i}^\dagger S_{ij} dB_j + ( S_{ij} - \delta_{ij} I)d\Lambda_{ij} \Big\} U(t), \quad {\rm with\ } \quad U(0)= \Isf,
\end{align}
where $L_{i}$ is the system operator that couples to the $i$th input mode, $H$ is the system Hamiltonian, $S_{ij}$ are scattering operators that are constrained by the identities: $S_{ik}S_{jk}\dg= \delta_{ij} I$ and $S_{ki}\dg S_{kj} = \delta_{ij} I$, see \cite[
Appendix A]{GougGohmYana08} and \cite[Sec. IV]{GougJame09a} and the references therein. The input quantum noise increments and gauge process are defined as:
\begin{align}\label{eq:multi_io}
dB_i	=\int_t^{t + dt} ds \, b_i(s),\,\, ~~{\rm and}~~ \,\, d\Lambda_{ij}=\int_t^{t + d t} ds \, b_i^\dag(s) b_j(s).
\end{align}
Now it is possible to give a more general interpretation of the gauge process. $d\Lambda_{ij}$ represents direct scattering of photons from mode $i$ to mode $j$ during the time increment from $t$ to $t+dt$ that \emph{are not} mediated by energy exchange with internal degrees of freedom of the localized components (\eg ``beamsplitter''-like scattering).  When $i=j$, it represents a phase shift of mode $i$, as in the single mode case. $S_{ij}$ is a system operator that reflects the effect on the system when a photon is directly scattered from mode $j$ to mode $i$. 
%

Once again, the localized component is completely specified by a collection of $S$ and $L$ matrices, and an $H$ matrix. For conciseness this can be specified in vector notation by the triple  
$\mathbf{G} = (\mathbf{S}, \mathbf{L}, H)$ 
\begin{align}
\mathbf{S} =   \left( \begin{array}{ccc}
S_{11} & \hdots & S_{1n} \\
 \vdots & \ddots & \vdots \\
 S_{n1} & \hdots & S_{nn}
\end{array} \right), \ \ \mathbf{L} =  \left( \begin{array}{c}
L_1 \\ \vdots \\ L_n
\end{array} \right) ,
\end{align}
where there are $n$ input-output ports and the operators $S_{i,j}$ and $L_i$ have dimension of the local system. The generalization of the scattering identities translate to a unitarity condition on $\mathbf{S}$; \ie $\mathbf{S}^\dagger \mathbf{S}= \mathbf{S} \mathbf{S}^\dagger = \mathbf{I}^I_n$ (see \cref{eqn:op_matrix_notation}), where we have defined 
\begin{align}
\mathbf{I}^A_{n} &\equiv \left( \begin{array}{ccc}
A & \hdots & 0 \\
 \vdots & \ddots & \vdots \\
 0 & \hdots & A
\end{array} \right), ~~ \text{\ie an $n\times n$ matrix with the operator $A$ on the diagonal and zeros elsewhere},
\end{align}
for any operator $A$. Similarly, we can define vector notation for the input fields and gauge processes:
\begin{align}
\mathbf{B} =  \left( \begin{array}{c}
B_1 \\ \vdots \\ B_n
\end{array} \right),
\ \
\mathbf{\Lambda} = \left( \begin{array}{ccc}
\Lambda_{11} & \hdots & \Lambda_{1n} \\
 \vdots & \ddots & \vdots \\
 \Lambda_{n1} & \hdots & \Lambda_{nn}
\end{array} \right) .
\end{align}
We will often refer to such systems as having $n$ input and output \emph{ports}.
Let us specify some notation for these operator-valued matrices. Let 
\begin{subequations}\label{eqn:op_matrix_notation}
\begin{align}
 \mathbf{A} = (\mathbf{a}_{ij}), \text{with $\mathbf{a}_{ij}$ being arbitrary operators.}\\
 \text{Then, } \mathbf{A}\dg \equiv (\mathbf{a}_{ji}\dg ),\, \mathbf{A}^{\mathsf{T}} \equiv \{a_{ji}\},\, \mathbf{A}^* \equiv ( a_{ij}\dg ).
\end{align}
\end{subequations}

This vector notation allows \cref{eq:QSDE_multi} to be written in concise form as
\begin{align}
\label{eq:QSDE_vector}
d U(t) =  \Big\{- ( \smallfrac{1}{2} 
\mathbf{L}^\dagger \mathbf{L}  + i H )dt   + d\mathbf{B}^\dagger ~\mathbf{L}  -\mathbf{L}^\dagger \mathbf{S} ~d\mathbf{B}  +\mathrm{tr}[(\mathbf{S}-\mathbf{I}_n^I)d\mathbf{\Lambda}^{\mathsf T}]  \Big \} U(t), \quad {\rm with\ } \quad U(0)= \Isf.
\end{align}

An important point to emphasize is that specifying a triple $(\mathbf{S},\mathbf{L},H)$, along with an initial state of the local components and input fields in a network, completely specifies all properties of a network. This is because the way in which the operators \vSLH\ couple to the field is fixed by \cref{eq:QSDE_vector}, which prescribes the time evolution of all network components.

At this point, we state some useful rules for working with the stochastic increments present in \cref{eq:QSDE_multi}. Since second order terms in the increments $dB(t)$ can be non-zero we must specify all multiples of stochastic and deterministic increments up to second order. This is conventionally given in the form of an It\={o} table. When the underlying fields ($b_i(t)$) are in the vacuum state, the corresponding It\={o} table is 
\begin{align} \label{eq:ItoTable}
\begin{tabular}{l |c c c c c}
		$\times $        		& $dB_k$ & $d\Lambda_{kl} $ & $dB_k^{\dag }$ & $dt$ \\ \hline
		$dB_i$           		& 0      & $\delta_{ik} dB_l$               & $ \delta_{ik} dt$      & $0$ \\
		$d\Lambda_{ij} $	& 0      & $\delta_{jk} d\Lambda_{il}$  & $\delta_{jk} dB_i^{\dag }$  & $0$ \\
		$dB_i^{\dag }$ 		& 0        & 0                      & 0                       & 0\\
		$dt$                		& 0         & 0                      & 0                       & 0 
\end{tabular} ,
\end{align}
where we take the row and multiply by the column (row $\times$ column) to obtain the resulting product under vacuum expectation. All increments are at the same time -- \ie $dB_i \equiv dB_i(t)$, $d\Lambda_{ij} \equiv d\Lambda_{ij}(t)$. All increments at different times multiply to zero. \\

\begin{aside}[SLH descriptions of phase shifters, beamsplitters and cavities]\label{aside:SLHex}
A phase shifter for a single itinerate mode (i.e. one input and one output field) has an SLH triple:
\begin{align}
\left( e^{i\phi}, 0, 0\right)
\label{slh:phaseshift}
\end{align}
where $\phi$ is the phase shift angle. Notice the phase shifter does not have internal degrees of freedom and thus no system Hamiltonian $H_{\rm sys}$ or coupling operators $L$.

 Similarly a beamsplitter, which combines two itinerate modes, and also has no internal degrees of freedom has the SLH triple
\begin{align}
\left( \left[\begin{array}{cc}r_{11} & t_{12} \\ t_{21} & r_{22}\end{array}\right], \left[\begin{array}{cc}0\\0\end{array}\right],0 \right),
\label{eq:SLH_BS}
\end{align}
using the convention that the reflected fields are the output pairs to the input fields. Recall that
the entries of the scattering matrix must satisfy constraints stemming from the unitarity; \ie $\mathbf{S}\dg \mathbf{S} = I$. A 50-50 beamsplitter would be
\begin{align}
\left(\frac{1}{\sqrt{2}}\left[\begin{array}{cc}1&1\\-1&1\end{array}\right],\left[\begin{array}{cc}0\\0\end{array}\right],0 \right).\nonumber
\end{align}
Note that since a beam-splitter has no internal degrees of freedom, the entries of the scattering matrix are scalars as opposed to operators acting on the internal degrees of freedom.

The SLH triple for a one-sided cavity with a single resonant mode, described in \cref{eq:CavIOT}, is 
\begin{align}
\left(I,\left[\sqrt{\gamma} a\right],\Delta a\dg a\right),\label{eq:SLH_cav}
\end{align}
where $\gamma$ is the power decay rate from the cavity, and $\Delta$ is the cavity mode detuning.
 Finally, a cavity that couples to two itinerant modes, which only couple to each other through the resonator (\eg a Fabry-Perot cavity), has the SLH triple
\begin{align}
\left(\mathbf{I}^I_2,\left[\begin{array}{c}\sqrt{\gamma_1}a\\\sqrt{\gamma_2}a\end{array}\right],\Delta a\dg a\right).\label{eq:SLH_2cav}
\end{align}
{We also emphasize that the elements of $\mathbf{S}$ matrices may be operator valued, rather than c-numbers.  This typically occurs in SLH models in which additional approximations (e.g. adiabatic elimination, Sec. \ref{sec:elimination} ) have been employed.  Physically, operator-valued $\mathbf{S}$ elements indicate instances such as Faraday interactions \cite{Bout07} or qubit meausurement \cite{KercBoutSilb09,KercNurdNurd10}, in which direct field-scattering is dependent on the state of the localized system.}
\end{aside}

\subsection{SLH composition rules}\label{sec:slh}
The SLH composition rules, developed by Gough and James  \cite{GougJame09,GougJame09a} are algebraic prescriptions for combining SLH triples of individual components whose asymptotic free fields are connected in various manners. These algebraic rules tell us how to simply compose networks of SLH components and can be considered the heart of the SLH framework.

Three critical physical assumptions underly these rules: (i) the Markov approximation is valid for the interaction between localized components and propagating fields, (ii) that the fields interconnecting the components propagate in a dispersionless, linear medium, and further, that the time for propagation between localized components is negligible, and (iii) the input fields into the network are in the vacuum state. Assumption (iii) might seem overly restrictive but we will see in Section \cref{sec:generalizations} that non-vacuum states of the propagating fields can be introduced into the network using various extensions. Under these assumptions, the SLH composition rules are derived in Refs. \cite{GougJame09a, GougJame09}. We will summarize the rules here, and then \Cref{eg:slh_comp_rules,eg:SLHapp2} illustrate the application of the rules.

\begin{figure}[]
\includegraphics[width=\columnwidth]{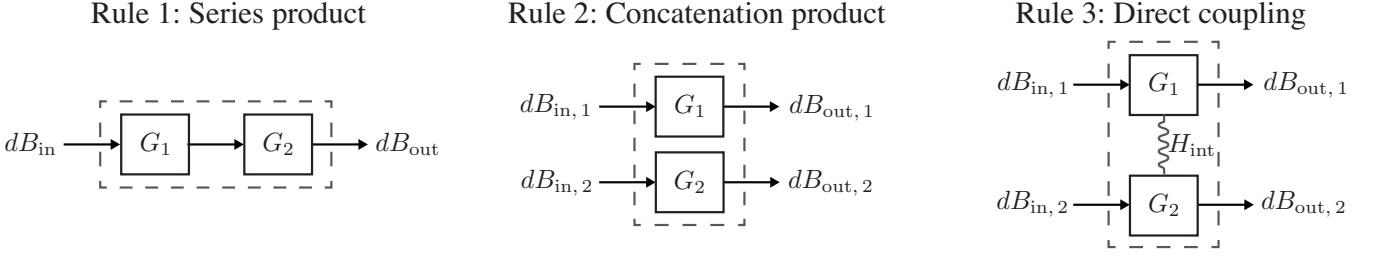}
\caption{Schematic representations of the series product $ G_2 \triangleleft G_1$, concatenation product $G_1 \boxplus G_2 $, and direct coupling  $G_1 \bowtie G_2$ which is special case of the concatenation product.}\label{fig:series_concat}
\end{figure}

\begin{SLHrule}[\bf Series product or Cascade rule] \label{SLHrule:series}
We begin with the cascading of the output from one localized network, represented by $G_1 = (\mathbf{S}_1,\mathbf{L}_1,H_1)$, into the input of another, represented by $G_2 = (\mathbf{S}_2,\mathbf{L}_2,H_2)$, see \Cref{fig:series_concat}. Both systems must have the same number of input fields and the $i$th output field of $G_1$ is the $i$th input to $G_2$. We explain how to relax both of these assumptions shortly. The result of this connection, referred to as the \emph{series product} of $G_1$ and $G_2$~ \cite{GougJame09}, and denoted as $G_{2} \triangleleft G_{1}$, is given by
\begin{align}    
 G_{\rm T}
 &= (\mathbf{S}_{\rm T},  \mathbf{L}_{\rm T},H_{\rm T}) \nonumber\\
&= G_{2} \triangleleft G_{1} \nonumber\\
&= (\mathbf{S}_{2},  \mathbf{L}_{2},H_{2}) \triangleleft  (\mathbf{S}_{1},  \mathbf{L}_{1},H_{1}) \nonumber\\
&=\left ( \mathbf{S}_{2}\mathbf{S}_{1},  \mathbf{L}_{2}+\mathbf{S}_{2} \mathbf{L}_{1}, H_{1}+H_{2}+\frac{1}{2i}( \mathbf{L}_{2}^\dag \mathbf{S}_{2} \mathbf{L}_{1}- \mathbf{L}_{1}^\dag \mathbf{S}_{2}^\dag  \mathbf{L}_{2}) \right).\label{eq:SLH_cascade}
\end{align}
Note that the effective Hamiltonian for the combined system has picked up a dependence on the coupling operators for the component blocks.  Note that the Hamiltonian term makes $G_{2} \triangleleft G_{1} \neq G_{1} \triangleleft G_{2} $, due to the fact that the fields linking the localized components are directional.
\end{SLHrule}

\begin{SLHrule}[\bf Concatenation product] \label{SLHrule:concat}
Now we examine the parallel grouping of two components with independent input-output fields and SLH triples $G_1 = (\mathbf{S}_1,\mathbf{L}_1,H_1)$ and  $G_2 = (\mathbf{S}_2,\mathbf{L}_2,H_2)$; see \Cref{fig:series_concat}. The result of this parallel grouping, referred to as the {\em concatenation product} of $G_1$ and $G_2$~ \cite{GougJame09a}, and denoted $G_{1} \boxplus G_{2}$, is given by
\begin{align}
\label{eq:concatenation}
 G_{\rm T}
 &= (\mathbf{S}_{\rm T}, \mathbf{L}_{\rm T},H_{\rm T}) \nonumber\\
&= G_{1} \boxplus G_{2} \nonumber\\
&= (\mathbf{S}_{1}, \mathbf{L}_{1},H_{1}) \boxplus  (\mathbf{S}_{2}, \mathbf{L}_{2},H_{2}) \nonumber\\
&=\left( \left[\begin{array}{cc}\mathbf{S}_1 & 0 \\0 & \mathbf{S}_2\end{array}\right],
\left[\begin{array}{c}\mathbf{L}_1 \\ \mathbf{L}_2 \end{array}\right], H_{1}+H_{2} \right).
\end{align}
\end{SLHrule}	
 
\begin{SLHrule}[\bf Direct coupling] \label{SLHrule:direct}
The next composition rule is {\em direct coupling}, which is a generalization of the concatenation product. Consider $G_1$ and $G_2$ in parallel, and then add a direct Hamiltonian interaction between the two systems, $H_{\rm int}$, which is an operator on $\mathcal H_1\otimes \mathcal H_2$, the tensor product of Hilbert spaces of $G_1$ and $G_2$. 
\begin{align}
\label{eq:directcoupling}
 G_{\rm T}
 &= (\mathbf{S}_{\rm T}, \mathbf{L}_{\rm T},H_{\rm T}) \nonumber\\
&= G_{1} \boxplus G_{2} \nonumber\\
&= (\mathbf{S}_{1}, \mathbf{L}_{1},H_{1}+H_{\rm int}) \boxplus  (\mathbf{S}_{2}, \mathbf{L}_{2},H_{2}) \nonumber\\
&=\left( \left[\begin{array}{cc}\mathbf{S}_1 & 0 \\0 & \mathbf{S}_2\end{array}\right],
\left[\begin{array}{c}\mathbf{L}_1 \\ \mathbf{L}_2 \end{array}\right], H_{1}+H_{2} +H_{\rm int} \right).
\end{align}
 Then there is ambiguity in how to specify the direct coupling product. $H_{\rm int}$ can be associated with $G_1$ or $G_2$ or symmetrically into both. We prefer to adopt the convention that the coupling resides in the first system.  Recently, this composition rule has also been denoted as a ``bowtie" product $G_{\rm T}= G_{1} \bowtie G_{2} $~\cite{ZhanJame12}.
 \end{SLHrule}
 
\begin{figure}[]
\includegraphics[width=\columnwidth]{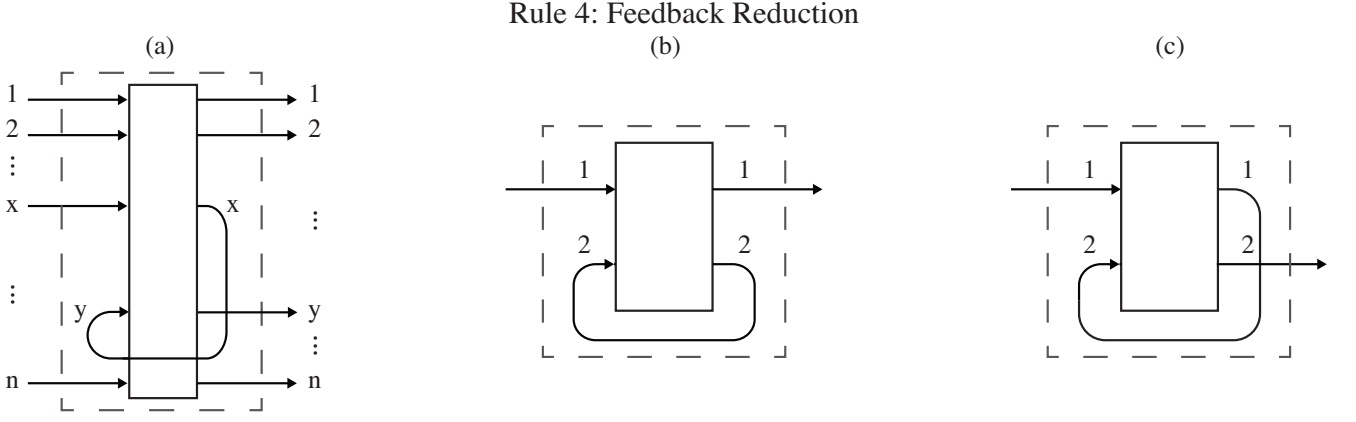}
\caption{Schematic representations of the feedback reduction~\cite{GougJame09a}. Notice that any output port can be connected to any input port.}\label{fig:feedback}
\end{figure}

\begin{SLHrule}[ \bf Feedback reduction and network interconnection] \label{SLHrule:feedback}
Finally, we examine the most complicated interconnection: the feedback reduction. 
Given a component described by an SLH triple $G = (\mathbf{S},\mathbf{L},H)$, the feedback reduction computes the SLH triple that results from interconnecting an output of $G$ to an input of $G$. Let the original system $G$ have $n$ input and output ports. Let $x$ be the output port that is connected to the input port $y$, we denote this interconnection by $x \rightarrow y$ (see  \cref{fig:feedback}(a)). The feedback reduction rule eliminates this internal link and results in a new system $G_{\rm red} = (\mathbf{S}_{\rm red}, \mathbf{L}_{\rm red}, H_{\rm red})$, where \cite{GougJame09}
\begin{subequations}\label{eq:SLH_feedback}
\begin{align}
\mathbf{S}_{\rm red} &= \mathbf{S}_{\bar{x}\bar{y}}+\mathbf{S}_{\bar{x}y}(I-S_{xy})^{-1}\mathbf{S}_{x\bar{y}} 
 \\
\mathbf{L}_{\rm red} &= \mathbf{L}_{\bar{x}}+\mathbf{S}_{\bar{x}y}(I-S_{xy})^{-1}L_{x} 
 \\
H_{\rm red} &= H+\frac{1}{2i} \left( \mathbf{L}\dg \mathbf{S}_{:,y}(I-S_{x,y})^{-1} L_x - L_x\dg (I-S\dg_{x,y})^{-1} \mathbf{S}\dg_{:,y} \mathbf{L} \right)
 \end{align}
 {and the identity $I$ has the dimesion of the reduced Hilbert space. }
\end{subequations}
Here the subscripts on $\mathbf{S}$ and $\mathbf{L}$ with overbars denote matrices with certain rows or columns removed. Explicitly,
\begin{subequations}\label{eq:mat_def}
\begin{align}
\mathbf{S}_{\bar{x}\bar{y}}&\equiv \left(\begin{array}{cc}
\mathbf{S}_{1:x-1;~1:y-1} & \mathbf{S}_{1:x-1;~y+1:n} \\
\mathbf{S}_{x+1:n;~1:y-1}& \mathbf{S}_{x+1:n;~y+1:n} 
\end{array}\right)\\
\mathbf{S}_{\bar{x}{y}}&\equiv\left(\begin{array}{c}
\mathbf{S}_{1:x-1;~ y}  \\
\mathbf{S}_{x+1:n;~y} 
\end{array}\right)\\
\mathbf{S}_{x\bar{y}}&\equiv\left(
\mathbf{S}_{x;~ 1:y-1} \quad  
\mathbf{S}_{x;~y+1:n} 
\right)\\
\mathbf{S}_{:,y}  &  \equiv\mathbf{S}_{1:n;~ y}   \\
\mathbf{L}_{\bar{x}} &\equiv\left(\begin{array}{c}
\mathbf{L}_{1:x-1}  \\
\mathbf{L}_{x+1:n} 
\end{array}\right).
\end{align} 
\end{subequations}
Or in words: $S_{x,y}$, and $\mathbf{S}_{:,y}$, are the $(x,y)$ element, and the $y$th column, of $\mathbf{S}$, respectively. $\mathbf{S}_{\bar{x},\bar{y}}$ is $\mathbf{S}$ without the $x$th row and $y$th column (the $(x,y)$ minor of $\mathbf{S}$). $\mathbf{S}_{\bar{x},y}$ is the $y$th column of $\mathbf{S}$ with the $x$th row deleted, and similarly, $\mathbf{S}_{x,\bar{y}}$ is the $x$th row of $\mathbf{S}$ with the $y$th column deleted. Lastly, $\mathbf{L}_{\bar{x}}$ is $\mathbf{L}$ without the $x$th row, and $L_{x}$ is the $x$th row of $\mathbf{L}$.
Importantly, if one is eliminating multiple ports, {\ie $x$ and $y$ are sets}, the order in which the eliminations are done does not matter. However, one must be careful about tracking indices in this case since once an input or output is eliminated by connecting it to another output or input, the correspondence between the ordering of the inputs/outputs and the row and column numbers of 
$\mathbf{S}$ and $\mathbf{L}$ must be reconsidered. {We elaborate on this important issue in \cref{SLHremark:vec_elim}, which also contains an example that illustrates how to interconnect a network by simultaneously eliminating multiple internal wires.}

The concatenation product and the feedback reduction rule are sufficient to compose any number of components and construct arbitrary networks. The basic procedure to follow for an arbitrary network is to (i) form the concatenation product of all components in the network as if all components are independent and unconnected, and then (ii) apply the feedback reduction rule to implement all connections in the network. Thus the series product is a special case of the feedback reduction. However, we specify it as a separate rule since it is so commonly used.

{It is instructive to examine some abstract examples to illustrate the feedback reduction rule.  Our example explores the difference between \Cref{fig:feedback}(b) and (c).} Consider the feedback network in \Cref{fig:feedback}(b), whereby output 2 is connected to input 2. An application of the mathematical prescription of the \emph{feedback reduction} yields
\begin{align}
 \widetilde G_{\rm T}
 &= (\widetilde S_{\rm T},\widetilde L_{\rm T},\widetilde H_{\rm T})\nonumber\\
 &= [G]_{2\rightarrow 2} \nonumber\\
&=\left [\left( \left[\begin{array}{cc}S_{11} & S_{12} \\ S_{21} & S_{22}\end{array}\right],
\left[\begin{array}{c}L_1 \\ L_2 \end{array}\right], H \right)\right ]_{2\rightarrow 2}\\
&=\left(S_{11}+S_{12}(I -S_{22})^{-1}S_{21}, L_1+S_{12}(I -S_{22})^{-1}L_2,\right.\nonumber\\ 
&\quad\quad H+ \left.\frac{1}{2i}(L_{2}^\dag S_{22}(I-S_{22})^{-1}L_{2}+L_{1}^\dag S_{12}(I-S_{22})^{-1} L_{2}-\rm h.c.) \right).\label{eq:fb1}
\end{align}	
Now consider connecting the output of port 1 to the input of port 2, as depicted in \cref{fig:feedback}(c). This results in:
\begin{align}
 \widetilde G_{\rm T}
  &= [G]_{1\rightarrow 2} \nonumber\\
&=\left [\left( \left[\begin{array}{cc}S_{11} & S_{12} \\ S_{21} & S_{22}\end{array}\right],
\left[\begin{array}{c}L_1 \\ L_2 \end{array}\right], H \right)\right ]_{1\rightarrow 2} \nonumber \\
&=\left(S_{21}+S_{22}(I -S_{12})^{-1}S_{11}, L_2+S_{22}(I -S_{12})^{-1}L_1,\right.\nonumber\\ 
&\quad\quad H+ \left.\frac{1}{2i}(L_{2}^\dag S_{22}(I-S_{12})^{-1}L_{1}+L_{1}^\dag S_{12}(I-S_{12})^{-1} L_{1}-\rm h.c.) \right).\label{eq:fb2}
\end{align}	
The difference between \cref{eq:fb1} and \cref{eq:fb2} demonstrates that the reduced system depends on which ports are connected in feedback. 

Some intuition for the physics captured in the feedback reduction rule can be gained by examining the form of the coupling term in \cref{eq:fb2}: $L_{\rm T} = L_2+S_{22}(I -S_{12})^{-1}L_1 = L_2+S_{22}(I + S_{12} + S_{12}^2 + S_{12}^3 + ...)L_1$, where in the second expression we have Taylor expanded $(I-S_{12})^{-1}$. This expansion makes it clear that the output at port 2 under the feedback reduction is a combination of $L_2$ and $L_1$ after it has gone an indeterminate number of scatterings from port 1 to port 2. The modifications to the other members of the triple under the feedback reduction can be interpreted in a similar manner.

\end{SLHrule}

\begin{Remark}[\bf Padding: combing systems with unequal numbers of input-output ports] \label{SLHremark:padding}
Combining systems that have different numbers of input and output ports is often required when composing SLH networks. While \cref{SLHrule:series} does not strictly allow this, \cref{SLHrule:concat} comes to the rescue. The general problem is to cascade an $M$ input-output port network $G_1 = (\mathbf{S}_1,\mathbf{L}_1,H_1)$  into a $N>M$ port network   $G_2 = \left (\mathbf{S}_2,\mathbf{L}_2,H_2\right )$ and supposing that $N-M=n$. We begin by defining a trivial SLH component, called a padding element, that simply scatters input fields directly to output fields for $n$ modes.
Generally the padding element of dimension $n$ is denoted \cite{TezaNiedPavl12} 
\begin{align}
\mathbb{I}_{n} &= \left(I_n,\mathbf{0},0\right),
\end{align}
where $\mathbf{0}$ is the length $n$ zero vector and $I_n$ is the $n\times n$ identity matrix. Now we concatenate and then cascade
\begin{align}
G_2 \triangleleft (\mathbb{I}_{n} \boxplus G_1 )
\end{align}
\end{Remark}	

\begin{Remark}[\bf Permuting input-output channels: rewiring between network nodes] \label{SLHremark:rewire}
Strict cascading in the SLH framework leads to the $i$th output field of $G_1$ being the $i$th input to $G_2$. If the actual 
interconnections are more complex it may be difficult to construct a SLH model of the network that obeys the true mapping.
This difficulty can be lifted by inserting a trivial component that reorders the output fields, described the SLH triple~ \cite{TezaNiedPavl12}:
\begin{align}
\mathbb{P}_{\sigma} &= \left( \mathbf{P}_\sigma,\mathbf{0},0\right).
\end{align}
here $\sigma$ denotes the permutation in relational notation, and $\mathbf{P}_\sigma$ is the permutation matrix with elements $P_{j,k}=\delta_{j,\sigma(k)}$ where $\delta_{j,k}$ is a Kronecker delta.
Importantly this definition ensures correct composition of permutations i.e. $P_{\sigma_2\circ \sigma_1}=P_{\sigma_2} \triangleleft P_{ \sigma_1} $. One can think of this component as effectively modeling the rerouting of fields between two components.
\end{Remark}

\begin{figure}[]
\includegraphics[width=\columnwidth]{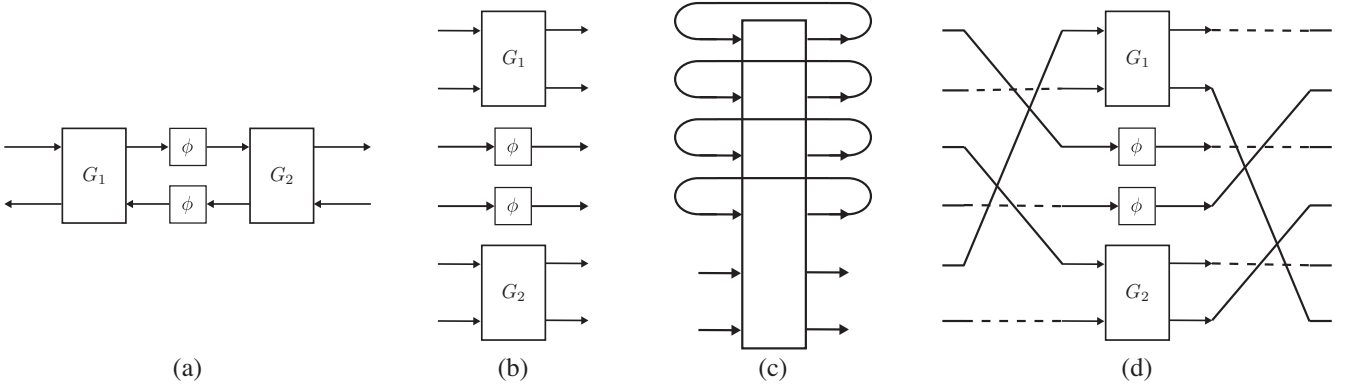}
\caption{ { Network interconnection using the feedback reduction rule ~\cite{GougJame09a}. (a) The network to be modeled. Note that we have a loose interpretation of which ports are considered in and out ports in this figure. (b) The same components concatenated. Here all the inputs are on the left and all the output ports are on the right. (d) To eliminate multiple internal nodes at once using the formula given in \cref{eq:SLH_feedback} one must bring networks into a form where the eliminated nodes are {\bf block contiguous}. We choose to have all eliminated nodes at the top of the circuit. (d) the required permutations of the input and ouput ports for the circuit given in (c) to bring it into the correct form so that feedback connections in (d) can be applied, i.e. by connecting  $i_{\rm out}\rightarrow i_{\rm in}$ for $i\in\{1,2,3,4\}$ we obtain (a). } }\label{fig:networkconnect}
\end{figure}
\begin{Remark}[\bf Network interconnection and eliminating multiple ports] \label{SLHremark:vec_elim}
{
There are some subtleties in applying the vector form of the rule \cref{eq:SLH_feedback}. We now give an example to illustrate how to interconect a network while simultaneously eliminating multiple internal wires. We wish to wire up the network given in \cref{fig:networkconnect} (a). We specify the SLH triples 
\begin{align}
 G_{1} = \left (\mathbf{I}^I_2,\left[\begin{array}{c}L_1 \\L_2\end{array}\right]  ,H_{1}\right ),\quad
 G_{2} = (e^{i\phi},0,0),\quad
 G_{3} = (e^{i\phi},0, 0),\quad
  G_{4} = \left (\mathbf{I}^I_2,\left[\begin{array}{c}L_5 \\L_6\end{array}\right]  ,H_{2}\right ) .
\end{align}
We form the concatenation product
\begin{align}
G_{T}&= (\mathbf{S}_{\rm T}, \mathbf{L}_{\rm T},H_{\rm T})=  G_{1} \boxplus G_{2} \boxplus G_{3}\boxplus   G_{4}
\end{align}
which is shown in the \cref{fig:networkconnect} (b). Notice that we have consistently put all input ports on the left and all output ports on the right.

To wire up \cref{fig:networkconnect} (b) to look like \cref{fig:networkconnect} (a) we would connect the following in and out ports:
\begin{center}
\begin{tabular}{|c|c|}
\hline
out & in\\
\hline 1 &3\\
\hline 3 &5\\
\hline 6 &4\\
\hline 4 &2\\
\hline
\end{tabular}
\end{center}
If one tries to na\"{\i}vely apply \cref{eq:SLH_feedback} the result is nosense. In order for \cref{eq:SLH_feedback} to be applicable, we need to bring the network input and output ports and internal ports into a block contiguous form. One such form is depicted in \cref{fig:networkconnect} (c). The remainder of this remark illustrates how to do this for this particular example, the basic technique involes permuting the input and output ports using \cref{SLHremark:rewire}.

The correct permutations for the input ports are  $\mathbf{P}_{[5,2,1,4,3,6]}$ [depicted on the left hand side of \cref{fig:networkconnect} (d)], while the ouput ports are  $\mathbf{P}_{[1,4,3,6,5,2]}$ [depicted on the right hand side of \cref{fig:networkconnect} (d)]. The corresponding scattering matricies are
\begin{align}
S_{\rm in}=\left(\begin{array}{cccccc}
0 & 0 & 0 & 0 & 1 & 0 \\
0 & 1 & 0 & 0 & 0 & 0 \\
1 & 0 & 0 & 0 & 0 & 0 \\
0 & 0 & 0 & 1 & 0 & 0 \\
0 & 0 & 1 & 0 & 0 & 0 \\
0 & 0 & 0 & 0 & 0 & 1\end{array}\right)\quad {\rm and}\quad
S_{\rm out}=\left(\begin{array}{cccccc}
1 & 0 & 0 & 0 & 0 & 0 \\
0 & 0 & 0 & 1 & 0 & 0 \\
0 & 0 & 1 & 0 & 0 & 0 \\
0 & 0 & 0 & 0 & 0 & 1 \\
0 & 0 & 0 & 0 & 1 & 0 \\
0 & 1 & 0 & 0 & 0 & 0\end{array}\right).
\end{align}
The rewired system, depicted in figure \cref{fig:networkconnect} (a), is 
\begin{align}
 G_{\rm T}'
 &= (S_{\rm out},0,0)\lhd (\mathbf{S}_{\rm T}, \mathbf{L}_{\rm T},H_{\rm T})\lhd (S_{\rm in},0,0).
\end{align}
where the new network SLH operator are
\begin{align}
\mathbf{S}_{\rm T}'= S_{\rm out}\mathbf{S}_{\rm T}S_{\rm in}=
\left(\begin{array}{cccccc}
0 & 0 & 0 & 0 & I & 0 \\
0 & 0 & 0 & e^{i\phi} & 0 & 0 \\
e^{i\phi} & 0 & 0 & 0 & 0 & 0 \\
0 & 0 & 0 & 0 & 0 & I \\
0 & 0 & I & 0 & 0 & 0 \\
0 & I & 0 & 0 & 0 & 0\end{array}\right), \quad
\mathbf{L}_{\rm T}'=  S_{\rm out} \mathbf{L}_{\rm T}=
\left(\begin{array}{c}L_1 \\0 \\0 \\L_6 \\L_5 \\L_2\end{array}\right),
\quad
H_{\rm T}= H_{1}+H_{2}.
\end{align}

With respect to the new wiring, we connect the following ports to correctly wire the system:
\begin{center}
\begin{tabular}{|c|c|}
\hline
out & in\\
\hline 1 &1\\
\hline 2 &2\\
\hline 3 &3\\
\hline 4 &4\\
\hline
\end{tabular}.
\end{center}
This obeys the wiring convention in \cref{fig:networkconnect} (c). In the interest of being very explicit about how to do the elimination we specify the following operators from \cref{eq:mat_def}
\begin{subequations}
\begin{align}
\mathbf{S}_{x,y}&=\left(\begin{array}{cccccc}
0 & 0 & 0 & 0 \\
0 & 0 & 0 & e^{i\phi}  \\
e^{i\phi} & 0 & 0 & 0  \\
0 & 0 & 0 & 0  \\
\end{array}\right)\!, 
\mathbf{S}_{\bar{x}\bar{y}}= \left(\begin{array}{cc}
0 & 0\\
0& 0
\end{array}\right)\!,
\mathbf{S}_{\bar{x}{y}}=\left(\begin{array}{cccc}
0 & 0 & I & 0  \\
0 & I & 0 & 0  \\
\end{array}\right)\!,
\mathbf{S}_{x\bar{y}}=\left(\begin{array}{cc}
I  &0\\
0 &0\\
0 &0\\
0 &I\\
\end{array}
\right)\!, \\
\mathbf{S}_{:,y}  &=
\left(\begin{array}{cccc}
0 & 0 & 0 & 0  \\
0 & 0 & 0 & e^{i\phi}  \\
e^{i\phi} & 0 & 0 & 0  \\
0 & 0 & 0 & 0  \\
0 & 0 & I & 0  \\
0 & I & 0 & 0 \end{array}\right) ,
(I-\mathbf{S}_{x,y})^{-1}=\left(\begin{array}{cccccc}
1 & 0 & 0 & 0 \\
0 & 1 & 0 & e^{i\phi}  \\
e^{i\phi} & 0 &1 & 0  \\
0 & 0 & 0 & 1  \\
\end{array}\right)\!, 
\mathbf{L}_{\bar{x}} =\left(\begin{array}{c}
L_5  \\
L_2
\end{array}\right),
\mathbf{L}_{x} =\left(\begin{array}{c}
L_1\\
0\\
0\\
L_6
\end{array}\right).
\end{align} 
\end{subequations}
Substituting these operators into \cref{eq:SLH_feedback}, and after some matrix algebra, we find
\begin{subequations}\label{eq:network_red_eg}
\begin{align}
\mathbf{S}_{\rm red} &=e^{i\phi}\mathbf{I}^I_2
 \\
\mathbf{L}_{\rm red} &= \left(\begin{array}{c}
L_5  +e^{i\phi}L_1\\
L_2 +e^{i\phi}L_6
\end{array}\right)
 \\
H_{\rm red} &= H_1+H_2+\frac{1}{2i} \left( e^{i\phi} L_1 L_5\dg -e^{-i\phi} L_1\dg L_5 + e^{i\phi} L_2\dg  L_6
 - e^{-i\phi} L_2 L_6\dg \right).
 \end{align}
\end{subequations}
The resulting system is quite abstract, however, we will see later that this system can be used as a way to model counter propagation in IOT, see \cref{ex:counterprop} and \cref{eq:slh_twosite_counter}.
}
\end{Remark}

\begin{aside}[SLH composition rules]
\label{eg:slh_comp_rules}	

\hspace{0.8pt} {\em 1) Series product.}  We first reexamine the cascaded system treated in \cref{ex:cascade}, but using the SLH triples.  Consider two optical cavities cascaded as in \cref{fig:coupled_cavities}.  Individually, these cavities may be described using \cref{eq:SLH_cav}: $\left(I,\left[\sqrt{\gamma_i} a_i\right],\Delta_i a_i\dg a_i\right)$ where $i\in\{1,2\}$ specifies the cavity index.  From these triples, we derive the effective cascaded network depicted in \cref{fig:coupled_cavities} using \cref{eq:SLH_cascade}
\begin{align}
G_{\rm cascade} &= \left(I,\left[\sqrt{\gamma_2} a_2\right],\Delta_2 a_2\dg a_2\right)\triangleleft\left(I,\left[\sqrt{\gamma_1} a_1\right],\Delta_1 a_1\dg a_1\right)\nn\\
&=\Big(I,\left[\sqrt{\gamma_1} a_1+\sqrt{\gamma_2} a_2\right],\Delta_1 a_1\dg a_1+\Delta_2 a_2\dg a_2+\frac{\sqrt{\gamma_1\gamma_2}}{2i}(a_2\dg a_1-a_1\dg a_2) \Big).
\end{align}

{\em 2) Concatenation product.}  Consider the same two cavities, but now placed in parallel, such that the inputs, outputs, and internal degrees of freedom remain independent of each other.  From \cref{eq:concatenation} we have   
\begin{align}
G_{\rm concatenate} &= \left(I,\left[\sqrt{\gamma_1} a_1\right],\Delta_1 a_1\dg a_1\right)\boxplus\left(I,\left[\sqrt{\gamma_2} a_2\right],\Delta_2 a_2\dg a_2\right)\nn\\
&=\left(
\mathbf{I}^I_2,\left[\begin{array}{c}\sqrt{\gamma_1}a_1\\\sqrt{\gamma_2}a_2\end{array}\right],\Delta_1 a_1\dg a_1+\Delta_2 a_2\dg a_2\right).
\end{align}

{\em 3) Direct coupling.}  Again, consider two cavities but now they are ``crossed", and the cavities are coupled via a cross Kerr nonlinearity $\chi   a_1\dg a_1 a_2\dg a_2$. The generalized concatenation gives 
\begin{align}
G_{\rm } &= \left(I,\left[\sqrt{\gamma_1} a_1\right],\Delta_1 a_1\dg a_1 +\chi   a_1\dg a_1 a_2\dg a_2\right)\boxplus\left(I,\left[\sqrt{\gamma_2} a_2\right],\Delta_2 a_2\dg a_2\right)\nn\\
&=\left(
\mathbf{I}^I_2,\left[\begin{array}{c}\sqrt{\gamma_1}a_1\\\sqrt{\gamma_2}a_2\end{array}\right],\Delta_1 a_1\dg a_1+\Delta_2 a_2\dg a_2+\chi   a_1\dg a_1 a_2\dg a_2\right),
\end{align}
using the convention that the coupling resides in the first system.

{\em 4) Feedback.}  Now consider the two sided resonator described by \cref{eq:SLH_2cav}, but with the slight modification $\sqrt{\gamma_2}\rightarrow i\sqrt{\gamma_2}$, indicating that the cavity couples to itinerant mode 2 with a phase shift of 90$^\circ$ relative to mode 1.  Now, imagine routing the output of port 1 to the input of port 2.  Using \cref{eq:SLH_feedback}, this network has the reduced SLH triple
\begin{align}
G_{\rm feedback} &= \left[\left(
\mathbf{I}_2^I,\left[\begin{array}{c}\sqrt{\gamma_1}a\\ i\sqrt{\gamma_2}a\end{array}\right],\Delta a\dg a\right)\right]_{1\rightarrow2}\nn\\
&= \left(I,[(\sqrt{\gamma_1}+i\sqrt{\gamma_2})a],(\Delta-\sqrt{\gamma_1\gamma_2})a\dg a\right).
\end{align}
Note that the feedback has reduced the total number of input-output fields (ports) to one.  Also, the cavity mode now couples to this itinerant mode with the effective amplitude $\sqrt{\gamma_1+\gamma_2}$ and phase angle $\text{atan}(\sqrt{\gamma_2/\gamma_1})$, and the effective detuning of the cavity mode from the reference frequency has been reduced by $-\sqrt{\gamma_1\gamma_2}$.

{\em 5) Padding.} 
Here we show how to cascade a single input output component $G_1 = (S_1,L_1,H_1)$ into a two input and two output component  $G_2 = \left (\mathbf{S}_2,\mathbf{L}_2,H_2\right )$, where $\mathbf{S}_2$ is a $2\times 2$ matrix and $\mathbf{L}_2$ is a $2\times 1$ vector. Now using \cref{SLHrule:concat} we can concatenate and then cascade in two ways:
\begin{align}
G_2 \triangleleft (G_1 \boxplus  \mathbb{I}_{1}) \quad {\rm or} \quad G_2 \triangleleft (\mathbb{I}_{1} \boxplus G_1 ).
\end{align}
The two different paddings correspond to where the output of $G_1$ is fed into input port 1 or 2 of $G_2$, respectively.

{\em 6) Channel permuting.} Suppose we wanted to cascade two systems $G_1$ and $G_2$, but we want to connect: the third output of $G_1$ to the first input of $G_2$, the first output of $G_1$ to the second input of $G_2$, and the second output of $G_1$ to the third input of $G_2$. 
In this case we the permutation desired is $\sigma= [3,1,2]$. To make things clear system $s$ , for $s\in[1,2]$, has input-output field increments $dB^{(s)}_{{\rm in}, p}$ and $dB^{(s)}_{{\rm out}, p}$ where $p$ labels the port \ie $p\in[1,2,3]$. We will shortly show, in \cref{eq:inputoutput_vector}, that a permuting component with $\mathbf{S}$ operator $\mathbf{P}_{[3,1,2]}$ maps the output fields of component 1 to the input fields of component 2 as desired 
\begin{align}
\left[\begin{array}{c}
dB^{(2)}_{\rm in, 1} (t)\\ 
dB^{(2)}_{\rm in, 2} (t)\\ 
dB^{(2)}_{\rm in, 3}(t)
\end{array}\right]
 = \left[\begin{array}{ccc}
0 & 0 & 1 \\
1 & 0 & 0 \\
0 & 1 & 0
\end{array}\right]
\left[\begin{array}{c}
dB^{(1)}_{\rm out, 1} (t)\\ 
dB^{(1)}_{\rm out, 2} (t)\\ 
dB^{(1)}_{\rm out, 3}(t)
\end{array}\right].
\end{align}

\end{aside}
	
\subsection{Network Heisenberg equation of motion and network input-output relations}\label{sec:Network_heis_IO}
The SLH framework represents each network component as an SLH triple and the network construction rules outlined in the previous subsection specifies how to combine different components according to the network connectivity. Moreover, given a description of a network of components in terms of an SLH triple, $G = (\mathbf{S}, \mathbf{L}, H)$, and the corresponding unitary propagator, \cref{eq:QSDE_vector}, we wish to compute the output fields in terms of the input fields. 

We define the output field processes as time-evolved Heisenberg operators, where the evolution is given by the network propagator, \ie
\begin{align}
	B_{\rm out}(t) &= U\dg(t) B_{\rm in}(t) U(t) \\
	\Lambda_{\rm out}(t) &= U\dg(t) \Lambda_{\rm in}(t) U(t),
\end{align}
where again, it should be kept in mind that the time label on the input processes does not indicate a Heisenberg operator at time $t$, but rather a temporal mode label. This definition of the integrated output field is consistent with the output field $b_{\rm out}(t)$ defined in \cref{sec:cascade}, as we will show in \cref{ex:inpout}.  We wish to derive QSDEs describing the evolution of these output field processes, and to do so, we define their increments, \eg $dB_{\rm out}(t) = B_{\rm out}(t+dt) - B_{\rm out}(t)$, and similarly for $d\Lambda_{\rm out}(t)$. Expressing this increment in terms of the input fields yields \cite{Barchielli:1999bj}:
\begin{align}
\label{eq:BoutBin}
	dB_{\rm out}(t) &= U\dg(t+dt)B_{\rm in}(t+dt)U(t+dt) - U\dg(t)B_{\rm in}(t)U(t) \nn\\
					&=U\dg(t)U\dg(t+dt,t)B_{\rm in}(t+dt)U(t+dt,t)U(t) - U\dg(t)U\dg(t+dt,t)B_{\rm in}(t)U(t+dt,t)U(t) \nn \\
					&=U\dg(t)U\dg(t+dt,t)dB_{\rm in}(t)U(t+dt,t)U(t),
\end{align}
and similarly for $d\Lambda_{\rm out}(t)$. In the second line we have defined an infinitesimal propagator $U(t+dt,t)$ from $t$ to $t+dt$, and decomposed $U(t+dt) = U(t+dt, t)U(t)$ (this decomposition is possible because the dynamics is Markovian). Further, we used the fact $U\dg(t)U\dg(t+dt,t)B_{\rm in}(t)U(t+dt,t)U(t) = U\dg(t)B_{\rm in}(t)U(t)$ because $[U(t+dt,t), B_{\rm in}(t)]=0$ since the propagator from $t$ to $t+dt$ does not depend on any input fields at times $\leq t$. To obtain the form of the infinitesimal propagator we expand the formal solution \cref{eq:qsde_formal_soln} to first order in $dt$ \cite{Barchielli:1999bj}:
\begin{align}
	U(t+dt, t) = \Isf - (iH +\half L\dg L)dt   + L d\Bin\dg(t) - L\dg  S d\Bin(t)+ (S - I )d\Lambda_{\rm in}(t).
\end{align}

Computing using these definitions, one arrives at the following input-output relations for the general multiple input/output case \cite{Goug12, ZhanJame12}:
\begin{align}
\label{eq:inputoutput_vector}
d\mathbf{B}_{\rm out}(t)
&= \mathbf{S}(t) d\mathbf{B}(t) + \mathbf{L}(t) dt\\
\label{eq:inputoutput_vector2}
d\mathbf{\Lambda}_{\rm out}(t) 
&= \mathbf{L}^*(t) \mathbf{L}^{\mathsf{T}}(t) dt+
  \mathbf{S}^*(t) d\mathbf{B}^*(t) \mathbf{L}^{\mathsf{T}}(t) + \mathbf{L}^*(t) d\mathbf{B}^{\mathsf{T}}(t) \mathbf{S}^{\mathsf{T}}(t)
  +   \mathbf{S}^*(t) d\mathbf{\Lambda}(t) \mathbf{S}^{\mathsf{T}}(t).
\end{align}
Here, and in the remainder of this subsection, the time index on the system operators, \eg $\mathbf{L}(t)$, is meant to indicate that these are operators in the Heisenberg picture -- \ie $\mathbf{L}(t) \equiv (L_1(t), ..., L_n(t))^{T}$, with $L_i(t) = U(t)\dg (L_i\otimes I_{\rm field}) U(t)$, where the $I_{\rm field}$ in the tensor product denotes an identity on all field degrees of freedom.
\cref{eq:inputoutput_vector} is a generalization of the input-output relation in \cref{eq:Binout}; it specifies the output field increments $d\mathbf{B}_{\rm out}$ as a scattering transformation of the input fields plus a contribution from the internal states of the localized components. Similarly, \cref{eq:inputoutput_vector2} specifies the photons scattered to the output fields during the time increment from $t$ to $t+dt$ in terms of a combination of input photons and contributions from localized components. The following example explicitly calculates the input-output relation for a single port component, and applies it to derive the form of the output field from a cascaded network.

\begin{aside}[Deriving single port input-output relations]\label{ex:inpout}
Let us calculate the output field increment for a single port system by applying \cref{eq:BoutBin} and the rules of quantum stochastic calculus:
\begin{align}
dB_{\rm out}(t) 
&= U\dg(t)U\dg(t+dt,t) dB_{\rm in}(t) U(t+dt,t)U(t)\\
&= U\dg(t)\left[\Isf - (-iH +\half L\dg L)dt   + L\dg d\Bin(t) - S\dg L d\Bin\dg(t)+ (S\dg - I )d\Lambda_{\rm in}(t)\right] d\Bin(t) \times \nn \\
& ~~~~~~~ \left[\Isf - (iH +\half L\dg L)dt   + L d\Bin\dg(t) - L\dg  S d\Bin(t)+ (S - I )d\Lambda_{\rm in}(t)\right] U(t) \nonumber \\
& = U\dg(t) dB_{\rm in}(t) \left[\Isf  + L d\Bin\dg(t) + (S - I )d\Lambda_{\rm in}(t)\right] U(t) \nn \\
&= U\dg(t) \left[ S dB_{\rm in}(t) + L dt \right] U(t) \nn \\
&=  S(t) dB_{\rm in}(t) + L(t) dt,
\end{align}
where we have used the fact that $U(t)$ commutes with all increments at time $t$ since it only depends on increments at times $<t$. If we return to \cref{eg:cav}, where $L = \sqrt{\gamma}a$ and $S=I$, we recover the It\={o} form of the input output relation in \cref{eq:CavIOT}.

Using this input-output relation the interpretation of cascading becomes particularly simple. Consider two cascaded components, and suppose system $i$ has internal Hamiltonian $H_i$, couples to the itinerant field through the operator $L_i$, and $S_i=I$. The input field increment to component 1, $dB_{\rm in}(1,t)$, is transformed to an output field increment $dB_{\rm out}(1,t)= L_1(t)dt +dB_{\rm in}(1,t)$. In the zero delay limit, the output of the first component arrives immediately at component two and becomes the input to that component. That is $dB_{\rm in}(2,t)= dB_{\rm out}(1,t)$. Hence the output field from component 2 is 
\begin{align}
dB_{\rm out}(2,t) &= L_2(t) dt +dB_{\rm in}(2,t) \nn \\
&= L_2(t)dt +  L_1(t)dt+ dB_{\rm in}(1,t).
\end{align}
This is the It\={o} form of the cascaded input-output relation in \cref{eq:cascade_ex} when $L_i=\sqrt{\gamma_i}a_i$.
\end{aside}

Next, one can derive equations of motion for operators of the localized systems in the network using the general form of the propagator, \cref{eq:QSDE_vector}, and the expression for the differential of a Heisenberg operator (in It\={o} form):
\begin{align} 
		dO(t) = d(U_t \dg O U_t) = d U_t\dg OU_t + U_t\dg O dU_t + d U_t\dg O d U_t.
\end{align}
For an arbitrary system operator, $X(t)$, within the network the equation of motion becomes (using Einstein summation with all sums ranging over $1,...,n$) \cite{Gough:2014if}:
\begin{align}
dX(t) &= -i[X(t), H(t)]dt + L_i(t)\dg X(t) L_i(t) -\frac{1}{2}\left(L_i\dg(t) L_i(t) X(t) + X(t)L_i\dg(t) L_i(t) \right) \nn \\
&+ [L_i\dg(t), X(t)]S_{ij}(t)dB_j(t) + S\dg_{ij}(t)[X(t), L_i(t)]dB\dg_j(t) + \left[ S\dg(t)_{ki}X(t)S_{kj}(t) - \delta_{ij}X(t) \right] d\Lambda_{ij}(t)
\label{eq:Xeqn}
\end{align}
We can write this equation in more compact form by overloading some notation \cite{GougJame09}:
\begin{align}
\label{eq:Xeqn_vectorized}
 dX(t) =&  -i [ X(t), H(t) ] dt +  \mathcal{L} \dg[\mathbf{L}(t)] X(t)dt + d\mathbf{B}^\dagger(t) \mathbf{S}^\dagger(t) [ X(t), \mathbf{L}(t)]
+ [\mathbf{L}^\dagger(t), X(t)] \mathbf{S}(t) d\mathbf{B}(t)
\nonumber \\
 & + \mathrm{tr}\left[ (\mathbf{S}^\dagger(t)  \mathbf{I}_n^{X(t)} \mathbf{S}(t) - \mathbf{I}_n^{X(t)} ) d\mathbf{\Lambda}^{\mathsf{T}}(t)\right],
\end{align}
with
\begin{align}
\mathcal{L} \dg[\mathbf{L}] X 
&\equiv \sum_{i=1}^n L_i\dg XL_i- \half \left( L_i\dg L_i X + X L_i\dg L_i  \right),\label{eq:LindbladMulti} \\
[ X(t), \mathbf{L}(t)] &\equiv \left(
\begin{matrix}{}
[X(t), L_1]  \\
 \vdots  \\
  [X(t), L_n] 
\end{matrix} \right),  \\
[\mathbf{L}\dg(t),X(t)] &\equiv \left( [L_1\dg,X(t)], ..., [L_n\dg, X(t)] \right). 
\end{align}

\begin{figure}[]
\includegraphics[width=\columnwidth]{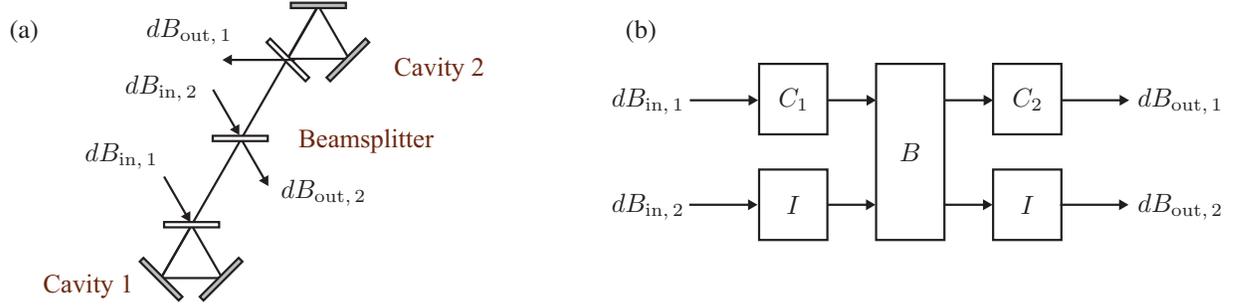}
\caption{a) Physical diagram of network $N$, modeled by \cref{eq:2CavDiag}.  b) SLH block diagram of network $N$, modeled by \cref{eq:2CavDiag}. The blocks labeled $I$ are padding components as discussed in \cref{SLHremark:padding}.}\label{fig:2CavDiag}
\end{figure}

For clarity, we also write the single-port versions of these input-output relations and the equation of motion:
\begin{subequations}
\begin{align}  
dB_{\rm out}(t) &= L(t) dt + S(t) dB_{\rm in}(t), \label{eq:dB}\\
d\Lambda_{\rm out}(t) &=   L\dg(t) L(t) dt +  L\dg(t) S(t) d\Bin(t) + S\dg(t) L(t) d\Bin\dg(t)  + S\dg(t) S(t) d\Lambda_{\rm in}(t), \label{eq:dLambda} \\
dX(t) &= -i[X(t), H(t)]dt + \mathcal{L}\dg[L(t)]X(t) dt   + [L\dg(t),X(t)]S(t) d\Bin(t) \nn \\
		& +S\dg(t)[X(t),L(t)] d\Bin\dg(t)  + (S(t)\dg X(t)S(t)-X(t)) d\Lambda_{\rm in}(t),\label{eq:dX}
\end{align}
\end{subequations}
where the single-port Heisenberg picture Lindblad operator is
\begin{align} 
\mathcal{L}\dg[L]X = L\dg XL- \half \left( L\dg L X + X L\dg L  \right). \label{eq:heis_pic_lindblad}
\end{align}
As a final point, we note that one may encounter the ``quantum flow" notation in literature where an operator $X$ at time $t$ is denoted in the Heisenberg picture by 
\begin{align}\label{eq:quantum_flow_notation}
j_{t}( X ) \equiv U_t \dg X U_t.
\end{align}
Although this notation is more cumbersome, it is more precise because it makes explicit the quantities in the Heisenberg picture, \eg in this notation the equation of motion in \cref{eq:dX} is 
\begin{align}  
dj_t(X) &= j_t(-i[X, H])dt + j_t(\mathcal{L}\dg[L]X) dt   + j_t([L\dg,X]S) d\Bin(t) \nn \\
		& +j_t(S\dg[X,L] )d\Bin\dg(t)  + j_t(S\dg XS-X) d\Lambda_{\rm in}(t).\label{eq:qflow}
\end{align}

\begin{aside}[SLH network input-output relations]
\label{eg:SLHapp2}

Consider the network depicted in \cref{fig:2CavDiag}a, with two optical cavities in series, but interrupted by a beamsplitter.  As will be discussed in \cref{sec:Loss}, inserting a beamsplitter between components is a common way to model transmission line loss in input-output networks.  The SLH block diagram of this network is depicted in \cref{fig:2CavDiag}b. The SLH triples involved in this network are 
\begin{align}
C_i &= \left(I,\left[\sqrt{\gamma_i} a_i\right],\Delta_i a_i\dg a_i\right)\nn\\
B &= \left(\left[\begin{array}{cc}\sqrt{1-\eta^2}&-\eta\\\eta&\sqrt{1-\eta^2}\end{array}\right],\left[\begin{array}{c}0 \\0\end{array}\right],0\right)\nn\\
\mathbb{I}_{1} &= \left(1,0,0\right)
\end{align} 
where $C_i$ models cavity $i$, $B$ models the beamsplitter with power reflectivity $\eta^2$, and $\mathbb{I}_{1}$ is a padding component that simply scatters an input to the output.  If we call this network $N$, we construct its SLH description using series and concatenation products   
\begin{align}
N &= \left( C_2\boxplus \mathbb{I}_1\right)\triangleleft B\triangleleft\left( C_1\boxplus \mathbb{I}_1\right)\nn\\
&=\left(
\mathbf{I}^I_2,\left[\begin{array}{c}\sqrt{\gamma_2}a_2\\0\end{array}\right],\Delta_2 a_2\dg a_2\right)\triangleleft\left(\left[\begin{array}{cc}\sqrt{1-\eta^2}&-\eta\\\eta&\sqrt{1-\eta^2}\end{array}\right],\left[\begin{array}{c}0 \\0\end{array}\right],0\right)\triangleleft \left(
\mathbf{I}^I_2,\left[\begin{array}{c}\sqrt{\gamma_1}a_1\\0\end{array}\right],\Delta_1 a_1\dg a_1\right)\nn\\
&= \left(\left[\begin{array}{cc}\sqrt{1-\eta^2}&-\eta\\\eta&\sqrt{1-\eta^2}\end{array}\right],\left[\begin{array}{c}\sqrt{\gamma_2}a_2+\sqrt{1-\eta^2}\sqrt{\gamma_1}a_1\\\eta\sqrt{\gamma_1}a_1\end{array}\right], \right.\nn \\
& ~~~~~~~~~~~~~~~~~~~~~ \left. \Delta_1 a_1\dg a_1+\Delta_2 a_2\dg a_2+\frac{\sqrt{1-\eta^2}\sqrt{\gamma_1\gamma_2}}{2i}(a_2\dg a_1-a_2a_1\dg)\right).\label{eq:2CavDiag}
\end{align} 
We see here that the role of padding (see \cref{SLHremark:padding}) for proper indexing of the fields at every stage of the network;  \eg although the first component that $dB_{{\rm in},2}$ has a non-trivial interaction with is $B$, in the first concatenation we need to specify that $dB_{{\rm in},2}$ interacts with the identity component $\mathbb{I}_1$ while $dB_{{\rm in},1}$ interacts with $C_1$.  

To obtain equations of motion for the cavity mode operators $a_1$ and $a_2$, we turn to \cref{eq:Xeqn_vectorized}. First, we evaluate the commutator with the Hamiltonian and Lindblad operator, \cref{eq:LindbladMulti}, for the annihilation operators $a_1$ and $a_2$ :
\begin{align}
-i\left[a_{1,2},~\Delta_1 a_1\dg a_1+\Delta_2 a_2\dg a_2+\frac{\sqrt{1-\eta^2}\sqrt{\gamma_1\gamma_2}}{2i}(a_2\dg a_1-a_2a_1\dg)\right]&= -i \Delta_{1,2}a_{1,2}\pm\frac{\sqrt{1-\eta^2}\sqrt{\gamma_1\gamma_2}}{2}a_{2,1},\nn\\
\mathcal{L}\dg\left[\mathbf{L}\right]a_1 &=-\frac{1}{2}(\gamma_1 a_1+\sqrt{1-\eta^2}\sqrt{\gamma_1\gamma_2}a_2)\nn\\
\mathcal{L}\dg\left[\mathbf{L}\right]a_2 &=-\frac{1}{2}(\gamma_2 a_2+\sqrt{1-\eta^2}\sqrt{\gamma_1\gamma_2}a_1)\nn
\end{align}
Next, the coefficients of the quantum noise driving terms are evaluated as:
\begin{align}
\left[\mathbf{L}\dg,a_1\right] \mathbf{S} & =  \left[\begin{array}{cc}-\sqrt{\gamma_1}\sqrt{1-\eta^2},&-\eta\sqrt{\gamma_1}\end{array}\right]\left[\begin{array}{cc}\sqrt{1-\eta^2}&-\eta\\\eta&\sqrt{1-\eta^2}\end{array}\right]\nn\\
&=\left[\begin{array}{cc}-\sqrt{\gamma_1},&0\end{array}\right],\nn\\
\left[\mathbf{L}\dg,a_2\right]\mathbf{S}&=\left[\begin{array}{cc}-\sqrt{\gamma_2},&0\end{array}\right]\left[\begin{array}{cc}\sqrt{1-\eta^2}&-\eta\\\eta&\sqrt{1-\eta^2}\end{array}\right]\nn\\
&=\left[\begin{array}{cc}-\sqrt{\gamma_2}\sqrt{1-\eta^2},&-\sqrt{\gamma_2}\eta\end{array}\right],\nn\\
\mathbf{S}\dg[a_i,\mathbf{L}]&=0, ~~~~~ i=1,2.\nn
\end{align}
Finally, the coefficients of the gauge process increment vanish since
\begin{align}
\mathbf{S}\dg \mathbf{I}_2^{a_i} \mathbf{S} - \mathbf{I}_2^{a_i} = 0 \nn
\end{align}
Putting these expression together, the equations of motion for the annihilation operators $a_i$ are
\begin{align}
da_1 &= -(i\Delta_1+\frac{\gamma_1}{2}) a_1dt-\sqrt{\gamma_1}dB_{{\rm in},1},\nn\\
da_2 &= -(i\Delta_2+\frac{\gamma_2}{2}) a_2dt-\sqrt{1-\eta^2}\sqrt{\gamma_1\gamma_2}a_1-\sqrt{\gamma_2}\left(\sqrt{1-\eta^2}dB_{{\rm in},1}+\eta dB_{{\rm in},2}\right)
\end{align}
Note that while $a_1$ is driven only by $dB_{{\rm in},1}$, $a_2$ is driven by $a_1$, $dB_{{\rm in},1}$, and $dB_{{\rm in},2}$, due to the cascade. The equations for the output fields $dB_{{\rm out},i}$ may be calculated using \cref{eq:inputoutput_vector}:
\begin{align}
\left[\begin{array}{c}dB_{{\rm out},1}\\dB_{{\rm out},2}\end{array}\right] &= \left[\begin{array}{cc}\sqrt{1-\eta^2}&-\eta\\\eta&\sqrt{1-\eta^2}\end{array}\right]\left[\begin{array}{c}dB_{{\rm in},1}\\dB_{{\rm in},2}\end{array}\right]+\left[\begin{array}{c}\sqrt{\gamma_2}a_2+\sqrt{1-\eta^2}\sqrt{\gamma_1}a_1\\\eta\sqrt{\gamma_1}a_1\end{array}\right]dt,
\end{align}
and similarly for the gauge processes $d\Lambda_{\rm out}$.  Thus the output field $dB_{{\rm out},1}$ now contain superpositions of the modes $a_1$ and $a_2$ (as well as superpositions of the input fields). 

\end{aside}

\subsection{Master equation description}\label{sec:Network_master_eq}
Once the SLH composition rules have been used to derive the form of a QION, it is sometimes useful to trace over the input-output fields, and derive an equation of motion for just the localized degrees of freedom. This will result in a statistical description of the dynamics of the localized systems since the effects of the propagating fields have been averaged over by the trace operation. Since the input fields are in the vacuum state, this description is easily given in terms of a Markovian master equation for the density matrix for the localized degrees of freedom. Explicitly, we wish to compute a dynamical equation for the quantity $\rho(t) = \tr_{\rm field}(\varrho(t))$, where $\varrho$ is the density matrix for the entire system, and $\tr_{\rm field}$ denotes a partial trace over the Fock spaces of all input-output fields. Here, consistent with the development of the SLH framework, we assume that the initial state of all the field modes is vacuum. Since the entire system evolves according to the propagator specified in \cref{eq:QSDE_vector},
\begin{align}
d\rho(t) &= d\tr_{\rm field}(U_t \varrho(0) U_t\dg) \nn \\
&= \tr_{\rm field}(dU_t \varrho(0) U_t\dg) + \tr_{\rm field}(U_t \varrho(0) dU_t\dg) + \tr_{\rm field}(dU_t \varrho(0) dU_t\dg)
\end{align} 
This computation can be carried out using the facts that the initial density matrix can be factored into density matrices for the localized components and the field modes, and that the field modes are in the vacuum state; \ie $\varrho(0) = \rho(0)\otimes\op{0}{0}$. Then using the It\={o} table in \cref{eq:ItoTable}, one arrives at the master equation corresponding to a general network parameterized by the SLH triple $\mathbf{G} = (\mathbf{S}, \mathbf{L}, H)$:
\begin{align}\label{eq:vac_master_eqn}
\frac{d\rho(t)}{dt} = -i[H, \rho(t)] + \sum_{i=1}^n \mathcal{L}[L_i]\rho(t),
\end{align}
where 
\begin{align}\label{eq:schro_pic_lindblad}
\mathcal{L}[L]\rho = L\rho L\dg - \frac{1}{2}\left(L\dg L \rho + \rho L\dg L \right),
\end{align}
which should be compared to \cref{eq:heis_pic_lindblad}. This is a deterministic equation of motion since the stochastic quantities have been averaged over.

In some instances not all of the output fields from the coherent quantum network are traced over. Instead, some may be monitored by detectors, and in such cases, one can write conditioned dynamical equations for the localized degrees of freedom, termed \emph{stochastic master equations}, \emph{quantum trajectory equations} or \emph{quantum filtering equations}. This topic is reviewed extensively in Refs. \cite{Carm93a,WiseMilb10}; there are great introductions in Refs.~\cite{Brun02,JacoStec06,WiseMilb10} and \cite{BoutHandJame07,BoutHandJame09} (the first group of references are from a physics perspective, while the second group are from a mathematical physics perspective). There is also extensive primary literature on this topic, \eg Refs.~\cite{BarcLanzPros82,Barc86,CaveMilb87,Bela89,Carm93a,DaliCastMolm92,GoetGrah94,Wise96,BarcPaga96,JackCollWall99,JackColl00}, and hence we do not discuss this topic further in this review.

Finally, note that the matrix $\mathbf{S}$ does not appear in \cref{eq:vac_master_eqn}. This is because the initial state of the field degrees of freedom was assumed to be vacuum. In \cref{sec:non_vac_input} we will discuss QIONs with non-vacuum input fields, and see that in this case the master equation for system degrees of freedom can depend on $\mathbf{S}$; see \cref{ex:coherent_state_me}.

\section{Linear quantum networks}\label{sec:linear}

Linear quantum networks, and linear quantum systems in general, are more experimentally accessible, especially in the optical regime, and thus have been extensively studied in quantum optics, \eg see Refs. \cite{Sim.Muk.etal-1994, WiseMilb10}. Linear quantum systems are most commonly encountered when dealing with collections of harmonic degrees of freedom and we will restrict our attention to this context here. In this case each degree of freedom is characterized by the annihilation and creation operators ($a_i(t), a_i\dg(t)$) with bosonic commutation relations: $[a_i(t), a\dg_j(s)] = \delta_{ij}\delta(t-s)$. In the context of QIONs, a linear quantum network is one where the localized components are composed of harmonic modes with quadratic Hamiltonians, and linear couplings to external, propagating fields. One can also include measurements of the output fields and retain a linear system description if the measurements are Gaussian, \eg homodyne measurement of an arbitrary quadrature or a heterodyne measurement.

 In this section we will formally specify linear QIONs and define useful alternative representations of such QIONs (\ie alternatives to the SLH triple representation). Linear systems are also extremely well studied in classical control theory and many  control theory techniques have been ported from the classical linear systems context to the quantum linear systems. In  \cref{sec:linear_review} we will present a review of some of these techniques and results.

\subsection{Passive linear quantum networks}
We begin by defining a sub-class of linear quantum networks, those containing only passive components. In the quantum optics context these are networks containing components such as beam-splitters, phase shifters, and empty cavities. Consider a QION with such components, represented by an SLH triple $\mathbf{G} = (\mathbf{S}, \mathbf{L}, H)$. We can collect the annihilation operators for all degrees of freedom in the network in a vector
\begin{align}
\mathbf{a}(t) = \left[ \begin{matrix}{}
  a_1(t) \\
  a_2(t) \\
  \vdots \\\
  a_m(t)
\end{matrix} \right]
\label{eq:passive_avec}
\end{align}
The condition that the QION is a passive linear network implies that \cite{Gough:2008vx} (i) the elements of $\mathbf{S}$ are scalars, (ii) all elements of $\mathbf{L}$ are linear in $a_i$, \ie there exist complex constants $\phi_{jk}$ such that $L_j = \sum_{k=1}^m \phi_{jk} a_k$, and (iii) $H$ is quadratic and conserves total photon number, \ie there exist complex constants $\omega_{jk}$ such that $H = \sum_{j,k=1}^m a_j\dg \omega_{jk} a_k$. 

Given the SLH triple, one can derive equations of motion for the internal degrees of freedom represented in the vector $\mathbf{a}(t)$ using \cref{eq:Xeqn} and also calculate the output fields from the QION using \cref{eq:inputoutput_vector}. Doing so yields a set of forced linear differential equations of motion for the internal degrees of freedom and a linear relationship between the inputs, outputs and $\mathbf{a}(t)$:
\begin{align}
\dot{\mathbf{a}}(t) &= A \mathbf{a}(t) + B \mathbf{\bin}(t) \label{eq:passive_adot} \\
\mathbf{\bout}(t) &= C \mathbf{a}(t) + D\mathbf{\bin}(t),
\label{eq:passive_ABCD}
\end{align}
where $\mathbf{\bin}(t)$ is defined as a vector of instantaneous input field annihilation operators -- \ie $\mathbf{\bin}(t) = \left[ b_1(t), b_2(t), .... b_n(t) \right]^{\mathsf{T}}$ (recall that $b_i(t) \equiv b_{{\rm in},i}(t)$) -- and $\mathbf{\bout}(t)$ a vector of output field annihilation operators. The matrices $A,B,C,D$ are defined in terms of the elements of the SLH triple as:
\begin{align}
A &= -\frac{1}{2} \Phi\dg \Phi  - i \Omega, &B &= -\Phi\dg \mathbf{S}, \nn \\
C &= \Phi, &D &= \mathbf{S},
\label{eq:passive_ABCD_SLH}
\end{align}
where $\Phi$ is an $n\times m$ matrix with elements $\phi_{jk}$, and $\Omega$ is an $m\times m$ Hermitian matrix with elements $\omega_{jk}$. For linear QIONs specifying the matrices $A,B,C,D$ is equivalent to specifying the networks using an SLH triple. \cref{eq:passive_adot,eq:passive_ABCD} strongly resemble the specification of a classical linear dynamical system \cite{Ste-1994}, in what is usually called the \emph{ABCD representation}. However, it is important to keep in mind one distinction: whereas in the classical linear systems case the matrices $A,B,C,D$ are independent and can be specified arbitrarily (as long as they are the appropriate dimensions), in the quantum case they are strongly dependent, as evidenced by their specification in terms of the SLH triple, \cref{eq:passive_ABCD_SLH}. Fundamentally, this is due to the constraints placed on quantum evolution placed by the uncertainty principle \cite[Chapter 6]{WiseMilb10}, or alternatively due to the fact that the evolution of the composite system is unitary.

We note that one can reverse the equalities in \cref{eq:passive_ABCD_SLH} in order to obtain an SLH triple given a linear system in the $ABCD$ representation, \ie
\begin{align}
\mathbf{S} = D, ~~~~~ \mathbf{L}=C \mathbf{a}, ~~~~~ H = \mathbf{a}\dg \left[i(A + \frac{1}{2}C\dg C) \right] \mathbf{a}
\end{align}

The linear differential equation for $\dot{\mathbf{a}}(t)$ in \cref{eq:passive_adot} can be solved by Laplace transform, and one can get an explicit expression relating the input and output fields for the QION in the Laplace domain \cite{GougGohmYana08}:
\begin{align}\label{eq:Linear_Laplace}
\hat{\mathbf{b}}_{\rm out}(s) = \Xi(s) \hat{\mathbf{b}}_{\rm in}(s) + \xi(s) \mathbf{a}_0,
\end{align}
where $\Xi(s) = D - C(s I_n - A)^{-1}C\dg D$, $\xi(s) = C(sI_n - A)^{-1}$, and $\mathbf{a}_0$ is the initial value of the internal degrees of freedom. In the above, we denote variables in the Laplace domain with a hat, \ie
\begin{align}
\hat{c}(s) = \int_0^\infty dt e^{-st} c(t),
\end{align}
for $c(t)$ a time domain quantity and $\Re\{ s \}>0$.
The matrix $\Xi(s)$ is referred to as the \emph{transfer function matrix}, again in analogy to classical linear systems, and it is sufficient to specify the input-output behavior of the QION.

\subsection{Active linear quantum networks}
The most general class of linear quantum networks admits components that are active in the sense that they do not conserve the total energy in the network (even in the absence of input and output ports). Some examples of such elements are squeezers (\eg optical parameter oscillators) and amplifiers. In this case the dynamics of the system can no longer be described by transformation of the annihilation operators given in \cref{eq:passive_avec}, and instead we must expand the state vector to include the conjugate creation operators, \ie
\begin{align}
\mathbf{\tilde{a}}(t) = \left[ \begin{matrix}{}
  a_1(t) \\
  \vdots \\
  a_m(t) \\
  a\dg_1(t) \\
  \vdots \\
  a\dg_m(t)
\end{matrix} \right]
\label{eq:active_avec}
\end{align}

An active linear QION has the following restrictions on its SLH triple \cite{Gough:2010in}: (i) the elements of $\mathbf{S}$ are scalars, (ii) all elements of $\mathbf{L}$ are linear in $a_i$ and $a_i\dg$, \ie there exist complex constants $\phi^-_{jk}, \phi^+_{jk}$ such that $L_j = \sum_{k=1}^m \phi^-_{jk} a_k + \phi^+_{jk} a\dg_k $, and (iii) $H$ is a general quadratic Hamiltonian that generates any symplectic transformation of the $m$ modes, \ie there exist complex constants $\omega^-_{jk}, \omega^+_{jk}$ such that $H = \sum_{j,k=1}^m a_j\dg \omega^-_{jk} a_k + a_j\dg \omega^+_{jk} a_k\dg + a_j \omega^{+*}_{jk} a_k$. 
Similar to the passive case we define the following matrices for later use: the $n\times m$ matrices $\Phi_\pm$ with elements $\phi^{\pm}_{jk}$ and the $m \times m$ matrices $\Omega_{\pm}$ with elements $\omega^{\pm}_{jk}$. We also define the following ``doubled up" matrices:
\begin{align}
\tilde{\Phi} = \left[ \begin{matrix}
 \Phi_- & \Phi_+ \\
 \Phi_+^* & \Phi_-^* \\
\end{matrix}
 \right], ~~~~~~ \tilde{\Omega} = \left[ \begin{matrix}
 \Omega_- & \Omega_+ \\
 -\Omega_+^* & -\Omega_-^* \\
\end{matrix}
 \right],
\end{align}
where $A^*$ for a matrix $A$ denotes element-wise complex conjugation.

Given an SLH triple for an active linear QION, the equation of motion for the state vector $\mathbf{\tilde{a}}$ and the input-output relation for the QION are also linear just as in the passive case:
\begin{align}
\mathbf{\dot{\tilde{a}}}(t) &= \tilde{A} \mathbf{\tilde{a}}(t) + \tilde{B} \mathbf{\tilde{b}}_{\rm in}(t) \label{eq:active_adot} \\
\mathbf{\tilde{b}}_{\rm out}(t) &= \tilde{C} \mathbf{\tilde{a}}(t) + \tilde{D}\mathbf{\tilde{b}}_{\rm in}(t),
\label{eq:active_ABCD}
\end{align}
where $\mathbf{\tilde{b}}_{\rm in}(t) = \left[b_1(t), ... b_n(t), b\dg_1(t), ... b\dg_n(t)\right]^{\mathsf T}$, and similarly for $\mathbf{\tilde{b}}_{\rm out}(t)$. The $ABCD$ matrices defining the linear system are in this case \cite{Gough:2010in}:
\begin{align}
\tilde{A} &= -\frac{1}{2} \tilde{\Phi}^{\flat} \tilde{\Phi}  - i \tilde{\Omega}, &\tilde{B} &= -\tilde{\Phi}^{\flat} \tilde{D} \nn \\
\tilde{C} &= \tilde{\Phi}, &\tilde{D} &= \left[ \begin{matrix}
 \mathbf{S} & 0 \\
 0 & \mathbf{S}^* \\
\end{matrix}
 \right],
\label{eq:active_ABCD_SLH}
\end{align}
where $\tilde{\Phi}^{\flat} = J_m \tilde{\Phi}\dg J_n$, with $J_n \equiv \left[ \begin{matrix}
  I_n & 0 \\
  0 & -I_n
\end{matrix} \right]
$ denotes an involution of the $2n \times 2m$ matrix $\tilde{\Phi}$ ($I_n$ is the $n\times n$ identity matrix). As with passive linear systems, although \cref{eq:active_adot,eq:active_ABCD} resemble the $ABCD$ representation of  a classical linear system, the $A,B,C,D$ matrices have additional constraints on them in the quantum context.

As in the passive linear network case, one can also define a transfer function matrix to capture input-output behavior in the Laplace domain. The expression for the transfer function matrix in this case is exactly the same as in the passive case but with all matrices replaced by their doubled up counterparts; \ie $A \rightarrow \tilde{A}$ and so on.

Gough \etal have specified network composition rules directly at the level of the $ABCD$ representation for linear QIONs \cite{Gough:2010in}, and so one could alternatively develop a model for a linear QION using this representation for each component if it is more convenient. We  note that sometimes the state of quantum linear systems is described using quadrature variables ($x_i \propto a_i + a_i\dg$ and $y_i \propto a_i - a_i\dg$) in the state vector. In this case the form of the linear equations in \cref{eq:active_adot}, \cref{eq:active_ABCD} is preserved, but the definitions of the $A,B,C,D$ matrices are modified and the input fields (that force the linear equations of motion) are also specified in quadrature form, \eg see \cite[Chapter 6]{WiseMilb10}. Finally, an important feature of linear quantum networks is that they preserve Gaussian states; \ie if all input modes are in Gaussian states, all output modes will also be in Gaussian states \cite{Weedbrook:2012tz} .

\begin{figure}[]
\includegraphics[width=\columnwidth]{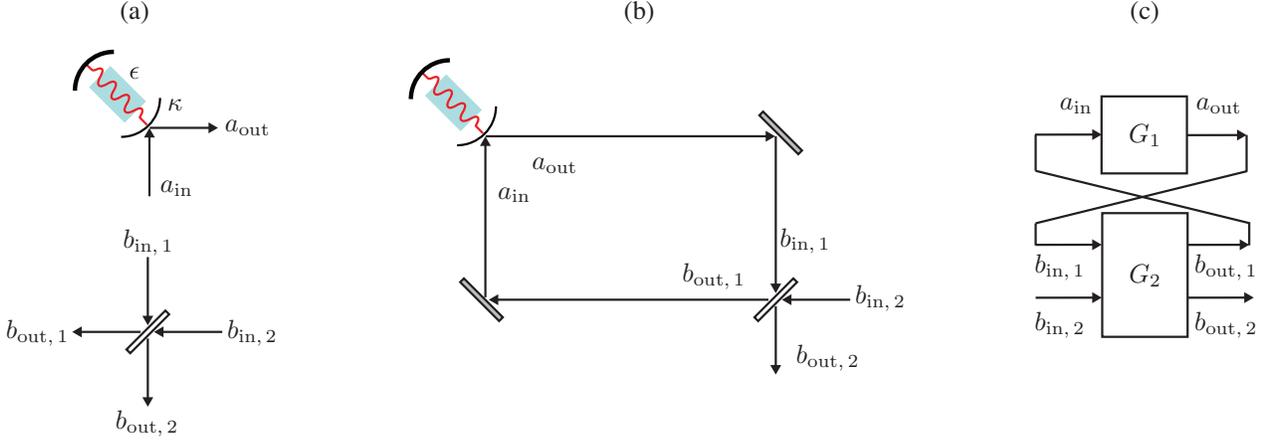}
\caption{Example of a linear quantum network. The SLH and $ABCD$ representations of this network are developed in \cref{ex:linearsys}. (a) shows the individual components in the network; $G_1$ is a degenerate optical parametric oscillator and $G_2$ is a beam-splitter. (b) shows the connected network with the two components in feedback configuration. (c) shows an equivalent block diagram representation of the connected network. Note that in (b) and (c), the fields labeled $a_{\rm in}, a_{\rm out}, b_{\rm in, 1}, b_{\rm out, 1}$ are not input and output fields of the connected network. We simply label the connecting links in order to clarify the relationship between the individual components in (a) and the connected network.}\label{fig:aside_linear}
\end{figure}
\begin{aside}[Enhanced squeezing via coherent feedback]\label{ex:linearsys}
The enhancement of squeezing of an optical field through coherent feedback has been examined by several authors \cite{WiseMilb94,Yan.Kim-2003a,Gough:2009wg,Iida12,Cris13}. The simplest experimental configuration for achieving such enhancement is sketched in  \cref{fig:aside_linear}, where a degenerate optical parameter oscillator (OPO) is assembled in feedback with a beam-splitter. The result is a linear quantum network, and here we develop the SLH and $ABCD$ representations of this network.

The SLH triples for the two components are specified as: 
\begin{align}
G_1 &= \left( I , ~[\sqrt{\kappa} a], ~i\varepsilon({a\dg}^2 - a^2) \right) \nn \\
G_2 &= \left( \left[\begin{matrix}{}
-\sqrt{1-\eta^2} & \eta \\
 \eta & \sqrt{1-\eta^2} 
\end{matrix} \right], \left[\begin{matrix}{}
 0  \\
 0 
\end{matrix} \right], 0 \right),
\end{align}
where $a$ is the annihilation operator for OPO cavity mode, $\varepsilon$ parameterizes the OPO nonlinearity, and $\eta$ is the transmission coefficient of the beam-splitter. Note the slight change of convention with \cref{eg:SLHapp2}, where $\eta$ was the reflection coefficient.
 
The first step in developing the SLH representation of the connected network is to form the concatenation product
\begin{align}
G_{\rm unconnected} = G_1 \boxplus G_2 = \left( \left[ \begin{matrix}{}
 I & 0 & 0\\
 0 & -\sqrt{1-\eta^2} & \eta\\
 0 & \eta & \sqrt{1-\eta^2}
\end{matrix} \right], \left[ \begin{matrix}{}
 \sqrt{\kappa} a \\
 0 \\
 0 
\end{matrix} \right], ~i\epsilon({a\dg}^2 - a^2)
\right).
\end{align}
Note that the ordering of the input and output ports of $G_{\rm unconnected}$ in terms of the physical fields denoted in  \cref{fig:aside_linear} are: port 1: $c_{\rm in/out}$, port 2: $b_{\rm in/out, 1}$, port 3: $b_{\rm in/out, 2}$. Next, we connect the output of port 1 to the input of port 2 and the output of port 2 to the input of port 1; \ie apply the feedback reduction rule \cref{eq:SLH_feedback} to connect $1\rightarrow 2$ and $2 \rightarrow 1$ (see  \cref{fig:aside_linear}(c)). 

Applying the feedback reduction rule $1 \rightarrow 2$ results in a network described by the SLH triple
\begin{align}
G_{1\rightarrow 2} = \left( \left[ \begin{matrix}{}
 -\sqrt{1-\eta^2} & \eta\\
 \eta &  \sqrt{1-\eta^2}
\end{matrix} \right], \left[ \begin{matrix}{}
 -\sqrt{1-\eta^2} \sqrt{\kappa} a \\
\eta \sqrt{\kappa} a 
\end{matrix} \right], ~i\epsilon({a\dg}^2 - a^2)
\right).
\end{align}
Notice that we have reduced the number of ports by performing this connection since one of the outputs has been routed to an input. Therefore, the next feedback reduction, which was $2 \rightarrow 1$ according to the port labeling for $G_{\rm unconnected}$ is now $1 \rightarrow 1$ for the system $G_{1 \rightarrow 2}$. Performing this reduction yields a system with a single input-output port and described by the SLH triple:
\begin{align}
G = \left( I, ~[l\sqrt{\kappa}a], ~i\varepsilon({a\dg}^2 - a^2)\right),\, l \equiv\frac{\eta}{1+\sqrt{1-\eta^2}} 
\end{align}
Thus, the effect of feedback is essentially to rescale the cavity decay $\kappa$ by $l^2\le1$.

Given this SLH representation of this active linear component, we can follow  \cref{eq:active_ABCD} to obtain the $ABCD$ representation. The state vector is $\mathbf{\tilde{a}}= \left[ a, a\dg \right]^{\mathsf{T}}$, and the input and output state vectors in terms of the original fields defined in  \cref{fig:aside_linear} are: $\mathbf{\tilde{b}}_{\rm in/out} = \left[b_{\rm in/out,2}, b\dg_{\rm in/out,2} \right]^{\mathsf{T}}$. Carrying out the computations prescribed in \cref{eq:active_ABCD}, we obtain the system matrices:
\begin{align}
\tilde{A} = \left[ \begin{matrix}{}
 -\frac{l^2 \kappa}{2} & \varepsilon \\
 \varepsilon & -\frac{l^2 \kappa}{2} 
 \end{matrix}
\right], ~~~~~ \tilde{B} = -l\sqrt{\kappa} ~I_2, ~~~~~ \tilde{C} = l\sqrt{\kappa}~I_2, ~~~~~ D = ~I_2,
\end{align}
where $I_2$ is the $2\times 2$ identity matrix.  

The relevant squeezing dynamics are more clearly seen in the quadrature basis $\mathbf{\tilde{p}}= 1/\sqrt{2}\left[ a+a\dg, i(a\dg-a) \right]^{\mathsf{T}}$, $\mathbf{\tilde{x}}_{\rm in/out} = 1/\sqrt{2}\left[ b_{\rm in/out,2}+b\dg_{\rm in/out,2}, i(b\dg_{\rm in/out,2}-b_{\rm in/out,2}) \right]^{\mathsf{T}}$, in which the system matrices diagonalize
\begin{align}
\tilde{A}_q = \left[ \begin{matrix}{}
 \varepsilon-\frac{l^2 \kappa}{2} & 0 \\
0 & -\varepsilon-\frac{l^2 \kappa}{2} 
 \end{matrix}
\right], ~~~~~ \tilde{B}_q = -l\sqrt{\kappa} ~I_2, ~~~~~ \tilde{C}_q = l\sqrt{\kappa}~I_2, ~~~~~ D_q = ~I_2.
\end{align}
Then, using \cref{eq:Linear_Laplace}, we can derive the transfer function $\Xi(s)$ that relates $\mathbf{\tilde{x}}_{\rm in}(s)$ to $\mathbf{\tilde{x}}_{\rm out}(s)$ (assuming $\mathbf{\tilde{p}}_0=0$)
\begin{align}
\mathbf{\tilde{x}}_{\rm out}(s)=\left[ \begin{matrix}{}
 \frac{(s-\varepsilon)-\frac{l^2 \kappa}{2}}{(s-\varepsilon)+\frac{l^2 \kappa}{2}} & 0 \\
0 & \frac{(s+\varepsilon)-\frac{l^2 \kappa}{2}}{(s+\varepsilon)+\frac{l^2 \kappa}{2}}
 \end{matrix}\right] \mathbf{\tilde{x}}_{\rm in}(s).
\end{align} 
For simplicity, just consider the steady state input-output response, \ie $s\rightarrow0$ in the above equation.  For $0<\varepsilon<l^2\kappa/2$, the $i/\sqrt{2}(b\dg_{\rm in,2}-b_{\rm in,2})$ the deamplification of the input quadrature is enhanced as $\eta\rightarrow0$, \ie as the beamsplitter becomes increasingly opaque.  In contrast, the other, $1/\sqrt{2}(b_{\rm in,2}+b\dg_{\rm in,2})$ input quadrature quadrature is amplified by the same amount.  Because deamplification of one quadrature is perfectly matched by the amplification of the other, the quadrature phase space of any scattered incident field is increasingly ``squeezed'' as $\eta\rightarrow0$, reducing the deamplified quadrature while preserving total area.
\end{aside}

\subsection{Survey of results regarding linear quantum networks}
\label{sec:linear_review}
Due to the mathematical simplicity of linear quantum networks and their formal similarity to classical linear systems, many results concerning their dynamics and control have been derived. Summarizing all of these  is out of the scope of this review, however, in the following we attempt to survey the major results. For another perspective, we refer the reader to a recent review of linear quantum networks from a control theory perspective by Petersen \cite{Petersen:2016uu}.

Some of most basic characterizations of classical linear systems are their stability, controllability and observability. Most of these characterizations carry over to linear quantum networks with little modification. For example, the notion of Hurwitz stability, captured by the eigenvalues of the $A$ matrix, is the same in the classical and quantum regimes \cite{Gough:2010in}. Controllability and observability in the quantum regime are captured by matrix rank conditions \cite[Chapter 6]{WiseMilb10}\cite{Gough:2013ti} that resemble the Grammian rank conditions for classical linear systems \cite{Ste-1994}.

Many of the most powerful control theory techniques in classical linear systems theory relate to optimal and robust feedback control.  To understand the feedback problem, consider the sketch in  \cref{fig:aside_linear}(c), where some subset of outputs of a quantum network component ($G_1$) are processed by another ($G_2$) and fedback as inputs to the original component. The fundamental question in feedback theory is how to design and realize the ``controller" $G_2$ to achieve some control goal related to the internal variables or outputs of the \emph{closed-loop} system consisting of $G_1$ and $G_2$. 
An important issue that arises in the design of coherent feedback controllers is the \emph{realizability} of the controller. That is, given a specification of a controller, is it physically possible to realize it in hardware using standard optical components? This is not usually an issue in classical control theory since any controller is assumed to be realizable (or can at least be approximated) using digital and analog electronics. In the quantum regime, a linear system specified in linear form, \ie \cref{eq:active_ABCD}, is realizable if and only if the following conditions are met \cite{Gough:2013ti, Jame08, WiseMilb10}:
\begin{align}
\tilde{A} + \tilde{A}^{\flat} + \tilde{C}^\flat \tilde{C}=0, \nn \\
\tilde{B} = -\tilde{C}^\flat \tilde{D}, \nn \\
\tilde{D}^\flat \tilde{D} = I_{2m}
\end{align}
A linear quantum system that meets these conditions is guaranteed to preserve the canonical commutation relations of the underlying system degrees of freedom, thus meeting that fundamental requirement for physical realizability. 

In the quantum context, very little is known about how to design such coherent controllers. Especially challenging is optimal or robust design where the closed-loop system behaves optimally according to some criteria or has guarantees of performance robustness. When $G_1$ and $G_2$ are both linear systems, Yanagasiwa and Kimura proposed to approach the problem of controller design using the transfer function matrix description of linear quantum networks \cite{Yan.Kim-2003, Yan.Kim-2003a}. This was followed by two notable extensions of powerful techniques from classical linear systems control to linear quantum networks: (i) the notion of optimal feedback controllers for the linear-quadratic-Gaussian (LQG) problem \cite{NurdJamePete09}, and (ii) the notion of $H^\infty$ robust control \cite{Jame08}. 

LQG control for a linear quantum system with Gaussian inputs aims to minimize a quadratic function of the integrated outputs, and possibly a quadratic function of the control inputs, of the closed-loop system; \eg the cost function $J_t = \int_0^t ds \expect{\mathbf{\bout}\dg(s) \mathbf{\bout}(s)}$. Such a controller design problem is common in classical linear control theory, where it is solved by a simple convex optimization or more commonly, by determining the solution to matrix Riccati equations \cite{Ste-1994}. The solution to the quantum LQG is complicated by the realizability conditions on the controller, which are not easily incorporated into a convex optimization. However, to overcome this obstacle Nurdin \etal transform the quantum LQG controller design problem into a computationally tractable rank constrained linear matrix inequality (LMI) problem \cite{NurdJamePete09}.

The optimal controller determined LQG controller design does not have any stability or robustness guarantees. In particular, if the model for the system $G_1$ is inaccurate or has uncertainties, the feedback controller may not perform as expected. A major success of modern classical linear control theory is the formulation of robust control, where the feedback controller can be designed to be robust to such model uncertainties. In Ref. \cite{Jame08} James \etal generalize one of the key tools from classical robust control, $H^\infty$ control to the quantum linear systems case. As with the extension of LQG control design, the key innovation by James \etal is a formulation of the $H^\infty$ controller design problem that incorporates controller realization conditions so that the resulting coherent feedback controller $G_2$ is guaranteed to be realizable.

Another direction in which there has been significant progress over the past few years is the controller synthesis problem. As mentioned above, there are strict realizability conditions on linear quantum systems. The coherent controller design methods described above incorporate these conditions, but even if the resulting controller is realizable, how does one construct it from basic optical components? This is the topic of \emph{controller synthesis} or \emph{realization} theory. Nurdin \etal established that an arbitrary linear quantum system can be synthesized by a chain of cascaded harmonic oscillator modes (\eg cavities) with some direct, \ie Hamiltonian, interactions between some modes, and provided a constructive procedure to determine the particular network required \cite{NurdJameDohe09}. {Later, Nurdin developed a scheme for removing the direct interactions and effectively implementing them through more complex, but completely field mediated, connections \cite{Nurdin:2010ws}. If the synthesis problem is restricted to realizing the transfer function (as opposed to the particular $ABCD$ matrices), then Nurdin has established that this can be achieved through a purely cascaded harmonic oscillator network, for passive linear systems \cite{Nurdin:2010vs}, and for arbitrary linear quantum systems \cite{Nurdin:2016kb}. It was also recently shown that several other network topologies of harmonic oscillators can be used to synthesize passive linear systems \cite{Gough:2013ti}. }
Finally, some other techniques that have been ported from classical linear systems theory to linear quantum networks are a variation of balanced truncation for linear system model reduction \cite{Nurdin:2014bm, Techakesari:2015ia}, and system identification for passive linear networks \cite{Guta:2016hy}.

\section{Extensions to the SLH framework}\label{sec:generalizations}

In this section we describe some extensions to the SLH framework that enable one to model commonly encountered experimental arrangements, phenomena, and imperfections. The extensions discussed involve applications of the standard SLH building blocks to capture more complex behavior such as back-reflection from interfaces, while preserving the modular network structure. In many instances, the extension boils down to approximating the more complex behavior as an interaction of freely propagating fields with a sequence of customized components.
Such extensions and applications of SLH are an active area of research, and therefore the extensions we discuss are not meant to be all-encompassing. Instead, the following sections are intended to give the reader some intuition about how to model more complex phenomena using the SLH framework. 
 
\subsection{Non-vacuum input states via source models}\label{sec:non_vac_input}
The SLH framework relies on all field input states into the network being in the vacuum state. In particular, the network composition rules were derived using this assumption. However, in most cases encountered in practice the input fields will be in non-vacuum states. Fortunately, there are simple extensions to the framework that accommodate these situations. 

\begin{figure}[h]
\includegraphics[width=\columnwidth]{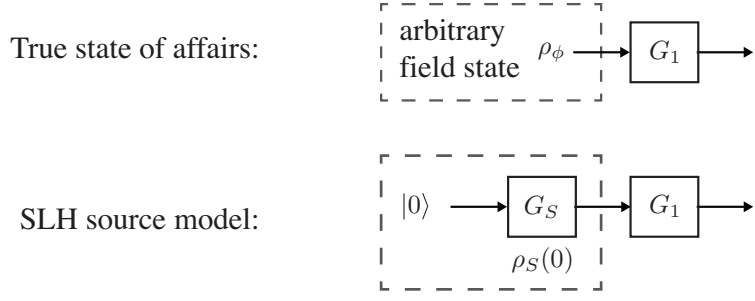}
\caption{
To model the driving a system with a field with arbitrary statistics we introduce a fictitous engineered source system. The source has a particular initial state $\rho_S(0)$ at $t=0$ and a description in terms of an SLH triple.
}\label{figSourceModels}
\end{figure}

The most commonly used method for accommodating non-vacuum input states is to explicitly model a network component that produces the input field state from vacuum input; in most instances this component is a minimal model for an idealized physical apparatus that produces the desired field states.
The general approach is to replace an arbitrary (possibly mixed) state of the field with a system with a particular initial state and then drive it with vacuum as depicted in \cref{figSourceModels}. In particular we wish to engineer some fictitious ``source'' system $G_S=(S_S,L_S,H_S)$ and initial state of the system $\rho_S(0)$ such that another system $G_1$ behaves as if it was driven by the arbitrary field state $\rho_\phi$. The combination of $G_S$ and $\rho_S(0)$ is often referred to as a \emph{source model}. We note that developing source models and modeling a system driven with light of arbitrary statistics was Gardiner's original motivation for developing the theory of cascaded systems \cite{Gard93}. Early work by Gardiner and Parkins analyzed simple two-system cascades to model driving an atom with thermal or finite-bandwidth squeezed light, and specified source models for these light sources \cite{GardPark94}. 
We now summarize some of the source models that have been constructed to generate commonly encountered field states. 

\subsubsection{Coherent states} 
Continuous-mode coherent states provide an accurate description of pulsed laser light and are mathematically defined  by \cite{BlowLoudPhoe90}
\begin{align} \label{eq:WPcoher}
\ket{ \alpha_{\xi} } =D(\alpha_{\xi})\ket{0}=\exp \left[ B\dg(\alpha_\xi) - B(\alpha_\xi) \right] \ket{0},
\end{align}
where $D(\alpha_{\xi})$ is the symmetrically ordered displacement operator and $B\dg(\alpha_\xi)$ is a wave packet creation operator
\begin{align}
B\dg(\alpha_\xi) = \alpha \int_{-\infty}^\infty ds \, \xi(s) b\dg(s),
\end{align}
and $\ket{0}$ is multimode vacuum.  The square normalizable function  $\xi(t)$ defines the wave packet temporal profile and 
the mean photon number in the wave packet is $\bar{n} = |\alpha|^2\int_{-\infty}^\infty ds |\xi(s)|^2$ \cite{BlowLoudPhoe90}. For convenience we define $\alpha(t) \equiv \alpha \xi(t)$. Continuous-mode coherent states have the eigenvalue relations
\begin{align}
dB(t) \ket{ \alpha_\xi } = \alpha (t) dt \ket{ \alpha_\xi }.
\end{align}
This definition of continuous-mode coherent states includes single-mode coherent states ($\alpha (t)=\alpha$ ) and vacuum ($\alpha (t)=0$) as special cases.  The displacement operator in \cref{eq:WPcoher}, also known as the \emph{Weyl operator}, has its own QSDE \cite{BoutHandJame07}
\begin{align} \label{eq:weylQSDE}
dD(\alpha_\xi) =\left[-\half |\alpha(t)|^2dt + \alpha(t) dB(t)\dg -\alpha^*(t)dB(t) \right]D(\alpha(t)).
\end{align}
This QSDE will provide an intuitive crutch for understanding the source models below.

Consider the simple, but non physical, source model:
\begin{align}\label{eq:coher_source_model}
G_{\rm coherent}= (1,\alpha(t),0). 
\end{align}
Driving a target system $G_{1}$ is simply a mater of performing the series product $G_{1}\lhd G_{\rm coherent}$. This model can be understood by subsituting the above SLH triple into \cref{eq:QsdeU}. Doing so yields $dU(t) = \{ -\half |\alpha(t)|^2dt + \alpha(t) d\Bin\dg(t) -\alpha^*(t)d\Bin(t) \} U(t)$ and since $U(t=0)=I_{\rm field}$, we have exactly \cref{eq:weylQSDE}. Thus when $G_{\rm coherent}$ is driven by vaccum it coherently displaces the field and the output of the component is the state \cref{eq:WPcoher}. This is consistent with a  Mollow transformation \cite{Moll75} of the output field, see \cref{eq:mollow_trans}. The source model in \cref{eq:coher_source_model} allows one to easily derive the master equation for a SLH network driven by a coherent state, see \cref{ex:coherent_state_me}.

A physical source model that produces this state as its output is specified by a cavity prepared in the input state $\rho_S(0)$ with a vacuum input field and SLH triple \cite{GougJameNurd12,GougZhan15} 
\begin{align}
{G}_{\alpha}= (I,\lambda(t)a,\Delta(t)a\dg a)  \quad {\rm and}\quad \rho_S(0)= \op{\alpha}{\alpha}
\end{align} 
with 
\begin{align}
\Delta(t)&=0 \quad {\rm and}\\ 
\lambda(t) &= \frac{1}{\sqrt{W(t)}}\xi(t), \quad {\rm where}\quad W(t)= \int_t^\infty ds |\xi(s)|^2. \label{eq:lambda_model}
\end{align}
This SLH component yields exactly the same input-output behavior as $G_{\rm coherent}$ \cite{GougJameNurd12,GougZhan15}.
The first source model is usually preferred since it is numerically more efficient (the cavity degree of freedom does not need to be modeled). However, the second source model generalizes more easily and guides the development of source models for producing other field states.

{ Finally we note that in some cases it is important to model laser light that has finite bandwidth \cite{DrumReid88}. The usual assumption is field amplitude or  intensity is relatively well stabilized so it is phase diffusion that ultimately leads to the finite bandwith. In this model we can consider $\alpha(t)\mapsto \alpha(t) \exp[i\phi(t)]$ where $\phi(t)$ describes the phase diffusion as a function of time. If $\expt{\dot\phi(t)\dot\phi(s)}= \gamma \delta (t-s)$ then the laser has a Lorentzian spectrum with a FWHM bandwidth of $\gamma /2\pi$ Hz \cite{DrumReid88}. }

\begin{aside}[The coherent state master equation and the $S$ operator]\label{ex:coherent_state_me}
In this example, we derive a master equation describing the dynamics of a system driven by a multimode coherent state. We sketch two different methods to derive the same master equation. The first method is straightforward, but it only works for coherent states. The second method is more general and can be used to derive master equations describing the dynamics of QIONs driven by Fock~\cite{BaraCookBran12} or cat~\cite{GougJameNurd12} states. 

The first method cascades the coherent state source model (\cref{eq:coher_source_model}) into an arbitrary localized component and then calculate the standard vacuum master equation, \cref{eq:vac_master_eqn}, for this cascaded system. The cascaded system is:
\begin{align}
 G_{\rm T} = (S_{\rm T}, L_{\rm T},H_{\rm T})
&= (S, L,H) \triangleleft  (1, \alpha(t),0) =\left ( S, L+S\alpha(t), H+\frac{1}{2i}(L^\dag S\alpha(t)-\alpha^*(t) S^\dag L) \right).\label{eq:coh_drive}
\end{align}
From \cref{eq:vac_master_eqn} the vacuum master equation for this model is $d\rho = -i[H_{\rm T}, \rho(t)]dt + \mathcal{L}[L_{\rm T}]\rho(t)dt $, which is manifestly in Lindblad form. Another, equivalent form for this equation, which is often encountered in the literature, is:
\begin{align}  \label{eq:meS}
\frac{d}{dt}\rho(t) = & -i[ H,\rho(t)] + \mathcal{L}[L]\rho(t)   +\alpha(t) [S\rho(t),L\dg] +\alpha^*(t) [L,\rho(t) S] + |\alpha(t)|^2(S\rho(t) S\dg-\rho(t)),
\end{align}
This form highlights the fact that the $S$ operator can appear in the master equation when the system is driven by a non-vacuum field.

The second method for deriving this master equation proceeds directly from the Heisenberg equation of motion, \ie \cref{eq:dX}. From a Heisenberg picture description it is possible to obtain the Schr\"odinger picture evolution by noting that
\begin{align} 
		\expt{ X(t)}_{\alpha_\xi} & =\! \tr_{\rm sys + field}\left [   \left( \rho(0)\! \otimes \! \op{ \alpha_\xi}{ \alpha_\xi } \right) X(t) \right ]\nonumber  \\
		& =\! \tr_{\rm sys + field}\left [  {U(t) \left( \rho(0) \otimes \! \op{ \alpha_\xi}{ \alpha_\xi } \right)U\dg(t) X\! } \right ]\nonumber  \\
	& = \!\tr_{\rm sys} \left [ { \rho(t)  X } \right ] ,
	\end{align} 
where and define $\rho(t)= \tr_{\rm field}[U_t(\rho_0\otimes\op{\alpha_\xi}{\alpha_\xi}  ) U_t \dg]$. Recall that $dX(t)$ is really a notational short cut for the quantum flow in \cref{eq:qflow}. Thus if we take the trace of $dj_t(X)$ with the inital state $\rho_0\otimes\op{\alpha_\xi}{\alpha_\xi} $ and use the above manipulation, for every term in \cref{eq:qflow}, we can derive the master equation. However we will need to know the action of the quantum noise increments, $dB_{\rm in}$ \& $d\Lambda_{\rm in} $, on the input field state:
\begin{subequations}
\begin{align}
	dB_{\rm in}(t)\ket{\alpha_\xi} & = \, \alpha(t) dt\ket{\alpha_\xi},\\
	d\Lambda_{\rm in}(t)\ket{\alpha_\xi} & = dB_1\dg \alpha(t) \ket{ \alpha_\xi}.
\end{align}
\end{subequations}
For example, consider the term $j_t([L\dg,X]S )d\Bin(t)$ in \cref{eq:qflow}
\begin{align}  
 \tr \!\left [(\rho_0\otimes\op{\alpha_\xi}{\alpha_\xi}  ) U_t\dg [L\dg,X]S U_t d\Bin(t) \right ] 
 &= \alpha(t) dt\, \tr \!\left [(\rho_0\otimes\op{\alpha_\xi}{\alpha_\xi}  ) U_t\dg [L\dg,X]S U_t \right ] \\
 &= \alpha(t) dt\, \tr \!\left [ U_t(\rho_0\otimes\op{\alpha_\xi}{\alpha_\xi}  ) U_t \dg [L\dg,X]S  \right ] 
\end{align}
now we explicitly take the field trace to obtain
\begin{align}  
 \alpha(t) dt\, \tr _{\rm sys}\!\left [ \rho(t) [L\dg,X]S  \right ] 
 &=   \alpha(t) dt\, \tr_{\rm sys} \!\left [ [S  \rho(t) ,L\dg] X  \right ] 
\end{align}
which implies the term $\alpha(t) dt [S\rho(t),L\dg]$ should appear in the coherent state master equation.
Carrying this out for all the terms in \cref{eq:qflow}, we arrive at the full coherent state master equation i.e. \cref{eq:meS}.
\end{aside}

\subsubsection{Finite-bandwidth squeezed states}
The very first paper~\cite{CollGard84}, and early applications \cite{ParkGard88,RitsZoll88,RitsZoll88a}, of input-output theory were about driving systems with squeezed light. Squeezed light produced by realistic sources, \eg a degenerate optical parametric oscillator (OPO), is bandwidth limited, typically by the transitions linewidths of the atoms in the non-linear medium. A source model for such a source is given by a cavity model with SLH triple \cite{RitsZoll88,RitsZoll88a,GardPark94}:
\begin{align}
G_{\rm squeezed} = \left (I, \sqrt{\gamma}a, \frac{i}{2}(E {a\dg}^2 - E^* a^2) \right),
\end{align}
where $a$ is the cavity mode, $\gamma$ is the bandwidth of the squeezed light. Note that this source is explicitly modeling the light source (degenerate OPO), and $|E|$ is proportional to the amplitude of the pump field for this setup. 

The output field from this source is quadrature squeezed with some finite bandwidth. The normally ordered quadrature variances (when $E$ is chosen to be real and positive) are explicitly \cite{CollGard84,GardPark94,RitsZoll88a, RitsZoll88}:
\begin{align}
\expect{: X(t+\tau), X(t) :} &= \frac{\gamma E}{2} \frac{\exp \left( -(\frac{1}{2}\gamma - E)|\tau| \right)}{\frac{1}{2}\gamma - |E|} \nn \\
\expect{: Y(t+\tau), Y(t) :} &= -\frac{\gamma E}{2} \frac{\exp \left(-(\frac{1}{2}\gamma + E)|\tau| \right)}{\frac{1}{2}\gamma + |E|},
\end{align}
where the field quadratures $X$ and $Y$ are related to the anilation operatory by $a = X +iY$, $:O:$ denotes normal ordering of the expression $O$, and  $\expect{a,b} \equiv \expect{ab} - \expect{a}\expect{b}$. In this model the $Y$ quadrature of the output field is squeezed. Similar models can be constructed for two-mode squeezed states~\cite{ReidWall86}, which also implies that one can construct a source model for thermal states by simply tracing out (ignoring) one mode of the two-mode squeezed state source model \cite{Wang:2008hu}.

\subsubsection{Fock and $N$-photon states}\label{sec:fockstates}
A continuous-mode single-photon state is a single photon coherently superposed over many spectral modes with the spectral density function $\tilde{\xi}(\omega)$ determing the weight of the superposition. In the time domain, $\xi(t)$ is a square-normalized temporal wave packet, $\int dt \, |\xi(t)|^2 = 1$, that modulates the carrier frequency  \cite{BlowLoudPhoe90,Loud00}:
\begin{align}  \label{eq:singlephoton}
\ket{1_{\xi}} 
& =B\dg(\xi) \ket{0} =\int d\omega \, \tilde\xi(\omega) b\dg(\omega)\ket{0} = \int dt \, \xi (t) b\dg(t) \ket{0}
\end{align}
where $[B (\xi), B\dg (\xi)]=1$. The state $\ket{1_{\xi}}$ can be viewed as a superposition of instantaneous photon creation times weighted by the temporal wave packet. Continuous-mode Fock states in the wave packet $\xi (t)$ with $N$ photons can be constructed in the usual way \cite{BlowLoudPhoe90}:
\begin{align} \label{eq:Fock}
	\ket{N_{\xi}}  &=\frac{1}{\sqrt{N!}}\left [ B\dg(\xi)\right ]^{N}\ket{0},
\end{align}
and have the eigenvalue relation
\begin{align}
dB_t \ket{ N_\xi } =  \sqrt{N} \xi (t) \,dt  \ket{ N-1_\xi }. 
\end{align}
The temporal superposition present in Fock states means that there will be temporal correlations between different times for any system interacting with such a state. Thus systems driven by Fock states necessarily behave in a non-Markovian fashion. Using a clever source model this can be represented as a larger Markovian system.

The first cascaded model for a single photon was first discovered by Gheri \etal ~\cite{GherElliPell98}. This was subsequently generalized to any superposition or mixture of single photon and vacuum, ie. $\rho_\phi=\sum_{j,k=0}^1 \gamma_{kj}\op{\phi_j}{\phi_k}$ where $\ket{\phi_1}=\ket{1_\xi}$ and $\ket{\phi_0}=\ket{0}$, by Gough \etal ~\cite{GougJameNurd12}. This source model consists of a two level atom with the initial state $\rho_S(0) = \sum_{j,k=0}^1 \gamma_{kj}\op{j}{k}$ dipole coupled to the vacuum $(I, \lambda(t)\sigma_-,0)$, with $\lambda(t)$ given by \cref{eq:lambda_model}. The general source model for a Fock state is \cite[see Theorem 2]{GougZhan15}
\begin{align}
G_{\rm Fock}=(I,\lambda(t)a ,0)\quad {\rm with}\quad \rho_S(0) = \op{n}{n} .\label{eq:fock_source}
\end{align}
In many experimental settings one can create a state of light with a fixed photon number but it can not be written in the form of \cref{eq:Fock}. Such states have a definite number of photons but in an arbitrary spectral distribution function $\tilde\psi(.)$, and are called $N$-photon states. A general $N$-photon state is 
\begin{align}\label{eq:SDFnphoton}
\ket{\psi_N} =\int d\omega_1 & \dots d\omega_N \, \tilde\psi(\omega_1,\dots,\omega_N) b\dg(\omega_1)\dots b\dg(\omega_N) \ket{0}.
\end{align}  
Then, in the time domain a general $N$-photon state can be written as
\begin{align}\label{eq:TDFnphoton}
		\ket{\psi_N} =\int dt_1 & \dots dt_N \, \psi(t_1,\dots,t_N) b\dg(t_1)\dots b\dg(t_N) \ket{0}.
\end{align} 
Master equations have been derived for systems driven by this kind of field~\cite{BaraCookBran12}. {Source models for such input states exists but are fairly complicated. See the work of Gough \etal, where they give a class of source models for a large family of field states termed \emph{continuous matrix product states} \cite{Gough:2014gj}. Continuous-mode $N$-photon states belong to this family and in this case the source models coincide with the ones described above.}  However there is an interesting special case that has been solved. Consider $N$ photons in different wavepackets, $\psi_i$, possibly overlapping in time (\eg a photon gun); \ie
\begin{align}\label{eq:sepTDF}
		\ket{\psi_N} \propto \int dt_1 & \dots dt_N \, \psi_1(t_1)\dots \psi_N(t_N) b\dg(t_1)\dots b\dg(t_N) \ket{0}.
\end{align} 
This input state can be mimicked by a source model that is a multimode cavity with $N$ different time dependent couplings, $\lambda_i(t)$, to the same input-output field. This source model is detailed in Theorem 3 of Ref. \cite{GougZhan15}.

\subsubsection{Cat states}
We shall often refer to superpositions of continuous-mode coherent states as (continuous-mode) cat states. The cat states we consider are
\begin{align}
\ket{\psi_{\rm cat}} = \sum_{j=1}^{n} s_j \ket{\alpha_j(t)},
\label{eq:super-state}
\end{align}
where $\ket{\alpha_j(t)}$ are coherent states, determined by complex-valued functions $\alpha_j(t)$ with $\alpha_j(t) \neq \alpha_k(t)$ if $j \neq k$. The superposition weights $s_j$ are complex numbers such that the state is normalized $\langle \psi | \psi \rangle =\sum_{j,k} s_j^* s_k\expect{\alpha_j(t)|\alpha_k(t)} =1$. Constructions for the source system are given in section IV. C of \cite{GougJameNurd12} and section 4 of \cite{GougZhan15}. In Ref.~\cite{GougJameNurd12} the source model is  a qudit with $n$ levels and the $L$ operators are projectors onto the $j$th qudit level with time dependent couplings given by $\alpha_j(t)$. The initial state of the qudit, $\rho_S(0)$, is carefully chosen and related to  $s_j $ and $\langle \alpha_k(t) | \alpha_j(t)\rangle$. In Ref.~\cite{GougZhan15}, the construction involves a multimode cavity instead of a qudit.

 \subsection{Alternatives to source models}\label{sec:alternative_source_mod}
In the following we will review two alternatives to source models for accommodating non-vacuum field input states. These are important because in some cases it may be difficult to construct a source model for the field driving a QION.

The first alternative to source models proceeds by decomposing an arbitrary input field into a basis that we can do quantum stochastic calculus in, \ie one in which we can derive a master equation. If necessary, the field can be approximated by truncating in that basis. There are three bases, so far, that we can work with (1) Fock states~\cite{BaraCookBran12,BaraComb17}, (2) $N$ photon states~\cite{BaraCookBran12}, {(3) a subset of multiphoton states in $M$ modes \cite{SongZhanXi16}, and (4) cat states~\cite{GougJameNurd12,GougJameNurd13}.} In these bases the ordinary SLH composition rules apply. For simplicity we restrict the following discussion of this approach to a single input output mode and to Fock states.

The second alternative for dealing with non-vacuum input states aims to extend the SLH framework itself to accommodate an important class of input states: Gaussian states. 
 
\subsubsection{Simulation in a Fock basis}\label{sec:fock_input}
Unentangled Fock states, \ie a state of the form \cref{eq:Fock}, span single mode Hilbert (Fock) space and form a basis for arbitrary states within the wave packet $\xi(t)$,
 \begin{align} \label{eq:fockdecomp}
 \rho_{\mathrm{field}} = \sum_{m,n=0}^\infty c_{m,n} \op{m_{\xi}}{n_{\xi}},
 \end{align}
 where $\rho _{\mathrm{field}} \geq 0$, $\mathrm{Tr}[\rho _{\mathrm{field}}]=1$ and $\rho _{\mathrm{field}\phantom{\dg}}=\rho _{\mathrm{field}}\dg$.  
 Using the techniques introduced by Baragiola \etal ~\cite{BaraCookBran12} we can describe the dynamics of an SLH network described by the triple $(S,L,H)$ when the input field is given by \cref{eq:fockdecomp}. The state of the SLH node at any time is
 \begin{align} \label{eq:totalstate}
 	\varrho_{\rm total} (t)=\sum_{m,n} c_{m,n} \varrho_{m,n}(t),
 \end{align}
 where the generalized state matrices $\varrho _{m,n}(t)$ are the solutions to a set of master equations. The set of coupled master equations are~\cite{BaraCookBran12}
 \begin{align} \label{eq:fockME}
 \frac{d}{dt} \varrho_{m,n}(t) 
 & = - i[H, \varrho_{m,n} ] + \mathcal{L}[L] \varrho_{m,n} \nonumber  \\
 & \!+\! \sqrt{m} \xi(t) [S \varrho_{m-1,n}, L^\dagger] \!+\!  \sqrt{n} \xi^*\!(t) [L,\varrho_{m,n-1} S^\dagger  ] \nonumber \\
 & \!+\! \sqrt{mn} |\xi(t)|^2\!\left(S \varrho_{m-1,n-1} S^\dagger - \varrho_{m-1,n-1}  \right) .
 \end{align}
 The initial conditions for these equations are: $\varrho_{n,n}$ should be initialized with the initial system state $\rho_{\rm sys}$, the off-diagonal equations, $\varrho_{m,n}$ for $m\neq n$, should be initialized to zero. Some special cases of \cref{eq:fockME} were first derived in Ref.~\cite{GherElliPell98} and extended in Ref.~\cite{,GougJameNurd12}.
  
 To compute expectations of observables it is helpful to define the expectation value,
 \begin{align}  \label{eq:fockexpt}
 \mathbbm{E}_{m,n}[ O ] &  \equiv  \tr_{\rm sys }[ \varrho\dg_{m,n} O].
 \end{align}
 where $O$ is a (possibly) {joint} operator on the system and field. Then an expectation value of a system operator $X$ is given by
 \begin{align}
 \mathbbm{E}_{\rm total}[X(t)] 
 &={\rm Tr}_{\rm sys+field}\left[ \varrho\dg_{\rm total}(t) X \right]=\sum_{m,n} c^*_{m,n} \mathbbm{E}_{m,n}[X(t)] \label{eq:opexptfock}.
 \end{align}
 This equation also allows us to compute output field quantities; \eg in the case of the output photon flux, taking expectations over Fock states using \cref{eq:opexptfock} yields an equation for the mean photon flux, 
 \begin{align} 
 \frac{d}{dt} \emn{m,n} [\Lambda^{\rm out}_t(t)] &= \emn{m,n} [ L\dg L]  + \sqrt{m} \xi^*(t) \emn{m-1,n} [  S\dg L ] 
  + \sqrt{n} \xi(t) \emn{m,n-1} [ L\dg S ] + \sqrt{mn} |\xi(t)|^2  .
 \end{align}
 The solution to this equation $\mathbbm{E}[\Lambda_t^{\rm out}(t)]$ gives the integrated mean photon number up to time $t$. This technique has been extended to multiple input-output modes and spectrally entangled input states in Ref.~\cite{BaraCookBran12}. Baragiola's thesis is a reference for this topic~\cite{Bara14}.

Finally, we note that if $\rho_{\mathrm{field}}$ has a large mean field component then the Mollow transformation \cite{Moll75}, 
 \begin{subequations}\label{eq:mollow_trans}
\begin{align}
dB_t & \mapsto dB_t  + \alpha(t)dt \\
dB_t\dg & \mapsto dB_t\dg + \alpha^*(t)dt \\
d\Lambda_t & \mapsto 	d\Lambda_t + \alpha^*(t)dB_t + \alpha(t)dB_t\dg + |\alpha(t)|^2 dt,
\end{align} 
 \end{subequations}
can be used to transform away the mean field and thus more efficiently simulate in a displaced Fock basis. In addition to master equaiton methods Monte Carlo methods, i.e. quantum trajectories, can be used to simulate these equatons see \cite{BaraComb17} and the references therein.

\subsubsection{General gaussian input states}\label{sec:gaus_input}
Gaussian states are a wide class of field states that are particularly important because many experimental sources of light produce Gaussian states, \eg coherent, squeezed, and thermal states. Because of their experimental relevance there are extensive reviews on Gaussian states in quantum optics and information, see Refs.~\cite{Ferraro:2005tp, Wang:2008hu, Weedbrook:2012tz}.
Here we discuss the feasibility of incorporating these field states as inputs to a QION. 

A Gaussian state in quantum theory, call it $\rho_G$, is a state where the (possibly complex) quasi-probability distribution, \eg Glauber--Sudarshan $P$ function, Wigner $W$ function, or Husimi $Q$ distribution, is Gaussian. For a single mode this is equivalent to a density operator that is the exponential of a quadratic in the annihilation and creation operators -- \ie $\rho_G \propto \exp \left (-c_0a\dg a -c_1a a\dg -c_2a^2-c_3a^{\dagger 2}-c_4 a -c_5 a\dg \right )$ for $c_i \in \C$~\cite[see Sec. 4.4.5]{GardZoll-2004}. An alternative way to characterize a complex Gaussian state is by the first and second order moments of $a,a\dg$ (the mean and covariance matrix). The relationship between these moments and the numbers $c_i$ is explained in ~\cite{GardZoll-2004}. A multimode field in a Gaussian state is also characterized by its first and second moments, which can be written explicitly as:
\begin{subequations}\label{eq:gauss_expect}
\begin{align}
 \expt{b(t)}_G &= \alpha(t)  \\
\expt{b(t)b(t')}_G &= M\delta(t-t')    \quad \quad \expt{b\dg(t)b\dg(t')}_G = M^*\delta(t-t')  \\
\expt{b\dg(t)b(t')}_G &= N\delta(t-t')     \quad \quad \,\,\,\, \expt{b(t)b\dg(t')}_G = (N+1)\delta(t-t') ,
\end{align}
\end{subequations}
where $\alpha(t),M \in \C$ and $N \in \R$. We have assumed here that the state has stationary second moments, while allowing the mean to be time-varying. The parameters $N$ and $M$ parameterize the covariance ellipse of the Gaussian, and are constrained by the inequality 
\begin{align}
N(N+1)\ge |M|^2,
\end{align}
which constrains the Gaussian state to have enough phase space area to be a valid quantum state satisfying the Heisenberg uncertainty relation. When $M=0$ the field is in a thermal state with $N= N_{\rm th}$ thermal photons. Non-zero $M$ indicates a squeezed state of the mode, and when $N(N+1)=|M|^2$ there is only sqeezing and no thermal photons~\cite{GardZoll-2004}. 

\begin{Remark}[\bf Squeezing parameters]
	When $M$ is non zero a more convenient parameterization is in terms of physical squeezing parameters, namely 
$M = e^{-2i \phi}\sinh(2r)(N_{\rm th} +\half)$
and $N= \cosh(2r)N_{\rm th} + \sinh^2(r)$,
where the parameters $r,\phi$ appear in the squeezing operator $S(r,\phi)= \exp\left [\half r (b^2 e^{-2i \phi} - b^{\dagger 2}e^{2i \phi} ) \right ]$, and are known as the squeeze factor and the squeeze angle, respectively. The parameter $N_{\rm th}$ denotes the number of thermal photons, \eg $N(N+1)=|M|^2$ implies $N_{\rm th}=0$. In experimental literature squeezing is usually calculated in decibels, and the conversion here is $r_{\rm dB}= 10 \log_{10}(e^{2r})= 20 r \log_{10}e $. The analysis we have presented is in the interaction picture, the relationship between this picture and the usual notion of side bands of the carrier frequency is presented in \cite{CaveSchu85} and  \cite[Sec. 10.2]{GardZoll-2004}.
\end{Remark}

The quantum It\={o} table corresponding to \cref{eq:gauss_expect} is
\begin{subequations}\label{eq:gass_ito}
\begin{align}
 \expt{dB_t} &= \alpha(t) dt \\
 dB(t)dB(t) &= M dt   \quad\quad  dB\dg(t)dB\dg(t) = M^* dt   \\
dB\dg(t)dB(t) &= N dt  \quad\quad  \,\,\,dB(t)dB\dg(t) = (N+1) dt   .
\end{align}
\end{subequations}
Typically the mean field component is removed via the Mollow transformation \cite{Moll75}, see \cref{eq:mollow_trans}. For this reason most authors consider $\alpha(t)= 0 $ unless explicitly stated.

\emph{Single components with Gaussian input fields.} The interaction of single localized components with white noise Gaussian fields has been extensively studied in the quantum optics literature \cite{CollGard84,Gard86,RitsZoll88,Cira92,DumParkZoll92,GardParkZoll92,TurcGeorHood98,WiseMilb10}. In fact, the description of Gaussian fields interacting with single quantum systems has been very successful; \eg Gardiner's predictions ~\cite{Gard86} of inhibited atomic phase decays in a squeezed light environment was recently verified experimentally~\cite{MurcWebeBeck13}. At the core of this description is the It\={o} QSDE that describes the system-field evolution (under the same interaction Hamiltonian and approximations described in \cref{sec:cascade}) when the itinerant single mode field that the system interacts with is in a Gaussian state \cite{GardColl85,DumParkZoll92,GardParkZoll92}:
\begin{align}   
dU(t) = \Big\{ - \left( iH +\half \left [(N+1)  L\dg L  +N LL\dg -M^*LL - ML\dg L\dg \right ]  \right )dt   + L d\Bin\dg(t) - L\dg  d\Bin(t)  \Big\} U(t),
\label{eq:gauss_prop}
\end{align}
with $U(0)=\Isf$,
and the increments $dB_{\rm in}, dB\dg_{\rm in}$ satisfy the It\={o} table \cref{eq:gass_ito}, $H$ is the localized component's Hamiltonian, and $L$ is the operator that coupled with the itinerant field mode. Generalization of this propagator to the case of multiple input-output modes is straightforward because the input fields are orthogonal; \ie it effectively amounts to adding an index ``$i$" to $L$, $d\Bin$, $M$, and $N$, and summing over $i$. Using the general relation between input and output fields, \cref{eq:BoutBin}, one can show that the input-output relations remain unchanged under this propagator. However, the equation of motion for a system operator $X$, is modified to (cf.~\cref{eq:dX}):
\begin{subequations}
\begin{align}  
dX(t) =& -i[X, H]dt \nn\\
&+ (N+1)\mathcal{L}\dg[L]X dt+N\mathcal{L}\dg[L\dg]X dt +M [L\dg,[L\dg,X]] dt +M^*[L,[L,X] ]dt    \nn \\
		&+ [L\dg,X] d\Bin(t) +[X,L] d\Bin\dg(t).  
\end{align}
\end{subequations}
This equation of motion gives rise to the master equation for localized degrees of freedom:
\begin{align}  
\dot\rho =& -i[H + i(\alpha^*(t)L-\alpha(t)L\dg),\rho] + (N+1)\mathcal{L}[L]\rho +N\mathcal{L}[L\dg]\rho  +M [L\dg,[L\dg,\rho]]  +M^*[L,[L,\rho]].    
\end{align}
While this master equation is not written in Lindblad form it can be brought into such form via diagonalization ~\cite{DumParkZoll92}. As for the vacuum input master equation, \cref{eq:vac_master_eqn}, there exist homodyne and heterodyne unravellings of this master equation, \ie stochastic master equations or quantum filters, see \cite[Sec. 4.4.1]{WisePHD94} and \cite[Sec. 4.8.2]{WiseMilb10}.

Note that there is no $S$ matrix in \cref{eq:gauss_prop}, and thus this equation does not capture pure scattering dynamics. This is partly because such dynamics were not of concern when the equation was derived \cite{GardColl85,DumParkZoll92,GardParkZoll92}, but as we will discuss further in the next section, there are some fundamental obstacles to incorporating pure scattering dynamics in the presence of arbitrary Gaussian input fields. 

\emph{Quantum networks with Gaussian input fields.} In the spirit of cascaded systems one can also seek to model the dynamics of a quantum network of localized components that is driven by Gaussian fields. Two studies that have examined this are Ref. \cite{GardPark94}, which considered cascading cavities, atoms, and beamsplitters driven by thermal and squeezed fields, and Ref. \cite{WiseMilb94}, which considered systems in series and feedback configuration driven by Gaussian fields.  However, these studies construct the network dynamics manually on a case-by-case basis, like we derived the dynamics of a cascaded system in \cref{sec:CascadeSys}, \ie by relating the output field of one component to the input field of another. Of course, it would be more desirable to have a general and systematic approach that prescribe algebraic rules for constructing network components. 

In response to this, Gough and James have examined the extension of the SLH framework to treat general Gaussian input fields \cite{GougJame16}. They demonstrate that one can model series and feedback connections using the standard SLH rules, see \cref{sec:slh}, even when input fields are arbitrary Gaussian fields. However, this comes at the cost of a reinterpretation of the dynamical equations implied by the resulting SLH triple for the network. Gough and James show that the SLH triple for the network, when the input fields are in non-vacuum Gaussian states, should be interpreted in terms of a corresponding Stratonovich QSDE. In other words, while in the vacuum input case, an SLH triple (for an arbitrary network of components) implies the It\={o} QSDE \cref{eq:QSDE_vector} for the system propagator, and a corresponding It\={o} QSDE for system operators within the network, \cref{eq:Xeqn}, when the network inputs include arbitrary Gaussian fields, the dynamical equations that correspond to the SLH triple for the network (constructed using the normal SLH composition rules) can only be written in Stratonovich form (the ``representation free form" of Ref. \cite{GougJame16}). Note that one can write down an It\={o} form of these dynamical equations (every QSDE has It\={o} and Stratonovich forms), but as shown in Ref. \cite{GougJame16} these It\={o} equations become dependent on the exact state of the input fields. More explicitly, in the It\={o} form the $L$ members of the SLH triple carry information about the state of the input fields. This runs counter to the modular philosophy of the SLH framework, which requires the description of network components to be independent of input fields fed into them -- these descriptions should capture \emph{intrinsic} properties \footnote{This philosophy is motivated by electrical circuit theory where circuit component (\eg resistor) descriptions are independent of the input signals.}. In fact, this state of affairs is already hinted at by the form of the propagator in \cref{eq:gauss_prop}: writing down an SLH triple that generates this propagator would lead to a dependence of the $L$ operator, which is meant to be property of the system alone, on the field state (parameterized by $N,M$).

A further restriction that one encounters when accommodating non-vacuum Gaussian input fields directly into the SLH framework is that the network components cannot include arbitrary scattering matrices, \ie $S \neq I$. Gough and James demonstrate an approach for effectively modeling simple static beamsplitter scattering that is consistent with non-vacuum Gaussian inputs (also see Refs. \cite{GardPark94,WiseMilb94} for prior work on this topic), but arbitrary scattering components are not compatible with the approach developed in Ref. \cite{GougJame16}. In other words, there is no generalization of \cref{eq:gauss_prop} that captures arbitrary scattering dynamics. 

To summarize, the results of Ref. \cite{GougJame16} imply that if one requires (i) a modular description with components capturing only intrinsic properties, (ii) general composition rules for these descriptions, and (iii) a direct representation of non-vacuum Gaussian input fields (\ie not through source models), then one must interpret the SLH triples in terms of corresponding Stratonovich dynamical equations.

Finally, we note that the results in Ref. \cite{GougJame16} are consistent with observations made by Gardiner and Collett on the limitations of It\={o} QSDEs \cite[Section III.D]{GardColl85} . Specifically, these authors mention that defining an It\={o} QSDE requires knowledge of the input fields to the network, while Stratonovich QSDEs are independent of the input fields. 

\subsection{Emission and propagation losses}\label{sec:Loss}
In the context of waveguides or free space experiments we call field modes that interact with the network components ``guided modes'', and imperfections that couple quanta into ``non-guided modes'' {\em losses}. The usual technique to account for losses is to introduce a fictitious mode to represent all the non-guided modes and trace over that mode at the end of the analysis. For example, while an ideal single-port cavity is represented by the SLH triple $G_{\rm
cav}=(I,\sqrt{\gamma}a, \Delta a\dg a)$, a cavity with losses is modeled by the addition of a fictitious port (with vacuum input). This is captured by the concatenation product $G_{\rm total} =G_{\rm system} \boxplus G_{\rm loss}$, where $G_{\rm loss} = (I,
\sqrt{\lambda} a, 0)$, with $\lambda$ being the rate of loss from the principal cavity mode.

While this introduction of a fictitious mode to capture losses is sufficient for many situations, it should be noted that one has to still be careful about modeling choices. For example, an atom coupled to a cavity field could emit into non-guided modes directly (spontaneous emission) or via a cavity mode, or via both mechanisms. In such cases, the fictitious mode (or modes) should be introduced in a way that is consistent with the physics.

Furthermore, loss in waveguides or during free-space propagation can often occur in a distributed manner. We discuss the modeling of distributed properties in more detail in  \cref{sec:continuum}, but note here that such losses are nearly always effectively captured by the incorporation of one or a collection of fictitious beam splitters with vacuum input (which is effectively introducing fictitious output ports to the propagation channel).

\subsection{Circulators} 
As we have previously discussed, circulators (or isolators) are common components in QIONs that enforce the unidirectional propagation of fields. Since  input-output theory and the SLH framework assume unidirectional fields, ideal circulators are implicitly present on many connections. However, real circulators have many non-idealities, including loss, imperfect isolation and finite bandwidth. 

In this section we will develop an SLH model for a symmetric and lossless 3-port circulator. The lossless characteristic means that the total input power is a conserved quantity, \ie all input power is transmitted to one of the output ports. Loss can be  incorporated into this model by appending fictitious beam-splitters at each output port, for example, see  \cref{sec:Loss}. The circulator non-idealities we consider include imperfect impedance matching (resulting in backreflections) and imperfect isolation (routing of the signal to the wrong port of the circulator). 

In the infinite bandwidth limit, a general (potentially non-ideal) circulator can be modeled by an SLH component of the form $(\mathbf{S}, 0, 0)$, where the three port unitary S-matrix is
\begin{align}
S = \left[\begin{array}{ccc}
S_{11} & S_{12} & S_{13} \\
S_{21} & S_{22} & S_{23} \\
S_{31} & S_{32} & S_{33}
\end{array}\right]
\end{align}
where  the subscripts on $S_{j,i}$ label the scattering from port $i$ to $j$. In the case of an ideal three port circulator this matrix becomes 
\begin{align}\label{eq:ideal_3_port_circ}
S_{\rm ideal} = \left[\begin{array}{ccc}
0 & 0 & 1 \\
1 & 0 & 0 \\
0 & 1 & 0
\end{array}\right].
\end{align}
The ideal circulator maps the input fields to output fields in the following way
\begin{align}
\left[\begin{array}{c}
b_{\rm out ,1} (t)\\ 
b_{\rm out ,2} (t)\\ 
b_{\rm out ,3}(t)
\end{array}\right]
 = \left[\begin{array}{ccc}
0 & 0 & 1 \\
1 & 0 & 0 \\
0 & 1 & 0
\end{array}\right]
\left[\begin{array}{c}
b_{\rm in, 1} (t)\\ 
b_{\rm in, 2} (t)\\ 
b_{\rm in, 3}(t)
\end{array}\right]
=
\left[\begin{array}{c}
b_{\rm in, 3} (t)\\ 
b_{\rm in, 1} (t)\\ 
b_{\rm in, 2}(t)
\end{array}\right].
\end{align}
If the circulator is symmetric but not perfect we have $S_{13}=S_{21}=S_{32}= t$, $S_{11}=S_{22}=S_{33}= r$, 
and $S_{12}=S_{23}=S_{31}= b$ \cite{Hage69}:
\begin{align}\label{eq:circnonideal}
S_{\rm non\ ideal} = \left[\begin{array}{ccc}
r & b & t \\
t & r & b \\
b & t & r
\end{array}\right]
\end{align}
with complex transmission, reflection, and isolation error coefficients $t$, $r$, and $b$, respectively. These coefficients 
must obey $|t|^2+|r|^2+|b|^2=1$ and $rt^*+tb^*+br^*=0$ as the S matrix is unitary \cite{Hage69}.  The non-idealities of the circulator are then captured by the parameters \cite{Ayas80}:
\begin{subequations}\label{eq:circloss}
\begin{align}
{\rm Reflection} &= |r|^2\\
{\rm Isolation\,error} &= |b|^2
\end{align}
\end{subequations}
Clearly  $|t|\gg |r|,|b|$ is desirable.

Another circulator non-ideality that has been modeled is finite bandwidth, since real circulators are only non reciprocal devices over a finite frequency bandwidth. SLH models for finite bandwidth 3 port~\cite{HabrStanLuki12}, 4 port~\cite{KercLaluChap15}, and more general circulators ~\cite{RanzAume15,LecoRanzPete17} have been given in the literature. The three port model consists of three coupled cavities \cite{HabrStanLuki12} and has the SLH triple:
\begin{align} 
&G_{\rm finite\ bandwidth} = \left (\mathbf{I}^I_3,
\left[\begin{array}{c}  \sqrt{\gamma} b_1  \\ \sqrt{\gamma} b_2 \\ \sqrt{\gamma} b_3\end{array}\right],
\sum_i \Delta_{\rm cav}  b_i\dg b_i + t( b_1\dg b_3 + b_2\dg b_1 e^{i\varphi}+b_3\dg b_1 + {\rm H.c}) \right ).\label{eq:finite_bandwidth_circ}
\end{align}
When $\varphi= - \pi/2$ and $t= \gamma/2$ this model behaves as a circulator for carrier frequencies close to the cavity frequency. As the magnitude of $\gamma$ is increased the circulator becomes higher bandwidth. After adiabatically eleminating the entire Hamiltonian, see \cref{sec:elimination}, one can show that \cref{eq:finite_bandwidth_circ} reduces to \cref{eq:ideal_3_port_circ} upto phases which can be absorbed into the input-output operators.

\subsection{Bi-directional waveguides, back-reflections, and finite length propagation}\label{subsec:bidirectional}
Input-output theory inherently describes one way propagation of fields.  Consequently the SLH framework is built upon the assumption of unidirectional propagation of fields between components in a network that are very close to each other. Unidirectional propagation is often also referred to as chiral propagation in the literature, \eg \cite{LodaMahmStob16}. However, many experiments have bi-directional propagation of fields, \eg from reflection, or impedance mismatches at node interfaces such as circulators, and finite propagation delays between components. Here we will describe how to construct an SLH model to capture bi-directional propagation on a waveguide and finite propagation distance in increasing generality.

\begin{figure}[]
\includegraphics[width=\columnwidth]{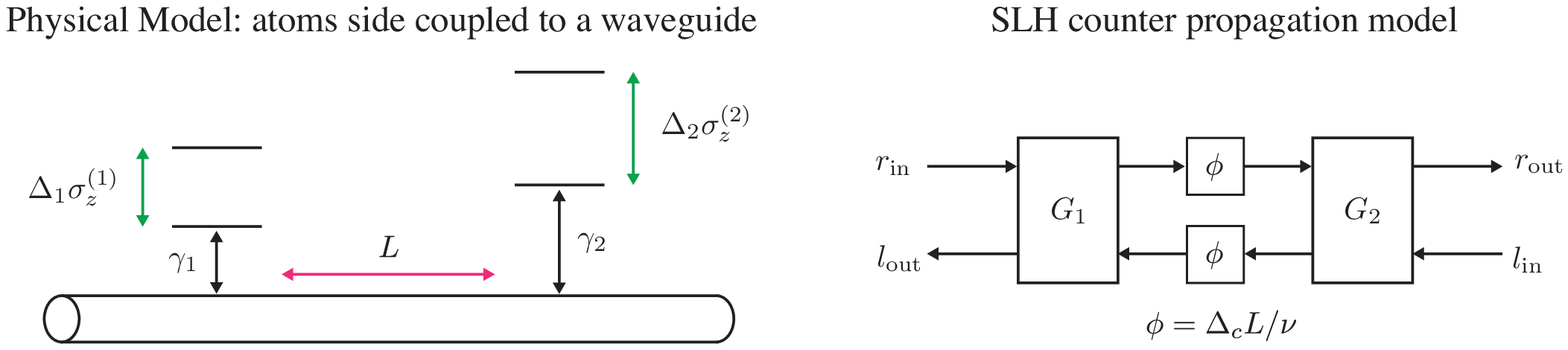}
\caption{A physical waveguide can have right and left propagating modes. In order to capture this in a SLH model we introduce two input-output modes. {To model the finite propagation length, $L$, between the stationary atoms we introduce a phase shift $\exp[i\phi]$ where $\phi = \Delta_c L /\nu $ which is proportional to the detuning from the carrier frequency $\Delta_c$ the length and the speed of light in the medium $\nu$. This simplification is only valid in the Markov approximation when the delay $\tau =L/\nu$ is neglibile compared to the atomic response time.} Not all two mode SLH models capture counter propagation \cref{ex:counterprop} discusses this distinction in more detail.\label{fig:CounterProp}}
\end{figure}

Naively cascading multiple input output networks results, typically, in fields that are co-propagating through a network. Consider the SLH construction where we cascade the right and left going modes and then concatenate these modes, \ie
\begin{subequations}\label{eq:counter}
\begin{align}
 G_{\rm R} &=\left ( G_{\rm R}^{(N)}\lhd\ldots \lhd G_{\rm R}^{(2)}\lhd G_{\rm R}^{(1)} \right )\\
  G_{\rm L} &=\left ( G_{\rm L}^{(1)}\lhd\ldots \lhd G_{\rm L}^{(N-1)}\lhd G_{\rm L}^{(N)} \right ),
\end{align}
\end{subequations}
with the $G_{\rm R}^{(i)}/G_{\rm L}^{(i)}$ modeling coupling of localized components to right-propagating/left-propagating fields. The total system is them composed as $G_{\rm sys} = G_{\rm R}\boxplus G_{\rm L}$. Note that one must be careful not to double-count the internal Hamiltonians of the components when forming this product. This construction lets one simulate systems such as the one depicted in \cref{fig:CounterProp} and other slightly more complicated arrangements \cite{BrodCombGeaB16,BrodComb16}. We illustrate this point in \cref{ex:counterprop}. In general one needs to use the method explained in \cref{SLHremark:vec_elim} to model counter-propagation.

{
In cascading quantum systems we are assuming that the systems are close enough that the propagation delay between then (\eg \cref{fig:coupled_cavities}) goes to zero, and the output field arrives at the next component with the same phase. One methods to account for finite propagation distance, call it $L$, between the cascaded elements is to introduce a phase shift between SLH components~\cite{LaluSandLoo13}. The phase shift element is 
\begin{align}
G_{\rm prop.\ length}=(e^{i\phi},0,0)\quad  \textrm{where} \quad \phi = \Delta_c L/ \nu ,
\end{align}
 $\Delta_c$ is the detuning from the carrier frequency, and the speed of light in the medium is $\nu$. For many components and different distances many phase shifts are required. This treatment is only valid for small delays in propagation, where the delay is small compared to the timescales of system dynamics (typically of order $1/\gamma$), or more precisely in a regime where $\gamma L \ll \nu$. See \cref{subsec:prop_delay} for solutions for modeling time delayed propagation outside this regime. The introduction of the phase shift can lead to nontrivial coupling between the cascaded systems, see \eg Ref. ~\cite{LaluSandLoo13}, we demonstrate this in \cref{ex:counterprop}.

The construction above is not general enough to account for counter propagation in a general network. Importantly,  the feedback reduction (\cref{SLHrule:feedback}) lets one connect networks with an arbitrary topology, thus we could also model counter propagation using this rule. }

\begin{aside}[Non-chiral propagation: counter propagation vs co-propagation ]\label{ex:counterprop}
{
Consider \cref{fig:CounterProp}, in which we depict two atoms coupled to a 1D waveguide with fields propagating in both directions. We choose to lump the Hamiltonian into the right propagating mode. 
To correctly model \cref{fig:CounterProp} we need to first cascade the components for each mode and then concatenate as done in \cref{eq:counter}. (Alternatively we can use the expression in \cref{eq:network_red_eg} with the appropriate substitutions.) In this particular example, the interactions with the right and left propagating modes are modeled as:}
\begin{subequations}
\begin{align}
G_R  &= G_R^{(2)}\lhd G_\phi \lhd G_R^{(1)}  
=  \left (I,\sqrt{\frac{\gamma_2}{2}}\sigma_-^{(2)},-\frac{\Delta_2}{2} \sigma_z^{(2)}\right) \lhd (e^{i\phi}I,0,0) \lhd \left(I,\sqrt{\frac{\gamma_1}{2}}\sigma_-^{(1)},-\frac{\Delta_1}{2} \sigma_z^{(1)} \right)  \\
G_L  &= G_L^{(1)}\lhd G_\phi \lhd G_L^{(2)}   
=  \left (I,\sqrt{\frac{\gamma_1}{2}}\sigma_-^{(1)},0 \right) \lhd (e^{i\phi}I,0,0) \lhd \left (I,\sqrt{\frac{\gamma_2}{2}}\sigma_-^{(2)},0\right).
\end{align}
\end{subequations}
After some algebraic simplification the total system becomes
\begin{align}\label{eq:slh_twosite_counter}
G_T&=G_R \boxplus G_L \quad {\rm (counter-propagation)}\\
&= \left (
e^{i \phi} \mathbf{I}^I_2,
\left[\!\begin{array}{c}
\sqrt{\frac{\gamma_2}{2}}\sigma_-^{(2)}+ e^{i \phi}\sqrt{\frac{\gamma_1}{2}}\sigma_-^{(1)}  \\ 
\sqrt{\frac{\gamma_1}{2}}\sigma_-^{(1)}+ e^{i \phi}\sqrt{\frac{\gamma_2}{2}}\sigma_-^{(2)}
\end{array}\!\right] , 
-\frac{\Delta_2}{2} \sigma_z^{(2)}-\frac{\Delta_1}{2} \sigma_z^{(1)} 
+ \frac{\sqrt{\gamma_1 \gamma_2}}{2}\sin(\phi) \left(\sigma_-^{(1)}\sigma_+^{(2)}+\sigma_+^{(1)}\sigma_-^{(2)}\right)\right ).
\end{align}
{There are a number of things that one can interpret from the form of this equation. Recall that the field for the right going mode after the first atom and the phase shift is $dB_{\rm out}^R= e^{i \phi}(\sqrt{\frac{\gamma_1}{2}}\sigma_-^{(1)} +dB_{\rm in}^R)$. This gets cascaded into the input of the second atom, which explains why the $L$ operator for the right mode has a phase shift associated with the first atom's coupling. The same argument applies for the left going mode but in the reverse order. This demonstrates that we have modeled counter-propagation.}  Further, the presence of the Hamiltonian term proportional to $\sin(\phi)$ couples the two atoms.  Tuning the effective distance between the atoms can give rise to interesting collective atom physics \cite{LaluSandLoo13,LooFedoLalu13}. In real space the coupling would appear as $\sin(k_c|z_1-z_2|)$ where $k_c$ is the propagation wavevector at the carrier frequency and $|z_1-z_2|$ is the distance between atom 1 and atom 2. There is no limitation to the number of impurities that can be cascaded in this way. It is easy to derive that the general coupling term for two distant SLH components $G_i$ and $G_j$ is $\sqrt{\gamma_i\gamma_j}\sin(\phi_{i,j})[L_i L_j\dg +L_i\dg L_j]$ where $\phi_{i,j}$ is the phase shift aquired in propagation between the components \cite{LaluSandLoo13,CaneManzShi15,CalaCiccChan16,HoodGobaAsen16,AsenHoodChan17}. Moreover with our construction we can introduce different phaseshifts in the left and right going modes, leading to richer physics. Note that this system is undriven, but it is simple to cascade drives from both sides, be it classical as in \cref{eq:coh_drive}, or quantum see \cref{eq:fock_source}.

We can contrast this with an alternate system where the two components interact via two co-propagating modes. This is modeled by first concatenating the interactions with the two modes for each component first, and then cascading them, \ie  
\begin{subequations}
\begin{align}
G_T^{(1)}  &= G_R^{(1)}\boxplus G_L^{(1)}  = \left (I,\sqrt{\frac{\gamma_1}{2}}\sigma_-^{(1)},-\frac{\Delta_1}{2} \sigma_z^{(1)} \right )\boxplus \left (I,\sqrt{\frac{\gamma_1}{2}}\sigma_-^{(1)}, 0 \right )\\
G_T^{(2)}  &= G_R^{(2)}\boxplus G_L^{(2)}  = \left (I,\sqrt{\frac{\gamma_2}{2}}\sigma_-^{(2)},-\frac{\Delta_2}{2} \sigma_z^{(2)}\right )\boxplus \left (I,\sqrt{\frac{\gamma_2}{2}}\sigma_-^{(2)}, 0 \right),\\
G_\Phi &=G_\phi\boxplus G_\phi = (e^{i \phi}I,0,0)\boxplus (e^{i \phi}I,0,0)
\end{align}
\end{subequations}
and
\begin{align}
G_T &=  G_T^{(2)} \lhd G_\Phi \lhd G_T^{(1)} \quad {\rm (co-propagation)}\\
&= \Bigg(
e^{i \phi}\mathbf{I}^I_2,\left[\!\begin{array}{c}\sqrt{\frac{\gamma_2}{2}}\sigma_-^{(2)}+ e^{i \phi}\sqrt{\frac{\gamma_1}{2}}\sigma_-^{(1)}\\ \sqrt{\frac{\gamma_2}{2}} \sigma_-^{(2)}+e^{i \phi}\sqrt{\frac{\gamma_1}{2}}\sigma_-^{(1)} \end{array}\!\right],   \\
&\quad\quad -\frac{\Delta_2}{2} \sigma_z^{(2)} -\frac{\Delta_1}{2} \sigma_z^{(1)}  + \frac{\sqrt{\gamma_1\gamma_2}}{2}\sin(\phi)\left(\sigma_-^{(1)} \sigma_+^{(2)} + \sigma_+^{(1)}\sigma_-^{(2)} \right)+\frac{\sqrt{\gamma_1\gamma_2}}{2i}\cos(\phi)\left(\sigma_-^{(1)} \sigma_+^{(2)} - \sigma_+^{(1)}\sigma_-^{(2)} \right) \Bigg)\nonumber
\end{align}
{ Here we see both coupling operators have the same phase shift appearing on the first atom, which illustrates that the fields are co-propagating. Moreover, notice the different effective Hamiltonians when compared to the counter-propagation cases. The co-propagating case has an additional effective imaginary term proportional to $\cos(\phi)$ times an anti-Hermitan operator. The co-propagating interaction could arise in chiral quantum optics~\cite{LodaMahmStob16} or with the use of circulators.

This analysis can be extended to many atoms along a waveguide~\cite{BrodCombGeaB16}. However for more complicated networks we must use \cref{SLHrule:feedback} as explained in \cref{SLHremark:vec_elim}. In fact the above counter-propagation example is exactly reproduced in \cref{SLHremark:vec_elim} with the appropriate substitutions.}

\end{aside}
 
\subsection{Model reduction by adiabatic elimination of fast degrees of freedom}\label{sec:elimination}

Model reduction is the process of approximating a complicated model by an analytic and or computationally simpler model. Adiabatic elimination is a form of model reduction applicable when there is a seperation in timescales for different system variables. For example, consider an input-output field coupled to a one sided cavity with the cavity mode coupled to a two level atom. If the cavity is very leaky, the cavity typically equilibrates to the input field and atomic state on a time scale faster than the time scale over which either the input field or atomic state varies.  Thus, the cavity state is primarily a dependent variable on the input and cavity states and need not be tracked for an accurate dynamical model.  In such cases, one says that the cavity may be \emph{adiabatically eliminated}. 

Adiabatic elimination has a long history in quantum and atom optics. As expressed by Gardiner, the aim is to find a ``method by which fast variables may be eliminated from the equations of motion in some well-defined limit" \cite{Gard84}. This is typically achieved by using a projection operator approach \cite{Gard84,GardStey84,SteyGard84,Coh.Dup.etal-1998}. With respect to the QSDEs for the propagator and the SLH framework adiabatic elimination was rigorously formulated in a series of papers by Bouten, Silberfarb and van Handel \cite{BoutSilb08} and \cite{BoutHandSilb08}.  The approach used in these papers is to first define a network node with some parameter $k$ that scales the fast dynamical rates $G^{(k)}=(S^{(k)},L^{(k)},H^{(k)})$.  Each $S^{(k)}$, $L^{(k)}$, and $H^{(k)}$ are operators on a Hilbert space $\mathcal H$.  Then, the adiabatic elimination procedure is applied which results in a node that approximates the original network without the $k$ dependence $G=(S,L,H)$, \ie operates at slow dynamical rates only.  Here, each $S$, $L$, and $H$ operate on $\mathcal H_0$, which is a subspace of $\mathcal H$ such that $\mathcal H_0=P_0\mathcal H$, where $P_0$ is a projection operation.  Specifically Bouten \etal prove that the unitary $U_t^{(k)}$ generated by $G^{(k)}$ converges to the unitary $U_t$ generated by the network $G$ in the following sense
\begin{align}
\lim_{k\rightarrow \infty}\sup_{{0\le t \le T}} ||(U_t-U_t^{(k)}) \ket{\psi}||=0,
\end{align}
for all {$\ket{\psi} \in \mathcal{H}_0\otimes {\cal F}$, where ${\cal F}$ is the symmetric Fock space for the fields coupled to the system}, provided certain -- yet to be stated -- preconditions hold. We remark below on the choice of $P_0$. 

Identifying $(S,L,H)$ first starts by defining a QSDE for $U_t^{(k)},$ which depends on the fast timescale $k$:
\begin{align}  \label{eq:pre_elim_qsdeunitary}
	 dU_t^{(k)} =& \Big\{ - (i H ^{(k)} +\half L_i^{(k)\dagger} L_i^{(k)} )dt+ L_i^{(k)} dB_i^\dagger - L_{i}^{(k)\dagger} S_{ij}^{(k)} dB_j 
			 \,\,+ ( S_{ij}^{(k)} - \delta_{ij} I)d\Lambda_{ij}  \Big\} U_t, 
\end{align}
ultimately this will limit to a propagator $U_t$ defined by:
 \begin{align} 
  dU_t =& \Big\{ - (i H  +\half L_i^{\dagger} L_i )dt+ L_i dB_i^\dagger - L_{i}^{\dagger} S_{ij} dB_j 
			 \,\,+ ( S_{ij} - \delta_{ij} I)d\Lambda_{ij}  \Big\} U_t,  \nonumber
\end{align}

For the adiabatic elimination procedure to hold, the pre-elimination operators $(S_{ij}^{(k)},L_{i}^{(k)},H ^{(k)})$ must have the following dependence on $k$
\begin{subequations}\label{eq:AE_koperators}
\begin{align}
K^{(k)} 
&= -\left (i H ^{(k)} +\half \sum_i L_{i}^{(k)\dagger} L_{i}^{(k)}\right ) = k^2 Y + k A +B\\
L_{i}^{(k)}&= k F_i+G_i\\
S_{ij}^{(k)}&=W_{ij},
\end{align}
\end{subequations}
for some operators $Y,A,B,F_i,G_i,W_{ij}$.
The physical interpretation of the $k$-dependance of these operators is as follows. Operators that depend on $k^2$ generate the fast dynamics that we wish to eliminate. The operators that have no dependence on $k$ generate the slow dynamics, and the operators that depend on $k$ couple the fast and slow timescales. 

Adiabatic elimination is allowable in the limit $k\rightarrow \infty $, \ie in the limit where the fast quantities tend to adiabatically ``follow'' the slowly varying quantities.  In this limit, the operators $(S_{ij}^{(k)},L_{i}^{(k)},H ^{(k)})$ limit to
\begin{subequations}
\label{eq:AE_operatorslimit}
\begin{align}
K 
&= -\left (i H  +\half \sum_i L_{i}\dg L_{i}\right ) = P_0(B-A\tilde Y A)P_0\\
iH&=-K -\half  \sum_i L_{i}\dg L_{i}\\
L_{i}&= (G_i-F_i\tilde Y A)P_0\\
S_{ij}&=  ( F_i \tilde Y F_l\dg+\delta_{il})W_{lj} P_0
\end{align}
\end{subequations}
The assumptions for this limit to hold are
\begin{align}
\label{eq:AE_assumptions}
\begin{split}
1. \,\,& \text{There exist $\tilde Y$ such that $\tilde Y Y = Y\tilde Y =P_1$, where $P_1=I -P_0$ is a projector onto the fast dynamics;\quad\quad}\\
2. \,\,& YP_0=0;\\
3. \,\,& F_i P_0 = 0 \text{ for all } i;\\
4. \,\,& P_0AP_0=0.
\end{split}
\end{align}
For more technical details on these conditions, ``Assumptions 2 (Structural requirements)" of Ref.~\cite{BoutHandSilb08} and Assumptions 3 and 4 of Ref.~\cite{BoutSilb08}. We note here that one has to make a judicious choice of $P_0$ for these conditions to hold, and this choice is often aided by physical insight into the dynamics of the network. 
Intuitively the projection operator $P_0$ acts on $\mathcal H$ and projects on to the slow dynamics $\mathcal H_0 = P_0 \mathcal H$, while $P_1=I -P_0$ is a projector onto the fast dynamics $\mathcal H_1 = P_1 \mathcal H$. Bouten and Silberfarb suggest that $\mathcal H_0$ should be thought of as the ground state subspace of $\mathcal H$ and  $\mathcal H_1$ is the excited state subspace. For a mechanical method of finding the relevant operators in terms of the projector $P_0$ see \cref{SLHremark:elimination}.

\begin{figure}[h]
\includegraphics[width=\columnwidth]{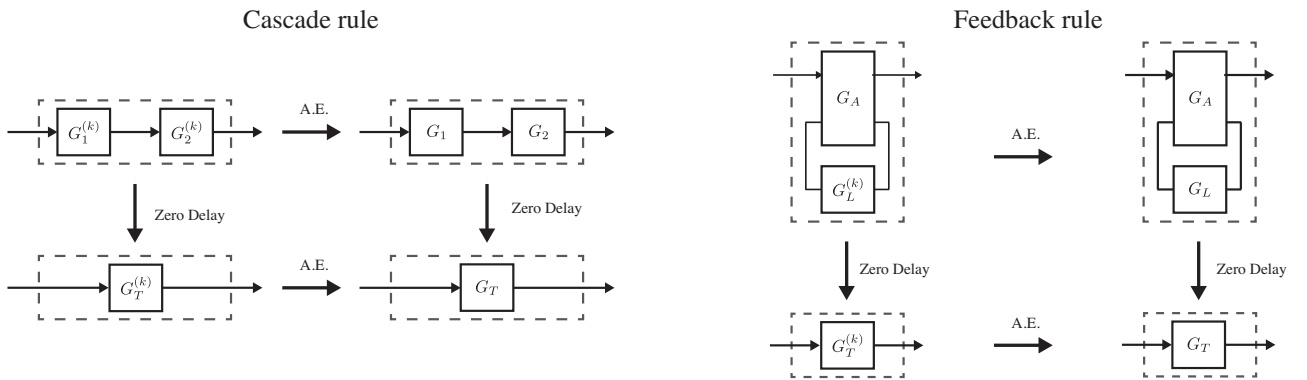}
\caption{Commutativity of adiabatic elimination (A.E.) and network composition rules.}\label{fig:Commute}
\end{figure}

In the context of a large QION, clearly if we can eliminate some parts the model will have reduced complexity model which should reduce simulation costs. The first application of adiabatic elimination within the context of a QION was by Warszawski and Wiseman~\cite{WarsWise00} to simplify the dynamics of a optical feedback network (although this work predates the SLH formalism and the adiabatic elimination technique outlined in this section, and therefore applied a less rigorous adiabatic elimination technique). An important question arises when applying adiabatic elimination to simplify QIONs. 
Are the network dynamics different if we (1) perform adiabatic elimination on individual network nodes and then apply the concatenation, series and feedback product rules or (2) apply the rules to compose the network and then adiabatically eliminate? 
The fact that these two approaches produce the same result was first established by Gough \etal ~\cite{GougNurdWild10}, and then in full generality by Nurdin and Gough \cite{NurdGoug12}. Intuitively these authors show that the two different pathways from the top left corner to the bottom right corner in the schematics shown in \cref{fig:Commute} result in the same SLH parameters. 

{There can be subtleties in the elimination process.  For example, when one attempts to use the above procedure to scale the amplitude of a coherent state field input (which corresponds to the parameter $k$ for this example), mathematical complications can arise. A procedure to overcome this difficulty was developed by Bouten (unpublished) and applied in \cite{KercBoutSilb09}, and generalized in the supplementary material of \cite[Sec. II C ]{KercPavlPavl11}.} For further explicit examples and applications of this adiabatic elimination technique see the Supplementary Information section of Ref. \cite{KercNurdNurd10} and Ref. \cite[Chapter 1]{Devoret:2014vz}, and Refs.~\cite{Mabu09,Mabu12,BoutSilb08,BoutHandSilb08}. \\

\begin{Remark}[\bf A quick way to find $Y,A,B,F_i$, and $G_i$ in \cref{eq:AE_koperators} ] \label{SLHremark:elimination}
In practice it can be tricky finding the operators $Y,A,B,F_i$, and $G_i$ in \cref{eq:AE_koperators}. Here we describe an intuitive method to find these operators. The method relies on using the projectors onto the ground space $P_0$ and the excited space $P_1=I -P_0$ to define the operators of interest. {This of course requires some intuition about $P_0$, perhaps obtained by physical insight or numerics. We begin by denoting operators before elimination (or any $k$ scaling ) with a bar e.g. $(\bar S, \bar L, \bar H)$. Now consider the following decomposition $\bar K = -\left (i \bar H +\half \sum_i \bar L_{i}^{\dagger} \bar L_{i}\right ) = Y + A +B$ (compare with \cref{eq:AE_koperators} which has the $k$ dependence). The operator $Y$ will be eliminated in the procedure so it is in the excited space, while the operator $A$ couples the excited space to the ground space, and $B$ is entirely in the ground space. This suggest the following definitions
\begin{align}\label{eq:slh_ae_k_ops}
Y \equiv P_1 \bar K P_1, \quad  A \equiv  P_1 \bar K P_0+P_0 \bar K P_1,\quad  \text{and } B\equiv P_0 \bar K P_0.
\end{align}
With similar reasoning for $ F_i $ and $G_i$ we define 
\begin{align}\label{eq:slh_ae_L_ops}
  F_i \equiv P_1 \bar L_{i} P_1+P_0 \bar L_{i} P_1,\quad G_i\equiv P_1 \bar L_{i} P_0+P_0 \bar L_{i} P_0.
\end{align}
Clearly the treatment of the term that scaled as $k$, in the original treatment, is different between \cref{eq:slh_ae_k_ops} and \cref{eq:slh_ae_L_ops}.} This is so that assumption (3) in \cref{eq:AE_assumptions} holds. Note that $W_{ij}$ does not depend on $k$ or $k^2$. One should also note that our choice for $F_i$ is different to that of the related method of \citet{ReitSore12}. 
\end{Remark}

\begin{aside}[Adiabatic elimination in a cavity QED model]\label{ex:adiabatic_elim}

The SLH model for a driven, two level atom coupled to a single sided optical resonator/cavity via the Jaynes-Cummings Hamiltonian is
\begin{align}
S^{(k)} = \left[\begin{array}{cc}I&0\\0&I\end{array}\right],\quad L^{(k)} = \left[\begin{array}{c}\sqrt{\kappa}a\\\sqrt{\gamma}\sigma_-\end{array}\right],\quad H^{(k)} = \Delta_r a\dg a+\Delta_a\sigma_+\sigma_-+g(a\dg\sigma_-+a\sigma_+)
\end{align}
where $I$ is the identity operator on the resonator-atom system, $a$ is the annihilation operator for the principal mode of the resonator, $\sigma_{+(-)}$ is the raising (lowering) operator for the two level atom, $\kappa$ is the decay rate of photons out of the resonator, $\gamma$ is the decay rate of the atom due to spontaneous emission into non-guided radiation modes, $g$ is the coupling strength between the atom and resonator mode, and $\Delta_{r(a)}$ is the energy detuning between the center frequency of the resonator (atom) and the reference frequency $\Omega$ (the model is in a rotating frame with respect to this frequency). The dependence of these parameters on the scaling parameter $k$ will be specified in the following paragraph. The are two input-output modes for this SLH component; the first one (that couples to the operator $L_1^{(k)}=\sqrt{\kappa}a$) corresponds to the guided mode that couples to the primary internal mode of the resonator, and the second one (that couples to $L_2^{(k)}=\sqrt{\gamma}\sigma_-$) corresponds to a fictitious single mode that represents the atom's spontaneous emission into all, non-guided, radiation modes, as discussed in \cref{sec:Loss}.
For simplicity, consider the case where the atom and resonator are on resonance with each other, \ie $\Delta_r=\Delta_a=0$.

Here, we will apply the adiabatic elimination theorem described in this section to calculate the much simpler, effective dynamics that emerge in the limit $\kappa,g\gg\gamma$.  To apply the theorem, we must specify how this limit arises due to the scaling of some dimensionless parameter $k$ that approaches infinity.  Therefore assume that 
\begin{align}
\kappa = k^2\kappa_0\text{, }\quad g = k^2g_0\text{, \quad and }\gamma = \gamma_0.  
\end{align}
Then, using  \cref{eq:AE_koperators}, we identify the operators in the $(S^{(k)},L^{(k)},H^{(k)})$ according to their scaling with $k$:
\begin{align}
Y &= -\frac 12\kappa_0 a\dg a-ig_0(a\dg\sigma_-+a\sigma_+),\quad A = 0,\quad B = -\frac 12\sigma_+\sigma_-\\\
F &= \left[\begin{array}{c}\sqrt{\kappa_0}a\\0\end{array}\right],\quad G = \left[\begin{array}{c}0\\\sqrt{\gamma}\sigma_-\end{array}\right],\\
W &= \left[\begin{array}{cc}I&0\\0&I\end{array}\right]
\end{align}
For this model, we choose $P_0 =  |0_r,0_a\rangle\langle0_r,0_a|$, \ie the projector on to the ground state of both the cavity and the atom, since in the limit of fast resonator decay, all excitations in the system will be damped. Next, we test to see if the assumptions required for the theorem are satisfied. Assumptions (2)-(4), that require $YP_0 = F_i P_0 = P_0AP_0 =0$ are easily verified by direct computation. Finding a $\tilde{Y}$ that satisfies the appropriate conditions is more complicated, and after some thought we find: 
\begin{align}
\tilde{Y}Y &= Y\tilde{Y} = I-P_0, ~~~~~ \text{ for}\nn\\
\tilde{Y}|0_r,0_a\rangle &= 0, \nn \\
 \tilde{Y}|n_r,0_a\rangle &= -\frac{2}{\kappa_0n_r}|n_r,0_a\rangle,\,~~ n_r>0\nn\\
\tilde{Y}|n_r-1,1_a\rangle &= \frac{ig\sqrt{n_r}}{g^2n_r+\frac14\kappa^2(n_r-1)n_r}|n_r,0_a\rangle +\frac{\frac12\kappa n_r}{g^2n_r+\frac14\kappa^2(n_r-1)n_r} |n_r-1,1_a\rangle,\,~~ n_r>0.
\end{align}
To compute the limiting, effective dynamical model $(S,L,H)$ we then apply \cref{eq:AE_operatorslimit}:
\begin{align}
K &= P_0(B-A\tilde{Y}A)P_0 = 0 \Rightarrow H = 0\nn\\
L &= (G-F\tilde{Y}A)P_0 = \left[\begin{array}{c}0\\0\end{array}\right]\nn\\
S &= (F\tilde{Y}F\dg+I)WP_0 = \left(\left[\begin{array}{cc}-2P_0&0\\0&0\end{array}\right]+\left[\begin{array}{cc}P_0&0\\0&P_0\end{array}\right]\right) = \left[\begin{array}{cc}-P_0&0\\0&P_0\end{array}\right]
\end{align}
Thus, we come up with an effective dynamical model in which the internal degrees of freedom are restricted to the state $|0_r,0_a\rangle$ with no effective Hamiltonian dynamics, nor effective coupling to either input-output mode coupled to the resonator, or the unmonitored modes that the spontaneous emission couples to. This reflects the fact that in the large $\kappa$, large $g$ limit, any excitations in the cavity or atom effectively decay instantly, restricting the internal dynamics to the $|0_r,0_a\rangle$ state.  The input-output relations are less trivial, using \cref{eq:inputoutput_vector} we find
\begin{align} 
d\mathbf{B}_{\rm out} = \left[\begin{array}{cc}-P_0&0\\0&P_0\end{array}\right]d\mathbf{B}_{\rm in}
\end{align}
which indicates that the first input mode into the network, corresponding to the real guided field mode, gets reflected with an additional $\pi$ phase shift, while the fictitious mode that models unmonitored radiation modes picks up no phase shift upon reflection.  This effective model reveals the fact that in this limit of vanishing $\gamma$ (equivalently, large $\kappa$ and $g$), the resonator looks on-resonant to the guided probe field, while the radiation modes experience no appreciable dynamical effect.  

\end{aside}

\subsection{Modeling distributed transformations}\label{sec:continuum}
The SLH framework is fundamentally based on a modular approach that models transformation of propagating fields by a network of discrete components. However, in some cases the properties and transformations we wish to model are distributed in space. 
For example, understanding the propagation of light through engineered nonlinear crystals \cite{Drum90,GlauLewe91,HuttBarn92,DrumHill99,LiscHeltSipe12,GrafKundReid17} requires modeling distributed transformations of propagating fields. In this section we discuss an approach to adapting the SLH framework to study the propagation of quantum fields through a continuous medium. This has been done many times in IOT, see for example the work of  \citet{CaveCrou87}. The essential strategy is to approximate the transformation as a large number of discrete components effecting infinitesimal transformations and then take a continuum limit of the cascaded model.

\begin{figure}[h]
\includegraphics[width=\columnwidth]{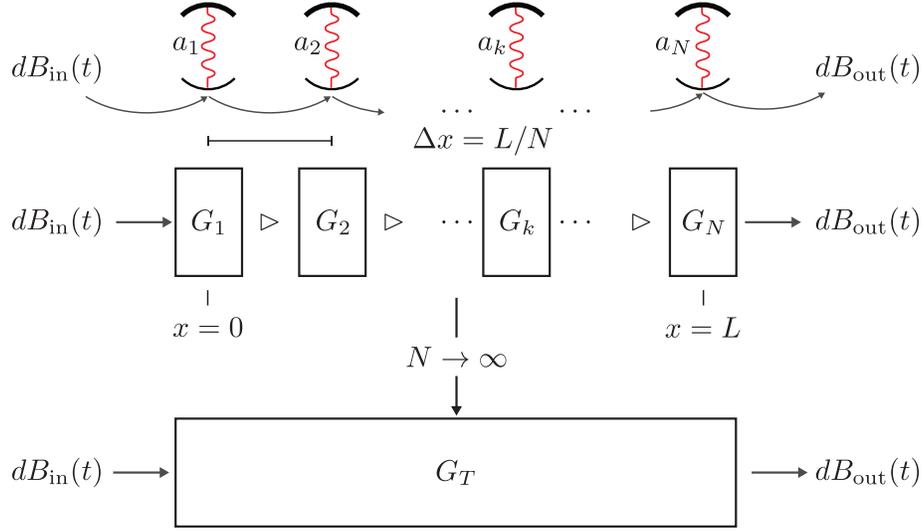}
\caption{An example of taking a continuum limit of an SLH model from Refs.~\cite{HushCarvHedg13,StevHushCarv14,FuCarvHush16}. In this example $N$ cavities with possibly different decay rates and resonance frequencies are cascaded. The top panel shows an experimental schematic, the first cavity is located at $x=0$ and the last at $x=L$. The second panel is the discrete SLH model for the top panel. In the third panel we have taken a continuum limit, to obtain an SLH model that can represent a continuous medium.
}\label{fig:ContinuumLim}
\end{figure}

We take as an example the work of Hush \etal~\cite{HushCarvHedg13}, which analyses a gradient echo memory using the SLH framework. A gradient echo memory is essentially a spatially distributed atomic ensemble. To model this with an SLH network one imagines the atomic ensemble as broken into thin slices, where the output of one slice is the input to the next slice, see  \cref{fig:ContinuumLim}. All slices contain a collection of atoms with different detunings, but the slices are considered so thin that there are no emission then re-absorption events within a single slice. In this weak excitation limit it can be argued that the resulting interaction with the atomic ensemble can be approximated as a coherent exchange with a bosonic degree of freedom and modeled using harmonic oscillator raising and lowering operators \cite{HushCarvHedg13,StevHushCarv14,FuCarvHush16}. Under this approximation, the slices formally resemble a collection of cascaded cavities, and the SLH triple that captures the dynamics induced by the $k$'th thin slice of the ensemble is:
\begin{align}
G_k = (I,\sqrt{\beta_k}a_k,\xi_k a_k\dg a_k),
\end{align}
where  $\beta_k$ is the coupling constant between neighboring slices/cavities and $a_k$ and $a_k\dg$ are the annihilation and creation operators for cavity mode $k$, and $[a_j,a_k\dg]=\delta_{jk}$. Cascading $N$ such components results in 
\begin{align}
G_T &= G_N \lhd\cdots  G_k  \ldots \lhd G_2 \lhd G_1 \nonumber\\
&= \left (I,\sum_{k=1}^N\sqrt{\beta_k}a_k, \sum_{k=1}^N\xi_k a_k\dg a_k +\frac{1}{2i}\sum_{j=2}^N \sum_{k=1}^{j-1} \sqrt{\beta_j \beta_k} (a_j\dg a_k -a_k\dg a_j)\right). \label{eq:GEM_SLH}
\end{align}
In principle we could now write down $dU$ for the entire system or work out an equation of motion for some operator $K$.  However, we are interested in the continuum limit of this approximation, and to take this limit we recall how this modular model relates to spatial co\"{o}rdinates. Imagine the first slice/cavity is located at $x=0$ and the last slice/cavity is located at $x=L$. Then the $k$'th cavity is located at $x(k) = k \times \Delta x$ where $\Delta x = L/N$.  In the continuum limit $N\rightarrow \infty$, $\Delta x$ becomes $dx$ and the sums in \cref{eq:GEM_SLH} become integrals:
\begin{align}
G_T &= (I, L_T, H_T )\nonumber \\
&= \left (I, \int_0^L dx \sqrt{\beta_x}a(x), \int _0^L dx\, \xi_x a\dg(x) a(x) +\frac{1}{2i} \int_0 ^L dx \int_0^x dy \sqrt{\beta_x \beta_y} [a\dg(x) a (y) -a\dg(y) a (x)]\right),\label{eq:GEM_contin}
\end{align}
where $[a(x), a\dg(x')] = \delta(x-x')$. To understand the transformation of the input field implemented by this continuum model, consider the input-output relation for the network, namely $dB_{\rm out}(t) = dB_{\rm in}(t) + L_T(t)dt$. In order to characterize the output field we need to solve for $L_T(t)$, which in turn involves solving for the local modes $a(x,t)$. Since the overall network model is linear we can easily write the equation of motion for $X= a(x,t)$ as \cite{HushCarvHedg13}
\begin{align}  
da(x,t) &=- i \xi_x a(x,t)dt - \frac{\sqrt{\beta_x}}{2}\int_0^x dx' \, \sqrt{\beta_{x'}} a(x',t) - \sqrt{\beta_x} dB_{\rm in}(t), 
\end{align}
and solve it for some initial conditions. This is done explicitly, using Laplace transform techniques, by Hush \etal to study gradient echo memories ~\cite{HushCarvHedg13}, single photon production~\cite{StevHushCarv14}, and cross phase modulation of photons in two gradient echo memories~\cite{FuCarvHush16}.

This notion of modeling material properties using the continuum limit of an SLH network is relatively unexplored and has significant potential.

\subsection{SLH and scattering theory}\label{sec:QFT_Smatrix}
The SLH framework is a route to modeling the internal dynamics of a QION and also to determine the relationship between the input and output field to the network. The output fields are specified by Heisenberg equations of motion for the canonical operators for these fields, \ie \cref{eq:inputoutput_vector}, and it is in principle possible to characterize the state of the output field by calculating moments of these canonical operators. However, one can exploit the connection between the SLH framework and scattering theory to make direct connection between the \emph{states} of the input fields and \emph{states} of the output fields. The central quantity that enables this in scattering theory is the scattering matrix, or S-matrix \footnote{Note that this S-matrix is distinct, but related to, the scattering matrix that forms the first element in any SLH triple. We overload this notation since the use of the symbol $S$ has become standard in both communities, and the type of scattering matrix referred to is usually clear from context.}, which is a unitary matrix that connects asymptotic input and output field states: $\ket{\omega}= S\ket{\nu}$, where $\ket{\omega}$ and $\ket{\nu}$ are asymptotic field states that are usually specified as energy eigenstates of the free Hamiltonian. In this section we briefly summarize the relationship between SLH and scattering formalisms. A detailed account of this relationship can be found in Refs.~\cite{DaltBarnKnig99,FanKocaShen10,RoyWilsFirs16} and a summary of recent scattering work can be found in Ref.~\cite{RoyWilsFirs16}.

We consider the interaction of a localized component with a single input-output mode, the generalization to many input-output modes is straightforward but cumbersome. 
 The elements of the S-matrix in the frequency domain is specified  by $S_{\omega  ,\nu } =  \bra{\omega ;{\rm out}}S\ket{\nu ;{\rm in}}$. The S operator is equivalently the propagator in the interaction picture with the following limits
\begin{align}\label{eq:singlephotonS}
S &=  \lim _{\substack{t_0\rightarrow -\infty \\t_1\rightarrow +\infty}} U_I(t_1,t_0)=  \lim _{\substack{t_0\rightarrow -\infty \\t_1\rightarrow +\infty}} e^{iH_0 t_1}e^{-iH( t_1-t_0)}e^{-iH_0 t_0},
\end{align}
in this expression $H_0 = H_B$ and $H = H_{\rm sys} +H_{\rm int}$  where $H_B, H_{\rm sys}, H_{\rm int}$ are from \cref{eq:HSysBath}.
An alternative way to describe the S operator is with the M\o ller wave operators $\Omega_\pm$, where $S=\Omega_-\dg\Omega_+$. The M\o ller operators map states of the system plus field to the infinite past or future and are denoted by: 
\begin{align}\label{eq:moller}
\ket{\mu^+} &=  \lim _{\substack{t_0\rightarrow -\infty }} U_I(0,t_0)\ket{\mu} \equiv \Omega_+\ket{\mu} \\
\ket{\mu^-} &=  \lim _{\substack{t_1\rightarrow +\infty }} U_I(0,t_1)\ket{\mu} \equiv \Omega_-\ket{\mu},
\end{align}
where $\ket{\mu}$ are eigenstates (of the field and system subsystems) of the free Hamiltonian and $\ket{\mu^{\pm}}$ are ``scattering eigenstates" of the interacting Hamiltonian, \ie $H_0 \ket{\mu}= \epsilon_\mu \ket{\mu}$ and $H\ket{\mu^{\pm}} = \epsilon_{\pm} \ket{\mu^{\pm}}$. A key assumption that we make is that the system is in its ground state in the asymptotic regime, \ie in the infinite past and future. Extensions to scattering theory that go beyond this assumption are possible, but we will not cover them.

The S-matrix elements in this notation becomes  
\begin{align}\label{eq:singlephotonS}
S_{\omega  ,\nu } &=  \bra{\omega}S\ket{\nu} = \expt{\omega^-|\nu^+}
\end{align}
 The next task is to relate this object to the input-output theory and more generally to the SLH framework. The time domain input and output fields in the asymptotic past or future, that is \cref{eq:bin} and \cref{eq:bout}, are the limit where $t_0 \rightarrow -\infty$ and $t_1 \rightarrow +\infty$. These can be related to the frequency domain representations $\bin(\omega),\bout(\omega)$ by a Fourier transform -- as was done in \cref{eq:binFourier}.  
The M\o ller operators act on Heisenberg-picture operators in the following way $\bin(\omega) = \Omega_+ b(\omega)\Omega_+\dg$, and $\bout(\omega) = \Omega_- b(\omega)\Omega_-\dg$.
In other words, the M\o ller operators propagate the field operators in the asymptotic past to the interaction region, or from the interaction into the asymptotic future.
Using this relation between the field operators and the M\o ller operators, we can deduce that:
\begin{align}
\bin\dg(\nu)\ket{0} &= \ket{\nu^+}\\
\bout\dg(\omega)\ket{0} &= \ket{\omega^-},
\end{align}
where we define $\ket{0}$ as the vacuum state of the field and the ground state of all components in the network. These expressions finally allow us to relate the input-output field operators in the frequency domain to the scattering matrix:
\begin{align}
S_{\omega  ,\nu } &= \expt{\omega^-|\nu^+}= \expt{0|\bout(\omega)\bin\dg(\nu)|0} . \label{eq:scat_realdeal}
\end{align}
This expression links the elements of the scattering matrix to the input and output fields specified by input-output theory, or more generally, the SLH framework. The following example illustrates how one can evaluate the right-hand-side of this relation to calculate the scattering matrix using the input-output relations derived from an SLH description of a QION.\\

\begin{aside}[A single photon scattering off a two level atom]

Consider the interaction of an input-output mode carrying a single photon with a localized component. We mostly follow the treatment in Ref. \cite{FanKocaShen10} in this example. In \cref{sec:fockstates} we defined frequency-domain wave packet in the asymptotic past as $\ket{1_\xi}\equiv\int d\nu \, \xi(\nu) b^{\dagger}(\nu) \ket{0}$. The scattered wavepacket is given by $S\ket{1_\xi}$, which can be evaluated as:
\begin{align*}
S \ket{1_\xi} &=\int d\nu \xi(\nu) S b\dg(\nu) \ket{0} = \int d\omega d\nu \xi(\nu) \ket{\omega}\!\bra{\omega} S \ket{\nu}  = \int d\omega d\nu S_{\omega,\nu} \xi(\nu) \ket{\omega} = \int d\omega \xi'(\omega) d\dg(\omega) \ket{0},
\end{align*}
where $\xi^{'}(\omega)\! \equiv \int d\nu \,  S_{\omega ,\nu } \xi(\nu)$. To obtain the second equality we have inserted a resolution of identity in terms of eigenstates of asymptotic output modes ($\int d\omega \ket{\omega}\bra{\omega}$), and $d\dg(\omega)$ are creation operators for these modes. We see from the final equality that the output photon is a wavepacket with a profile $\xi'(\omega)$, which is related to the input wavepacket profile by a deformation by the scattering interaction.

In order to evaluate the scattering matrix elements $S_{\omega, \nu}$, we turn to the expression in \cref{eq:scat_realdeal}. We first consider its Fourier transform to work with time domain quantities: 
\begin{align}\label{eq:1Smatrix0}
S_{\omega  ,\nu } = \frac{1}{\sqrt{2\pi}} \int dt \bra{0} \bout(t) \ket{\nu^{+}} e^{i \omega  t}.
\end{align}
By using the input-output relation, \ie \cref{eq:dB}, this becomes
\begin{align} \label{eq:SmatIO}
S_{\omega  ,\nu }  = \frac{1}{\sqrt{2\pi}} \int dt \bra{0} S(t) \bin(t) + L(t) \ket{\nu^{+}} e^{i \omega  t},
\end{align}
At this point $(S,L,H)$ is general. For concreteness, we now specialize to the case where the component is two level atom dipole coupled to the field, \ie $G_{\rm sys} =  (I, \sqrt{\gamma} \sigma_{-}, \frac{\Delta}{2} (I - \sigma_{z}) )$, and it is initially in the ground state. In this case, the expression for the scattering matrix element becomes:
\begin{align} \label{eq:SmatIO2}
S_{\omega  ,\nu }  = \frac{1}{\sqrt{2\pi}} \int dt \left[ \bra{0} \bin(t) \ket{\nu^{+}} e^{i \omega  t} + \bra{0} \sqrt{\gamma}\sigma_-(t) \ket{\nu^{+}} \right] e^{i \omega  t},
\end{align}
Consider the two terms in the integrand separately. The first term is $\bra{0} \bin(t) \ket{\nu^+} = \bra{0} \bin(t) \bin\dg(\nu)\ket{0}$. By Fourier transforming one of these $\bin$ operators and using the delta commutation relations between these operators, this expression evaluates to $e^{-i\nu t}/\sqrt{2\pi}$, and hence the integral of the first term simply reduces to $\delta(\omega - \nu)$. 

For the second term in the integrand, we need to evaluate $\bra{0} \sqrt{\gamma} \sigma_- (t)\ket{\nu^+}$. The SLH framework specifies equations of motion for system degrees of freedom, \ie \cref{eq:dX}. Sandwiching the equation of motion for $\sigma_-(t)$ between $\bra{0}$ and $\ket{\nu^{+}}$, we get: 
\begin{align}
 \bra{0} \frac{d \sigma_{-}}{dt} \ket{\nu^{+}} = & - \left(\frac{\gamma}{2} + i \Delta \right) \bra{0} \sigma_- \ket{\nu^{+}}  - \sqrt{\gamma} \bra{0} \sigma_z \bin(t) \ket{\nu^{+}},\label{eq:DEforSmat}
\end{align}
Recall that atom is in the ground state, so $ \bra{0}\sigma_z=\bra{0}$ and $\bra{0} \sigma_{z} \bin(t) \ket{\nu^{+}}  = \bra{0} \bin(t) \ket{\nu^{+}} =  e^{-i \nu t}/\sqrt{2\pi}$.
Therefore \cref{eq:DEforSmat} reduces to a simple first-order differential equation that we can solve for $\bra{0} \sigma_- \ket{\nu^{+}}$. Using this solution, we get
\begin{align} 
\frac{\sqrt{\gamma}}{\sqrt{2\pi}} \int dt e^{i \omega  t} \bra{0} \sigma_{-}(t) \ket{\nu^{+}} = - \frac{\gamma}{\frac{\gamma}{2} + i (\Delta-\omega)} \delta (\omega  - \nu),
\end{align}
Putting this together, we get an expression for the S-matrix element of interest:
\begin{align} \label{eq:1p1sscatt2}
S_{\omega  ,\nu } = - \frac{\frac{\gamma}{2} - i (\Delta-\omega)}{\frac{\gamma}{2} + i (\Delta-\omega)} \delta(\omega  - \nu).
\end{align}
This expression has been derived using various techniques in the past \eg \cite{YurkDenk84,CollGard84,FanKocaShen10}. 
We note that the above approach has been used to derive scattering matrix elements for other situations, including: scattering of two photons in one mode by an atom~\cite{FanKocaShen10}, scattering of two input-output modes with one or two photons by an atom~\cite{FanKocaShen10}, and scattering of coherent states in one or two input-output modes by an atom~ \cite{KocaRephFan12}. 
\end{aside}

The utility of casting the scattering calculation within SLH framework is that one can now calculate, in principle, the S-matrix representing scattering off an arbitrary network of quantum components described by an SLH triple \cite{XuFan15}. 
Indeed, recently a number of authors have recently used the SLH framework to analyze complex scattering calculations \cite{PanDongZhan16,BrodCombGeaB16}. 
Notably, Caneva \etal have recently shown how to include finite spatial distances between scattering elements in a SLH network, and include propagation delays discussed in \cref{subsec:prop_delay} \cite{CaneManzShi15}. {The simplest solution is to cascade propagation-length dependent phases between components as explained in \cref{subsec:bidirectional} and \cref{ex:counterprop}.}  The solution developed in Refs. \cite{CaneManzShi15,ShiChanCira15} shows that there is an intimate relationship between solving the scattering problem and the generalized state matrices defined in Ref. \cite{BaraCookBran12}.  More generally the scattering problem can become complicated when bound states are involved see e.g. \cite{Koca16}.

\subsection{Dispersive propagation}\label{sec:dispersion}

As stated in \cref{sec:slh}, two of the underlying assumptions behind the SLH framework is that the fields connecting localized components propagate in a dispersionless medium, and that the time for propagation is negligible.  

The the finite propagation delay part of the assumption is treated in  \cref{subsec:prop_delay}, but here we discuss dispersion. First we note that it is not a problem for the localized components to introduce dispersion; that is captured by IOT and the SLH framework. The SLH assumption is that there is negligible dispersion while \emph{propagating} between localized components, and while this assumption  is valid in free space, bulk optics setups, it can be violated in integrated implementations where guiding media can be dispersive. For example, silicon photonic waveguides can exhibit waveguide dispersion and material dispersion. The former is present if the waveguide's guiding properties depend on the light wavelength, and the latter arises from dependence of the material's refractive index on the wavelength. Both types of dispersion can be minimized by waveguide engineering, \eg \cite{Turner:2006gy, Zhang:10, Zhang:2012hz}, however, removing all dispersion can be challenging. 

Stace \etal have noted that dispersion can cause significant modifications to input-output theory, and have assessed the impact of this on quantum state transfer protocols \cite{Sta.Bar.etal-2004}. We revisit their analysis to understand the effects of dispersion on the dynamics of QIONs. First, let us return to the derivation of the dynamics of a cascaded network in  \cref{sec:CascadeSys}, and the Hamiltonian for the cascaded cavity example, \cref{eq:coupled_cavities_fullham}. This Hamiltonian is in the Markov approximation. For our purposes, let us write the interaction part of the Hamiltonian without this approximation:
\begin{align}
H_{\rm int, cascaded} =
i \int_{-\infty}^{\infty} d\omega \kappa_1(\omega) \left[ c_1 b\dg(\omega, x_0) - c_1\dg b(\omega, x_0)\right] + i \int_{-\infty}^{\infty} d\omega \kappa_2(\omega) \left[ c_2 b\dg(\omega, x_1) - c_2\dg b(\omega, x_1)\right]. 
\end{align}
Recall $c_i$ are operators acting cavity $i$ degrees of freedom. In addition to relaxing the Markov approximation (under which $\kappa_i(\omega) \rightarrow \sqrt{\gamma_i/2\pi}$), we have also been more explicit about the fields that the cavities interact with: cavity 1 (2) interacts with the field at location $x_0~ (x_1)$. In the dispersionless propagation case, $b(\omega, x_1) = b(\omega, x_0)e^{i \omega \tau}$, where $\tau = (x_1-x_0)/v$ and $v$ is the speed of propagation in the medium, and thus we could omit the spatial index. However, now that we are considering dispersion, we must be more careful and therefore explicitly denote the field's spatial index.

To understand the effect of dispersive propagation, we will express $b(\omega,x_1)$ in terms of $b(\omega, x_0)$ assuming quadratic dispersion, $\omega(k) = vk + \alpha k^2$, where $v$ is the speed of propagation in the medium and $\alpha$ is a constant. Inverting this relation we get: $k(\omega) = \frac{1}{2\alpha}\left ( -v + \sqrt{v^2 +4\alpha \omega }\right )$, and the group velocity of waves under this dispersion relation is given by
\begin{align} 
v_g = \left [\frac{d}{d\omega}k(\omega) \Big \vert_{\omega= \omega_c} \right ]^{-1}=  \sqrt{v^2 +4\alpha \omega_c }
\end{align}
where $\omega_c$ is the carrier frequency. Under quadratic dispersion, Stace has established that the effects of dispersion on the propagated field in $(t,x)$-space can be modeled by a delay and convolution with a \emph{channel transfer function} \cite{Sta.Bar.etal-2004, Stace:CA_LZXAD}:
\begin{align}
b(t, x_1) = H_L(t) * b(t-\tau, x_0), 
\end{align}
where $*$ denotes convolution and $L=x_1-x_0$. The transfer function $H_L(t)$ takes the form of a complex Gaussian \cite{Stace:CA_LZXAD}: 
\begin{align}
H_L(t) = \frac{v}{\sqrt{i4\pi\alpha \tau}}e^{\frac{it^2 v^2}{4\alpha \tau}}
\end{align}
Therefore performing a Fourier transform in time, we arrive at the following interaction Hamiltonian for the cascaded system under quadratic dispersive propagation between the two cavities:
\begin{align}
H_{\rm int, cascaded} =
i \int_{-\infty}^{\infty} d\omega \kappa_1(\omega) \left[ c_1 b\dg(\omega) - c_1\dg b(\omega)\right] + i \int_{-\infty}^{\infty} d\omega \kappa_2(\omega) \left[ H^*_L(\omega) c_2 b\dg(\omega)e^{-i\omega \tau} - H_L(\omega)c_2\dg b(\omega)e^{i \omega \tau}\right],
\end{align}
where as before, $b(\omega) \equiv b(\omega, x_0)$. Here we see that dispersion induces an $\omega$-dependent modulation of the interaction with the second cavity. Critically, this makes the Markov approximation of this interaction invalid because even if the physical interaction strength, $\kappa_2(\omega)$, is slowly varying across the frequencies of interest, the channel transfer function does not need to be. In fact, by noting that $H_L(\omega) \propto e^{i \omega^2/ \sigma^2}$, where $\sigma^2 \equiv \frac{v^2}{4\alpha \tau}$, we see that one would require one or a combination of, $v \rightarrow \infty, \alpha \rightarrow 0$ or $\tau\rightarrow 0$, for the Markov approximation to be feasible. All these conditions describe a dispersionless channel.

The above argument illustrates the fundamental incompatibility between dispersive propagation and the Markov approximation that forms one of the foundations of the standard SLH framework. More generally, it highlights the incompatibility between distributed transformations of propagating fields and the SLH framework, which assumes that all fields propagate freely apart from localized interactions with network components. In principle, it is possible to model distributed transformation using a large number of SLH components (or even a continuum), as discussed in  \cref{sec:continuum}. While non-Markovian dynamics can be captured through embedding in a larger Markovian model.

In this spirit, Stace and Wiseman have shown that quadratic dispersion can be captured using fictitious localized components that mimic the effect of dispersion on the propagating field \cite{StacWise06}. 
The field propagates freely between the localized components, and the relationship between the input and output fields of the fictitious localized components approximates dispersion of the input field by a dispersive waveguide of fixed length. For example, consider again the quadratic dispersion case. Stace and Wiseman show that propagation in a waveguide of length $L$ with this dispersion relation is approximated by assuming free propagation and inserting a fictitious cavity between the components connected by the waveguide described by the SLH triple
\begin{align} 
&G_{\rm dispersion} = (e^{i\phi},\sqrt{\gamma_d} a,\omega_d a\dg a),
\end{align}
and choosing
\begin{subequations}
\begin{align} 
 \gamma_d &= \sqrt{12} \Delta_d \\ 
\omega_d &= \omega_c -\Delta_d \\
\Delta_d &= \sqrt{\frac{\sqrt{3}v_g^2}{8\alpha \tau_p}}.
\end{align}
\end{subequations}
Here, $\tau_p = L/v_g$ is the propagation time over a length $L$,  $\tau_d = L^2/\alpha$ is the time for a pulse to disperse over the length $L$. The scattered field that arrives at the second component approximates a field that would have propagated along the original dispersive waveguide, provided $ \gamma_d , \Delta_d \ll \omega_c$  and $ \tau_p\ll \tau_d$. The phase shift imparted by the cavity, $\phi$ is fixed by the $\omega$ independent phase shift imparted by the dispersive medium, as explained in Ref. \cite{StacWise06}. Higher order dispersion could be modeled by adding more fictitious components.

Such approximate treatment of dispersive propagation is compatible with the SLH framework. However, this approach has limitations that are discussed in Ref. \cite{StacWise06}. Most seriously it is not valid in the regime where feedback loops between components exist in the network. Therefore, to date, there is no extension of the SLH framework that is fully compatible with arbitrary dispersive propagation. {An approach that might yield a solution to this, is to account for dispersion by explicitly modeling the medium causing the dispersion using a combination of SLH components, \eg see \cref{sec:continuum}. Alternatively one can directly solve the nonlinear Schr\"odinger equation as explained in supplementary material in Ref. \cite{KielCornWise11}, without explicit use of the SLH framework however. Nevertheless the results of such a procedure can then be used in SLH type calculations.}

\subsection{Time delayed field propagation}\label{subsec:prop_delay}

A key assumption made in the development of cascaded systems and the SLH framework is that of negligible time delay for propagation of the itinerant fields between network components (negligible at the timescales of the component dynamics). For example, this assumption was used explicitly in going from \cref{eq:coupled_cavities_langevin_finite_tau} to \cref{eq:coupled_cavities_langevin} in  \cref{sec:CascadeSys}. This negligible (or zero) time delay assumption becomes questionable in physically large quantum networks, \eg a network where the propagation time between components is comparable to the dynamical timescales within each component, or when there are significant delays in a coherent feedback loop, see \eg \cite{DornZoll02,NemePark16,KrafHeinLehn16}. 

\begin{figure}[]
\includegraphics[width=\columnwidth]{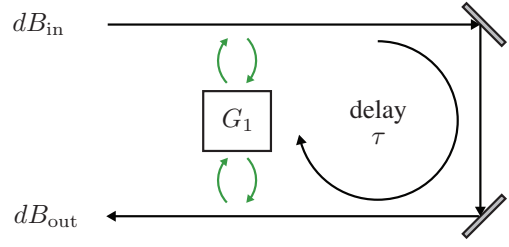}
\caption{Schematic of a system (\eg an atom) experiencing its own output field after a significant delay $\tau$.\label{fig:FeedbackDelay}}
\end{figure}

Consider two cascaded network components $G_1$ and $G_2$ that are a distance $L$ apart, this means the output of the first component arrives at component two delayed by $\tau =L/v$ seconds later, where $v$ is the speed of field propagation in the medium connecting the two components. The input field to component 1, $dB_{\rm in}(1,t)$, is transformed to an output field $dB_{\rm out}(1,t)= S_1(t)dB_{\rm in}(1,t)+L_1(t)dt$. This output arrives at component 2 with a time delay and acquires a phase proportional to the delay: $dB_{\rm in}(2,t)= e^{-i \Delta_c\tau}dB_{\rm out}(1,t-\tau)$. Hence the output field from component 2 is 
\begin{align}
dB_{\rm out}(2,t) &= S_2(t)dB_{\rm in}(2,t) + L_2(t) dt \nn \\
&= L_2(t)dt + e^{-i\Delta_c \tau} \left [S_2(t)L_1(t-\tau) dt+ S_2(t)S_1(t-\tau)dB_{\rm in}(1,t-\tau)\right ]
\end{align}
In this simple unidirectional cascade, and assuming the relative phase shifts between all relevant spectral components are negligible, one can absorb the phase factor in the retarded time for the second node \cite{Carm93,Gard93}, and the entire system can be described by Markovian QSDEs. { This is the approach taken in \cref{subsec:bidirectional} and \cref{ex:counterprop}. However, the situation is not so simple for time delays in more complex networks, \eg where the propagation time is long compared to the intrinsic dynamical timescales of the components, or networks with two way propagation of fields, or feedback loops with nonlinear components, see \cref{fig:FeedbackDelay} and \cite[Appendix 1]{KrafHeinLehn16}.} We now describe three recent attempts to model delays in more complex networks. Although none of these attempts represents a complete solution to modeling time delays, each tackles the problem with a different technical approach, and they are important advances towards overcoming this problem.

\subsubsection{Approach 1: introduce fictitious SLH components} 
Some intuition for the first approach, due to Tabak and Mabuchi, can be gained by realizing that time delays in propagation are a special case of dispersive propagation (time delay is linear dispersion). Therefore it is likely that one can find a method to model time delays similar to the approach taken by Stace and Wiseman for {approximately} modeling dispersive propagation in Ref. \cite{StacWise06} -- \ie by introducing fictitious SLH components, see  \cref{sec:dispersion} for details. Tabak and Mabuchi develop this intuition and provide a considerably more general solution for modeling time delays within the SLH framework \cite{TabaMabu16}. They begin by considering a large SLH network with a sub-network, possibly containing multiple components and multiple input and output ports, that induces a non-negligible time delay. In the case where this sub-network only contains linear, passive elements, Tabak and Mabuchi develop a method for constructing an effective model that approximates the dynamics and input-output behavior of the original sub-network within a frequency band of interest. Notably, this effective, fictitious model is fully compatible with SLH models, i.e. localized components interconnected by zero time-delay propagating fields. The Tabak and Mabuchi approach is possible because linear, passive sub-networks can be described by a transfer function $T(s)$, see  \cref{sec:linear} . The authors then approximate $T(s)$ by a phase factor and a finite product of poles and zeros (as the number of poles and zeros increases the approximation improves). Then this resulting approximate transfer function is realized using an SLH network of cavities with different resonance frequencies and linewidths. {We encourage the interested reader to look at section 4 of \cite{TabaMabu16} which contains a simple example consisting of a beamsplitter and a delay. Interestingly this can be show to be equivalent to a cavity \cite[section VII. B.]{Gough:2010in} }  {Two important practical issues to keep in mind when using this approach are (i) that the fictitious SLH network could have more components that the original network, and (ii) the fictitious SLH components will only provide an approximation to the time delayed dynamics, but the quality of this approximation can be improved by increasing the number of components in the fictitious network}.

\subsubsection{Approach 2: cascades from the past} 
The key insight in the approach develop by Grimsmo for incorporating time delayed propagation is that the state of the system plus field, in discrete time, can be represented using a structure similar to matrix product state (MPS) that he refers to as a super-operator product state \cite{Grim15} (also see the recent work by \citet{WhalGrimCarm16}). With this structure Grimsmo is able to show that the propagator for a network that includes time delayed feedback can be represented as a cascade of identical systems being driven by the output field of past systems. This approach defines a propagator for the entire system, which could be used to obtain QSDEs for system or field operators. Then one can trace out the auxiliary degrees of freedom (using an appropriately generalized notion of trace) to obtain a reduced state of the time delayed system. Although Grimsmo's approach is not integrated with the SLH framework for describing QIONs, it seems likely that such an integration is possible. In particular, the structure derived in the supplemental material of Ref. \cite{Grim15} is analogous to the linear fractional transformation used to derive the feedback reduction, see Rule \cref{SLHrule:feedback} in \cref{sec:slh}. Of course, the cost to modeling time-delays in this manner is that additional fictitious components must be included in the network, as in approach 1.
A numerical implementation of Grimsmo's approach is in the development branch of QuTiP~\cite{JohaNatiNori13,QuTiP_Team16}, see ~\cite{GrimsGITHUB}. As explained in Ref. \cite{TTM_GITHUB}, the long-time dynamics of QIONs with time delays can be approximated by using Grimsmo's technique in conjunction with the ``transfer tensor method'' introduced Ref.~\cite{CerrCao14}.

\begin{figure}[]
\includegraphics[width=\columnwidth]{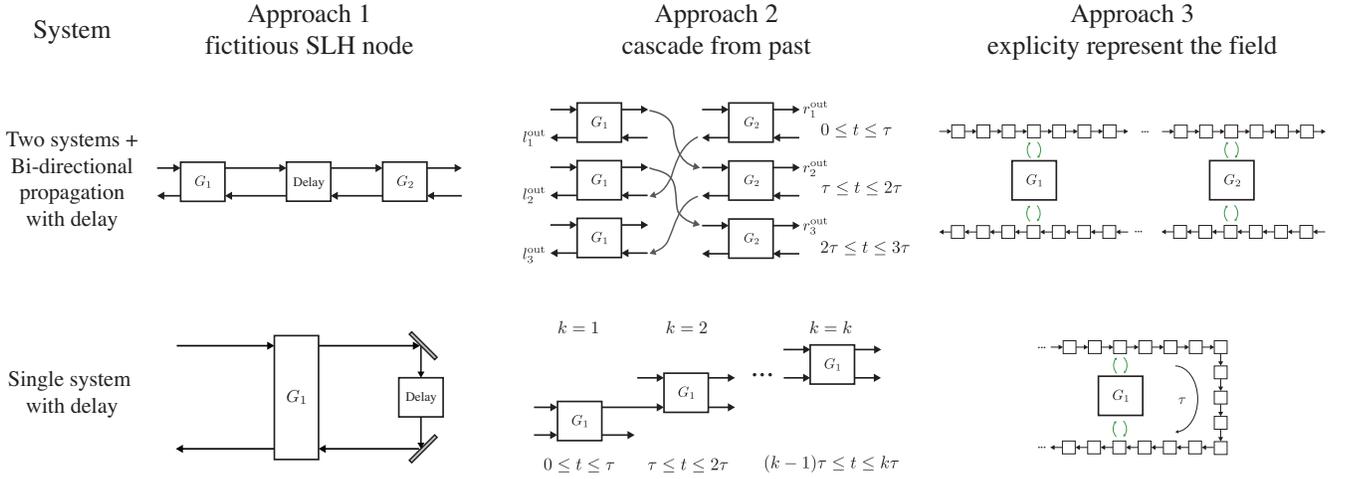}
\caption{ This figure explores the three approaches (the columns), explained in this section, to modeling systems with time delay.  The two example networks with delays are in the rows.  The first network (row 1) is depicted in \cref{fig:CounterProp} and represents two SLH systems with a finite delay between them and left and right propagating modes.  The second network (row 2) is depicted in \cref{fig:FeedbackDelay} and represents an SLH note experiancing its own output feedback with a finite delay.  In approach 1 (column 1) additional fictitious SLH components are introduced to model the delay. In approach 2 (column 2) the systems are driven by an cascaded version of their own outputs.  If one wants to simulate the network from time zero up to some integer multipule $k$ of the fundamental delay $\tau$, then $k$ cascades are required.  The first figure in this column was adapted from \cite{WhalGrimCarm16}.  In approach 3, one uses a discretized representation of the input, output and ``in-loop'' fields. This discrete representation is then simulated using tensor network methods. }\label{fig:MPSDelay}
\end{figure}

\subsubsection{Approach 3: explicit representation of the in-loop fields} 
Pichler and Zoller \cite{PichZoll16} use an MPS to explicitly and concisely represent the state of network components and fields. This is possible because the time bin modes in input-output theory can be related to spatial modes; recall that the field operator that interacts with the system at time $t$ is labeled $dB_{\rm in}(t)$, which could equally be written as a field mode that was originally distance $x$ from the origin, \ie $dB_{\rm in}(x/v)$, where the propagation speed is $v$ and $x = v t$. See \cref{fig:MPSDelay}. Further, the Markov nature of the system-field interaction restricts the amount of entanglement that can build between components and the propagating fields and thus enables an MPS description of the system. 
Because the wavefunction of the entire system is evolved in this approach, one can numerically determine expectation values or correlation functions of any subsystem, the in/out fields, and even in-loop fields (field propagating between network components, which are normally eliminated under an SLH treatment). 
Explicitly modeling all fields enables incorporation of arbitrary propagation time delays but one should note that this approach is somewhat counter to the QION and SLH framework philosophy of simplifying the description by eliminating intermediary fields from the model. However, one could use SLH to develop models for all the components in a large network and then use the Pichler and Zoller method to directly simulate system dynamics, including delays. This approach will clearly become infeasible as the size of the network grows because although the MPS description is concise, the number of degrees of freedom becomes intractable quickly. A related approach that uses MPS to explicitly represent the system and field in real space is used by Sanchez-Burillo \etal to solve a scattering problem \cite{SancZuecGarc14}.

\subsection{Software for automated modeling}
The modular description of networked quantum systems enabled by the SLH framework naturally opens up the possibility of modeling large networks using automated tools. Motivated by VHDL, a hardware description language used in electronic circuit design automation, Tezak \etal have developed the Quantum Hardware Description Language (QHDL) \cite{TezaNiedPavl12}, and associated design and analysis tools collected in the QNET package \cite{Anonymous:eYufDrLY}. Like VHDL, QHDL provides a syntax and language to describe QIONs in a standardized manner. This lays the foundation for automated tools for calculating SLH triples for arbitrary QIONs described via a text file or using a graphical layout of components and interconnections. This in turn enables hierarchical modeling of large networks; once the SLH models for base components are specified, these can be interconnected in arbitrary ways and QNET will derive the SLH triple description of the resulting network. Also, SLH triples for a library of commonly used network components are predefined in QNET. Finally, there is a suite of expanding tools for analysis of these networks, including: numerical simulation of master equations resulting from SLH triples, symbolic analysis and manipulation of input-output relations associated to an SLH triple, and automated layout tools to visualize QIONs as circuits. 

QHDL and QNET have been used to model and analyze complex QIONs, including optical circuits implementing quantum memories by autonomous subsystem quantum error correction \cite{KercNurdNurd10,KercPavlPavl11}, classical logic in large scale, low-power nanophotonic networks \cite{Sant14}, and to aid analysis of a coherent, non-linear superconducting microwave experiment \cite{Kerc12}.

\section{Integrated implementations of QIONs}\label{sec:challenges_integ_implem}

The notion of QIONs and the SLH framework were historically developed in the context of free-space propagating fields connecting bulk optical systems, where the physical assumptions listed in \cref{sec:slh} are generally valid. In order to build large scale QIONs one will inevitably have to turn to integrated technologies, where a huge number of components can be fabricated and networked together. Two promising integrated platforms for fabricating large-scale quantum coherent networks are silicon photonics and superconducting integrated circuits. Quantum coherent structures and high quality waveguides are routinely engineered on all of these platforms. However a number of issues arise when we consider modeling integrated coherent quantum networks on these platforms, and these make a direct application of the SLH framework to integrated systems non-trivial. In general terms, these issues are:
\begin{enumerate}
\item Integrated components can be significantly more lossy than bulk or free optical counterparts, frustrating coherent operation and quantum effects. 
\item A localized description of interactions, captured by SLH components, may not be accurate for some integrated circuits. Examples of this are material nonlinearity and waveguide dispersion, which manifest themselves as distributed properties of a waveguide.
\item It remains a technical challenge to fabricate high quality integrated circulators in both silicon photonics \cite{JalaPetrEich13} and superconducting electrical circuit technology \cite{DevoScho13}, which are often very useful to many QION implementations. 
\end{enumerate}
Some of these issues are partially addressed by the extensions to the SLH framework  discussed in  \cref{sec:generalizations}, but not all. In the following, we will discuss in more detail specific issues related to porting the SLH framework to silicon photonics and superconducting circuits.

\subsection{Integrated quantum coherent networks in silicon photonics}\label{sec:silicon}
The advantages and challenges to constructing QIONs in silicon photonics are discussed in detail in Ref. \cite{SaroSohCox16}. We briefly summarize this discussion here, and refer the reader Ref. \cite{SaroSohCox16} for more details.

Integrated photonics implementations of QIONs using silicon and silicon nitride at telecommunications wavelengths are particularly interesting because of the CMOS compatibility and relative maturity of integrated photonics on these platforms. A wide variety of linear optical elements are routinely fabricated on this platform, and there is an active research effort to produce low loss nonlinear components. The primary challenges to porting the SLH framework to this platform stem from the need to capture the range of optical phenomena resulting from electromagnetic field propagation in a nonlinear, dispersive medium. The dominant physical phenomena present in silicon and silicon nitride integrated photonics at 1550~nm, and absent in bulk-optics networks are (i) dispersion, (ii) scattering by the medium, including surface roughness scattering as well as Raman and Brillouin scattering, (iii) two-photon absorption and subsequent free carrier generation and heating in the medium. In the following we discuss each of these in turn.

Dispersion needs to be taken into account both in resonant structures (\eg cavities) and waveguides. In the former, it is largely an experimental design issue since it complicates phase matching, which subsequently makes the design of nonlinear elements such as OPOs difficult \cite{DuttLukeMani15}. Resonant structures must be engineered to have required phase matching properties and also be resonant for frequencies of the modes participating in the desired four-wave mixing process.  As long as the design of these elements accounts for dispersion, an SLH representation of these components is valid. For waveguides however, dispersion manifests itself as the dependence of the propagation velocity on the wavelength. As discussed in  \cref{sec:dispersion} this is incompatible with the assumptions of the SLH framework since it can invalidate the Markov approximation. Therefore, SLH modeling of QIONs implemented in integrated photonics will require engineered waveguides with minimal dispersion, that can be modeled by the perturbative approach covered in \cref{sec:dispersion}.

Surface roughness scattering leads to conversion of photons from modes of interest into other modes. This can be phenomenologically modeled as a linear loss mechanism that can be incorporated into the SLH description by the introduction of a fictitious beamsplitter for losses on waveguides, or the introduction of a fictitious input port with vacuum input for resonant structures. Nonlinear scattering phenomena such as Raman and Brillouin scattering are more difficult to incorporate due to the dependency of the loss coefficient on field intensity. Since the underlying scattering mechanisms ultimately arise from interactions with crystal phonons, they can be modeled fully quantum mechanically \cite[Secs. 6.4.1, 11.6]{DrumHill14}. As these models show, such scattering produces incoherent loss or gain of population in the modes of interest, as well as phase decoherence. Most significantly for the SLH framework, only in some special situations can these phenomena be modeled by a coupling to a Markovian reservoir \cite{DrumHill14}, which means that in most cases the effects of these nonlinear scattering processes cannot be modeled within the standard SLH framework. Accurately incorporating these nonlinear scattering processes within the SLH framework is an avenue for future work. 

Aside from these nonlinear scattering processes, the dominant nonlinear optical process of concern in silicon is two-photon absorption (TPA). At the optical powers typically circulating in coherent quantum networks, this nonlinearity is too weak to invalidate the assumption of linear propagation on waveguides \cite{SaroSohCox16}. However, in resonant structures, \eg ring resonator cavities, amplified field amplitudes can effectively enhance the nonlinearity. In such resonant structures the primary effect of TPA is to induce nonlinear (intensity dependent) loss. However, it also has secondary effects due to the associated creation of free carriers, whose concentration affects the refractive index of the material, which in turn changes its nonlinear and guiding properties (and causes dispersion if this change in refractive index is wavelength dependent). A natural approach to incorporating these effects within the SLH modeling framework is to apply input-output models for bulk-optical nonlinearities developed in quantum optics, \eg Refs. \cite{CollWall85,CollLevi91,LeviCollWall93}, and extend these to model integrated nonlinear processes using the approach of modeling distributed transformations detailed in \cref{sec:continuum}. However, this direct approach usually leads to SLH models of very large state space dimension that are difficult to simulate, and hence methods for alleviating this burden are required. 
Some noteworthy progress has recently been made in this direction through the formulation of a quantum model for free carrier dispersion in nanophotonic cavities \cite{HameMabu15}. This model is compatible with the SLH framework, but it is computationally difficult to directly simulate and analyze due to the large number of degrees of freedom (SLH components) that it introduces. As a result, Hamerly and Mabuchi adopt a semi-classical approximation of the dynamics to simulate and analyze their model. The approximate dynamical equations that Hamerly and Mabuchi derive from their SLH model represent a promising approach to simulation of large-scale quantum networks, and it would be fruitful to explore the full range of validity of the approximations used in Ref. \cite{HameMabu15}.

\subsection{SLH and superconducting microwave systems}\label{sec:SLH_microwave}

Superconducting microwave systems are another platform that show great promise for quantum engineering \cite{DevoScho13}.  Most, if not all, quantum optical components discussed in this article have excellent superconducting microwave analogs.  For example, high-Q, microwave transmission line or lumped element LC resonators coupled to transmission lines have very similar internal, input, and output dynamics to optical cavities coupled to guided wave, itinerant modes \cite{Devo97}\cite[Supplementary Information]{ClerDevoGirv10}.  Directional couplers and microwave hybrids act as asymmetric and symmetric optical beamsplitters, respectively.  Even nonlinear optical components ranging from nonlinear crystals to single, two-level atoms may be well-approximated by superconducting microwave circuits employing nearly lossless Josephson junction elements as the fundamental nonlinearity \cite{BlaiHuanWall04}.

SLH models excel at describing the dynamics of superconducting microwave systems that employ broadband, linear scattering components like directional couplers, high-Q resonant components (either linear or nonlinear), and transmission line interconnections \cite{MotzWhalSaro15,MotzHalpWang15,Kerc12,Kerc13}.  In these integrated systems, as in integrated photonics, one must take care to properly model inevitable back-reflections at the interfaces between components, as well as phase delays between components.  That being said, it can be easier to construct small scale, integrated superconducting microwave systems that do not suffer as much from transmission line dispersion, scattering, heating, and loss as integrated quantum photonics.  

Unfortunately, SLH models are only relevant to a very particular (albeit important) subset of superconducting microwave networks.  There are a number of important integrated, modular, coherent, quantum networks that are simply not expressible using IOT, let alone SLH.  Essential approximations, such as the Markov approximation and the assumption that components couple via asymptotically free fields, frequently break down in microwave circuits.  This is in part because microwave networks (operating at $\sim$100 MHz-100 THz frequency ranges) frequently bridge the gap between near- and far-field limits (\ie they contain feature sizes that range from $\mu$m to several cm).  For example, a lumped element capacitor whose leads are connected to a transmission line is not expressible in IOT.  For this circuit, one cannot approximate the interaction between the electric dipoles across the capacitor and the transmission line as narrowband (Markov approximation).  This interaction occurs at all frequencies, from DC up to what ever frequency the lumped element model breaks down.  Similarly, SLH is not the appropriate way to model the network of two, inductively coupled LC resonators, as these components are in each others' near field.  Increasing the physical separation between these two resonators causes the coupling strength to drop with separtation.  In contrast, two LC resonators coupled to a lossless transmission line (an SLH-compatible network) each couple to the transmission line with a fixed magnitude and phase that does not vary as the distance between the two resonators varies.  In this case, each entire LC resonator admits a modular description, with equations of motion for internal variables and fixed input-output relations that are constant regardless of whatever else is connected to the transmission line.  However, in the case of the inductively coupled, near-field resonators, the equations of motion for each LC resonator's variables vary with physical separation.  

Thus, while SLH models can be useful for modeling a wide range of superconducting microwave quantum networks, their applicability is limited, especially in general networks that operate in the lumped element, near-field limit.  Of course, the lumped element approximation of an electrical network is itself a modular modeling approach (\eg the equation of motion for charge across and current flux through a capacitor are constant regardless of whatever else is connected to the capacitor leads).  For this reason, since the 1980s, several authors have developed quite general methods for deriving the quantum dynamics of general, lumped element electrical networks \cite{YurkDenk84,Devo97,Burk04,WendShum05,Solg15}, typically for superconducting microwave applications.  Because these lumped element models are most applicable at a ``lower level'' in the hardware description of microwave networks than IOT and SLH (\eg considering each inductor and capacitor to be separate ``modules'', rather than identifying LC-resonator-type modules), such approaches tend to sacrifice modeling simplicity for accuracy.  While some connections have been made over the years \cite{YurkDenk84,BlaiHuanWall04,ClerDevoGirv10,Nigg12}, there is still much work to be done to smoothly bridge between these modeling regimes.  Quantum superconducting microwave circuits exist naturally in this intermediate regime, and will provide an excellent context to develop a more complete set of modeling techniques for quantum electromagnetic networks for years to come.

\section{Outlook}
\label{sec:outlook}

The theory and practice of QIONs have attracted steady and growing interest over several decades, starting with the development of IOT in the 1980s, then cascaded models in the 1990s, the development of the SLH framework in the 2000s, and extensions of SLH in the 2010s.  The popularity of the framework may be attributed to the conceptual clarity and computational simplifications that come from the modular approach of networking many quantum components (such as resonators, atoms or atom-like defects)  via asymptotically free fields.  We have attempted to summarize the development of QION up through the development of the SLH framework and some of its recent extensions.  Our aim is to attract a broader audience to \emph{apply} QION concepts in their own work.

While the theory of QIONs has become quite well-developed, and IOT has proven to be a perennially popular framework for analyzing quantum optical (and increasingly quantum microwave) experiments since the 1990s, fostering more experimental application of QION models to complement and guide the theoretical developments is perhaps the most pressing need for the field.  Indeed, many of the extensions to SLH considered in the past decade have focused on relaxing various assumptions in the original formulation, such as dispersionless waveguides and zero time delay propagation, to better reflect experimental reality, or developing methods for including common experimental non-idealities, such as component back reflections.  Similarly, the development of automated software tools like QNET for SLH modeling should also foster adoption of these techniques by the applied physics and engineering communities. Looking ahead, we identify three key areas of development for the SLH modeling framework:
\begin{enumerate}
\item Incorporation of non-Markovian coupling between localized components and propagating fields. As QION implementations migrate from free-space optics to the solid-state, as discussed in  \cref{sec:challenges_integ_implem}, many physical effects (\eg weak dispersion) will manifest as a non-Markovian coupling between localized components and input-output fields. Therefore relaxing the Markov approximation will be a critical need in such scenarios. {Progress in experiments has lead to renewed interest in this issue, with some recent works, by \citet{Dios12},  Zhang \etal \cite{ZhanLiuWu13} and Gough~\cite{Gough:2016ug}. The proposal by Zhang \etal \cite{ZhanLiuWu13} is interesting because they re-derive the cascade and concatenation product for non-Markovian systems. An SLH compatible approach is that taken by Xue \etal~\cite{XueHushPete16,XueNguyJame17}. These recent studies have built on the older works by Imamo\={g}lu \etal~\cite{Imam94,StenImam96}, Jack \etal \cite{JackCollWall99,JackColl00} and the pseudomodes approach \cite{DaltBarnGarr01,MazzManiPiil09} which represent first forays into non-Markovain input-output theory.  }

\item Development of analytic or numerical techniques or approximations for simulating the dynamics or steady-state properties of large-scale QIONs. Brute-force simulation of QION dynamics becomes intractable as the number of components in the networks becomes large. Therefore, useful approximation techniques, applied at the network level or at the level of approximating the state of the nodes of the network, are desirable to reduce the simulation burden and enable numerical analysis of large-scale QIONs. In \cref{sec:elimination} we explained one model reduction technique, namely adiabatic elimination. Some other techniques that have recently been explored are: generalized Schrieffer-Wolff transformations for dissipative systems\cite{Kess12}, semi-classical approximation of the Wigner function description of QION dynamics \cite{Sant14},  the kernel function approximation for nonlinear input-output models~\cite{ZhanLiuWu14}, a quasi-principal components approach to model reduction for nonlinear cavity degrees of freedom~\cite{ShiNurd16}, {and a method for unitary model transformation to represent systems using low dimensional manifolds ~\cite{TezaAminMabu17}.} Another promising direction is to use techniques from matrix product state and tensor network literature \cite{BridChub16,Orus14} to represent and simulate large QIONs; recent work aimed at modeling time-delays in QIONs (covered in \cref{subsec:prop_delay}) represent initial steps in this direction. Development of such approximations and a good understanding of their regimes of applicability will be a critical need for scaling up SLH-based analysis.
 
 \item The design and synthesis of nonlinear QIONs and feedback controllers. The analysis of nonlinear QIONs -- with or without model reduction -- is a unique strength of the SLH framework, but few tools exist within the framework for design of such networks (in contrast to \emph{linear} QIONs, for which, as discussed in \cref{sec:linear_review},  there exist several tools for analysis, design and synthesis). One would ideally like an algorithm or tool that accepts as input equations of motion specifying the behavior of a component, and produces a realizable QION built from a library of pre-specified components that approximates this behavior; something similar in spirit to synthesizing large unitary transformations out of a universal gate set in quantum computation \cite{mikeandike}.
Design and synthesis of non-linear systems is a notoriously difficult problem, so we expect that progress on this front will be difficult. However, any such progress will have high impact since most applications on QIONs require some nonlinear behavior. 

\end{enumerate}
Progress in any of these directions has the potential to expand the scope and applicability of QIONs.

On a more general note, we observe that many of the near term theoretical challenges for QIONs involve blurring the boundaries between what is and what is not a QION model.  The ideal ending point of this trend is analogous to the coexistence and frequent hybridization of lumped- and distributed-element component and network models in practical classical electronics. What might constitute a clear advance for QION modeling, for example, is a well-defined and easily applied method for systematically relaxing the zero time delay approximation with first, second, to $n$th order corrections that connect canonical SLH models to QION models where time delay is fully modeled. Recent attempts at extending the SLH framework to incorporate time delayed propagation and relax the Markov approximation represent some progress in this direction.

We hope that this review demonstrates that the QION is a concept for thinking about networked quantum systems. In addition to inherent modularity in representation this concept is compatible with many control theoretic and systems analysis tools, which we have attempted to survey. With increasing engagement from the applied physics, engineering and applied mathematics communities, we believe the maturity and applicability of the QION concept will only increase.

\begin{acknowledgments}
We gratefully acknowledge useful exchanges, on several topics, with (in alphabetical order): Ben Baragiola, Luc Bouten, Dainel Brod, Rob Cook, John Gough, Arne Grimsmo, Michael Hush, Matt James, Ekin Kocaba\c{s}, Anton Kockum, Clemens M\"uller, Hendra Nurdin, Ian Petersen, Daniel Soh, Tom Stace, Jason Twamley, Alkabtin Wadie, Simon Whalen, Howard Wiseman, Matt Woolley, Shibei Xue, and Guofeng Zhang. Special thanks goes to Ben and Rob for {\em many} useful suggestions, Jason and Ben for stressing the importance of adding \cref{SLHremark:vec_elim}, and Clemens for suggesting and then correcting  \cref{SLHremark:elimination}.
JC was supported in part by Perimeter Institute for Theoretical Physics and the Australian Research
Council through a Discovery Early Career Researcher Award (DE160100356) and via the Centre of Excellence in Engineered Quantum Systems (EQuS), project number CE110001013. Research at Perimeter Institute is supported by the Government of Canada through the Department of Innovation, Science and Economic Development and by the Province of Ontario through the Ministry of Research, Innovation and Science. 
MS was supported by the Laboratory Directed Research and Development program at Sandia National Laboratories.  Sandia National Laboratories is a multi-mission laboratory managed and operated by Sandia Corporation, a wholly owned subsidiary of Lockheed Martin Corporation, for the U.S. Department of Energy's National Nuclear Security Administration under contract DE-AC04-94AL85000.

\end{acknowledgments}

\appendix

\section{SLH representation of some basic components}

In this appendix we list SLH triples for some commonly encountered network components.

\begin{enumerate}
\item \textbf{Phase shifter:} A phase shifter has a single input and output field. SLH triple:
\begin{align}
\left( e^{i\phi}, 0, 0\right)
\label{slh:phaseshift}
\end{align}
where $\phi$ is the phase shift angle.

\item \textbf{Beam splitter:} This is a system with two inputs and two outputs. If we choose the convention that the reflected fields are the output pairs to the input fields, then the SLH triple for a beam splitter is:
\begin{align}
\left( \left[\begin{array}{cc}r_{11} & t_{12} \\ t_{21} & r_{22}\end{array}\right], 0, 0\right)
\label{slh:beamsplitter}
\end{align}
The entries of the scattering matrix must satisfy constraints stemming from the unitarity; \ie $S\dg S = I$.

\item \textbf{Coherent drive:} This element displaces the input state in the phase plane by $\alpha$. SLH triple:
\begin{align}
\left( 1, \alpha(t), 0 \right)
\label{slh:drive}
\end{align}

\item \textbf{One-sided cavity:} A perfectly reflecting mirror and a partially transparent mirror with photon decay rate $\kappa$. The primary quantized mode within the cavity has frequency detuning $\Delta_c$ and annihilation operator $a$. SLH triple:
\begin{align}
\left( I, \sqrt{\kappa}a, \Delta_c a\dg a \right)
\label{slh:onesided_cavity}
\end{align}

{
\item \textbf{One-sided cavity with a Kerr nonlinearity:} We take the previous model and add the Kerr nonlinearity to the cavity Hamiltonian $\chi a\dg a a\dg a$
\begin{align}
\left( I, \sqrt{\kappa}a, \Delta_c a\dg a+ \chi a\dg a a\dg a \right)
\label{slh:onesided_cavity_kerr}
\end{align}
}
\item \textbf{Fabry-Perot cavity:} A two-sided cavity with two partially transparent mirrors. SLH triple:
\begin{align}
\left( 
\mathbf{I}^I_2, \left[\begin{array}{cc} \sqrt{\kappa_1} a \\ \sqrt{\kappa_2} a \end{array}\right], \Delta_c a\dg a \right),
\label{slh:twosided_cavity}
\end{align}
where $\kappa_i$ are the photon decay rates of the two mirrors.

{
\item \textbf{Crossed cavites (or a two mode cavity) with a cross Kerr nonlinearity:}
Consider two one sided cavites with frequencies that are coupled via a cross Kerr nonlinearity $\chi   a_1\dg a_1 a_2\dg a_2$. The SLH parameterization for this system is 
\begin{align}
\left(
\mathbf{I}^I_2,\left[\begin{array}{c}\sqrt{\kappa_1}a_1\\\sqrt{\kappa_2}a_2\end{array}\right],\Delta_1 a_1\dg a_1+\Delta_2 a_2\dg a_2+\chi   a_1\dg a_1 a_2\dg a_2\right).
\end{align}
}

{
\item \textbf{Degenerate OPO:} A finite bandwith degenerate optical parametric oscillator (OPO) is modeled by a cavity model with SLH triple:
\begin{align}
\left (I, \sqrt{\kappa}a, \frac{i}{2}(E {a\dg}^2 - E^* a^2) \right),
\end{align}
where $a$ is the cavity mode, $\kappa$ is the bandwidth of the squeezed light, and  $E$ parameterizes the OPO nonlinearity.
}

{
\item \textbf{Two mode squeezing via a $\chi^{(2)}$ nonlinearity:} with a classical and undepleated pump approximation we have the SLH triple:
\begin{align}
\left (\mathbf{I}^I_2,  \left [\begin{array}{c} \sqrt{\kappa_1} a_1\\  \sqrt{\kappa_2} a_2 \end{array} \right], \frac{i}{2}( \epsilon e^{-i\Delta_p t}a_1\dg a_2\dg - \epsilon^*e^{i\Delta_p t} a_1a_2) \right),
\end{align}
where $\epsilon$  is the intensity of the classical pump i.e. the $\chi^{(2)}$ nonlinearity, and $\Delta_p$ is the pump frequency. By transforming to a rotating frame at half the pump frequency (i.e. $a_i \mapsto a_i e^{i \Delta_p t /2}$ we obtain
\begin{align}
\left (\mathbf{I}^I_2,  \left [\begin{array}{c} \sqrt{\kappa_1} a_1\\  \sqrt{\kappa_2} a_2 \end{array} \right], \frac{i}{2}( \epsilon a_1\dg a_2\dg - \epsilon^* a_1a_2) \right).
\end{align}

}
\item \textbf{Optomechanical system with a radiation pressure coupling:} 
{
a single sided optical cavity with a mechanically compliant mirror, the input field probes the cavity. SLH triple is:
\begin{align}
\left( \mathbf{I}^I_3, 
\left[\begin{array}{c} \sqrt{\kappa} a  \\ \sqrt{\Gamma(\bar n+1)}b \\ \sqrt{\Gamma \bar n} b\dg \end{array}\right]
, \Delta_{c} a\dg a +\Delta_{m} b\dg b - g a\dg a ( b\dg+ b )\right),
\label{slh:optomec_sys}
\end{align}
where $\kappa$ is the photon decay rate of the cavity, $\Delta_c, \Delta_m$ are detunings from the carrier frequency to resonant frequencies of the cavity mode and mechanical mode, respectively, and $g$ is the coupling between the electromagnetic and mechanical modes. The parameter $\Gamma$ is the coupling rate of the thermal phonon bath to the mechanics and $\bar n$ is the occupation number of thermal phonons in the bath. Extensions of this model are discussed in detail in Refs.~\cite{MilbWool11,BoweMilb15}.
}

\item \textbf{Linearized optomechanical system:} 
{
Linearizing the previous model see section 4 of~\citet{MilbWool11}. We have the SLH triple:
\begin{align}
\left( \mathbf{I}^I_3, 
\left[\begin{array}{c} \sqrt{\kappa} a  \\ \sqrt{\Gamma(\bar n+1)}b \\ \sqrt{\Gamma \bar n} b\dg \end{array}\right]
, \Delta_{c} a\dg a +\Delta_{m} b\dg b + g( a\dg+ a )( b\dg+ b )\right),
\label{slh:optomec_sys}
\end{align}
where $\kappa$ is the photon decay rate of the cavity, $\Delta_c, \Delta_m$ are detunings from the carrier frequency to resonant frequencies of the cavity mode and mechanical mode, respectively, and $g$ is the coupling between the electromagnetic and mechanical modes. The parameter $\Gamma$ is the coupling rate of the thermal phonon bath to the mechanics and $\bar n$ is the occupation number of thermal phonons in the bath. In the optomechanics literature it is common to use a series of approximations, judicious choice of the detunings, and frame changes to turn this model into a cooling, heating and QND measurement interactions. Extensions of this model are discussed in detail in Refs.~\cite{MilbWool11,BoweMilb15,MaPeteWool17}.
}

\item \textbf{Two level atom side-coupled to waveguide:} A two-level atom coupled to the waveguide modes with strength $\sqrt{\kappa_g}$ and to non guided modes with strength $\sqrt{\kappa_\perp}$. If $\sqrt{\kappa_\perp}=0$ the atom is perfectly coupled to the waveguide. SLH triple:
\begin{align}
\left( 
\mathbf{I}^I_2, \left[\begin{array}{cc} \sqrt{\kappa_g} \sigma_- \\ \sqrt{\kappa_\perp} \sigma_- \end{array}\right], \half \Omega \sigma_z  \right)
\label{slh:twolevelatom}
\end{align}

{
\item \textbf{Two level atom in a harmonic potential side-coupled to waveguide:} We have coupling operators associated with right moving $\sqrt{\kappa_r} $ and left moving $\sqrt{\kappa_l} $ waveguide modes as well as emmision in to nonguided modes with strength $\sqrt{\kappa_\perp}$ where we neglect recoil into nonguided modes. As before if $\sqrt{\kappa_\perp}=0$ the atom is perfectly coupled to the waveguide. The position and momentum operators of the center of mass motion for the atom obey $[\hat x,\hat p]=i$. The SLH triple is
\begin{align}
\left( 
\mathbf{I}^I_3, \left[\begin{array}{ccc} \sqrt{\kappa_r} \sigma_- e^{+i k_0 \hat x}\\ \sqrt{\kappa_l} \sigma_- e^{-i k_0 \hat x}\\ \sqrt{\kappa_\perp} \sigma_-\phantom{e^{-i k_0 \hat x}} \end{array}\right], \half \Omega \sigma_z + \frac{\hat p^2}{2m}+\frac 1 2 m \nu^2 \hat x^2 \right),
\end{align}
where  $\pm k_0$ is the wave vector associated with propagation in the left or right direction, the mass of the atom is $m$ ,and the harmonic trap frequency is $\nu$ in a frame rotating at the optical frequency.
}

\item \textbf{Open system atom-cavity systems:} A single sided cavity coupled to a two level atom.The SLH triple for the first example, the Rabi model is:
\begin{align}
\left( I, \sqrt{\kappa} a , \Delta_{c} a\dg a +\half \Omega \sigma_z + g\sigma_x (a \dg+ a ) \right),
\label{slh:rabi}
\end{align}
where $\kappa$ is the cavity mirror transmitivity, $\Delta_c$ and $\Omega$ are the cavity mode frequency detuning and atom transition frequency, respectively, and $g$ is the atom-field coupling strength. In the rotating wave approximation, this model becomes the Jaynes-Cummings model, with SLH triple:
\begin{align}
\left( I, \sqrt{\kappa} a , \Delta_{c} a\dg a +\half \Omega \sigma_z + g (\sigma_- a\dg+\sigma_+ a ) \right).
\label{slh:JC}
\end{align} 
{The generalization to the Tavis-Cummings model is obvious
\begin{align}
\left( I, \sqrt{\kappa} a , \Delta_{c} a\dg a +\half \Omega J_z + g (J_- a\dg+J_+ a ) \right).
\label{slh:TC}
\end{align}
}

\item \textbf{Circulators:} A three port circulator has the scattering matricies
\begin{align}
S_{\rm ideal} = \left[\begin{array}{ccc}
0 & 0 & 1 \\
1 & 0 & 0 \\
0 & 1 & 0
\end{array}\right] \quad \quad
S_{\rm non\ ideal} = \left[\begin{array}{ccc}
r & b & t \\
t & r & b \\
b & t & r
\end{array}\right]
\end{align}
If the circulator is symmetric but not perfect we have $S_{13}=S_{21}=S_{32}= t$, $S_{11}=S_{22}=S_{33}= r$, 
and $S_{12}=S_{23}=S_{31}= b$ \cite{Hage69} with complex transmission, reflection, and isolation error coefficients $t$, $r$, and $b$, respectively. These coefficients 
must obey $|t|^2+|r|^2+|b|^2=1$ and $rt^*+tb^*+br^*=0$ as the S matrix is unitary \cite{Hage69}.  The non-idealities of the circulator are then captured by the parameters \cite{Ayas80}:
${\rm Reflection} = |r|^2$, ${\rm Isolation\,error} = |b|^2$ and clearly  $|t|\gg |r|,|b|$ is desirable.

The four port circulator is described by the scattering matricies
\begin{align}
S_{\rm ideal} = \left[\begin{array}{cccc}
0 & 0 & 0 & 1 \\
1 & 0 & 0 & 0 \\
0 & 1 & 0 & 0 \\
0 & 0 & 1 & 0
\end{array}\right] \quad \quad
S_{\rm non\ ideal} = 
 \left[\begin{array}{cccc}
r & b & c & t \\
t & r & b & c \\
c & t & r & b \\
b & c & t & r
\end{array}\right] 
\end{align}
The coefficients $r, b, c, \&\ t$ must obey the conditions for the matrix to be unitary. The generalization to $N$ ports is straight forward.
\end{enumerate}

\pagebreak

\bibliography{QION_review_refs}
\end{document}